\documentclass[12pt]{article} 
\usepackage{graphicx}
\def\ba{\begin{eqnarray}}
\def\ea{\end{eqnarray}}

\def\lb{\label}
\def\be{\begin{equation}}
\def\ee{\end{equation}}

\def\C{\mathbb C}
\begin{document}

\title{Another infinite tri-Sasaki family and marginal deformations}
\author{O. P. Santillan \thanks{firenzecita@hotmail.com}}
\date {}
\maketitle

\begin{abstract}
Several Einstein-Sasaki 7-metrics appearing in the
physical literature are fibered over four dimensional
Kahler-Einstein metrics. Instead we consider here the natural
Kahler-Einstein metrics defined over the twistor space Z of any
quaternion Kahler 4-space, together with the corresponding
Einstein-Sasaki metrics. We work out an explicit expression for
these metrics and we prove that they are indeed tri-Sasaki.
Moreover, we present an squashed version of them which is
of weak $G_2$ holonomy. We focus in examples with three commuting
Killing vectors and we extend them to
supergravity backgrounds with $T^3$ isometry, some of them with
$AdS_4\times X_7$ near horizon limit and some others without this
property. We would like to emphasize that there is an underlying
linear structure describing these spaces. We also consider the effect
of the $SL(2,R)$ solution
generating technique presented by Maldacena and Lunin to these
backgrounds and we find some rotating membrane configurations reproducing
the E-S logarithmic behaviour.

\end{abstract}

\tableofcontents

\section{Introduction}

A duality between quantum field theories and strings was
proposed by t' Hooft in \cite{thooft}, where it was noticed that
any Feynman diagram of an $U(N)$ gauge theory with matter fields can
be drawn over a two dimensional surface. Few of these diagrams
corresponds to a plane or to an sphere, higher diagrams are drawn
over a surface with non trivial genus, such as the torus. Any closed
line contribute with a factor $N$ to the amplitude. The partition
function results in an expansion in terms of the form
$c_g(\lambda)N^{2-2h}$ being $h$ the genus (or the number of holes)
of the two dimensional surface and $c_g(\lambda)$ certain function
of the t´ Hooft coupling $\lambda=g^2 N$. In the large $N$ limit and
keeping fixed $\lambda$ the powers $N^{2-2h}$ goes all to zero
except for diagrams with $h=0, 1$. For this reason this limit is
called the planar limit. If $\lambda<<1$ then $g<<1$, this
corresponds to the perturbative regime. On the other hand for
$\lambda>>1$ the amplitudes have been found to be a sum of terms of
the form $A_g(\lambda)g^{h-1}$. In the practise $N=3$ can be
considered as a large number of colours and this expansion
corresponds to non perturbative phenomena
\cite{thooft}-\cite{Witten1}. If in addition the variable $\lambda$
became large then more diagrams contribute and become dense in the
sphere. It was suggested \cite{thooft} that this diagrammatic
expansion is describing a discrete version of an string theory in
which $A_g$ is interpreted as a closed string amplitude. These
string theory is defined as the result of summing all the planar
diagrams.

This duality between strings and gauge theories was developed
further in \cite{Juanma} and leaded to the AdS/CFT correspondence.
As is well known, D-branes are solitonic objects in superstring
theory which admit a gauge/gravity low energy interpretation. This
is because the low energy dynamic of massless open string states on
a Dp brane is, at first order, a (p+1)-dimensional supersymmetric
gauge theory, and in the closed string channel a Dp brane is a
solution of the low energy supergravity solution in presence of a
$C_p$ Rammond-Rammond p-form. The Yang Mills coupling constant is
related to the string coupling by $g_{YM}\sim g_s l_s^{p-3}$. This
suggests that is possible to make gauge theory calculations from
supergravity solutions and this is indeed one of the motivations of
the AdS/CFT correspondence \cite{Juanma}. The original statement of
AdS/CFT is that ${\cal N}=4$ super-Yang Mills theory is dual to type
IIB strings in $AdS^5\times S^5$. In fact $SU(N)$ ${\cal N}=4$
super-Yang Mills is the field theory on $N$ D3 branes at low
energies, and the near horizon limit of these branes is $AdS^5\times
S^5$. In addition, for the specific value $p=3$, we have that
$g_{YM}\sim g_s$ and one can take the limit $l_s\to 0$ and trust in
the supergravity approximation. For ${\cal N}=4$ super-Yang Mills
theory the beta function vanish at all orders and is therefore
conformally invariant, which means that the coupling constant is not
renormalized. This is reflected in the $AdS_5$ factor of conjectured
dual. The theory is also scale invariant quantum mechanically. As a
consequence of the conformal symmetry the number of supercharges of
the super Yang Mills side is 32, the same than IIB superstrings in
$AdS^5\times S^5$. The supergroups of both theories are the same.
Besides, both sides contains two parameters. For the Yang-Mills
theory they are $g_{YM}$ and $N$ and for the superstring side they
are the string tension $R^2/\alpha$ and $g_s$. In addition to the
identification $g_{YM}\sim g_s$ the AdS/CFT conjecture implies that
the t ´Hooft coupling is given by $\lambda=R^4/\alpha^2$.

The gravity description of string theory, which occurs when the size
of the graviton is much less than the radius of the space,
corresponds to the limit $\lambda>> 1$. In these limit non
perturbative phenomena of the super Yang-Mills side can be analyzed.
This implies that the AdS/CFT correspondence relates the week
coupling limit of one of the theories to the strong coupling of the
other and viceversa, which makes it a powerful tool in order to
study strongly coupled regimes in gauge theories. A precise
statement of the correspondence was developed further in
\cite{Klebanov}, where it was stated that composite operators of the
form $O_{i_1...i_k}(x)=Tr(\phi_{i_1}..\phi_{i_1})$ mix and are
renormalized, therefore they acquire anomalous dimensions. These
dimensions are identified with the energy eigenstates of IIB strings
over $AdS_5\times S^5$.

    A more recent advance in understanding AdS/CFT is the BMN
correspondence, in which the anomalous dimensions of large
R-operators were related to the spectrum of string theory on the
pp-wave limit of $AdS_5\times S^5$ \cite{Nastase}. This idea was
refined in \cite{Gubser} by stating that gauge theory operators with
large spin are dual to strings rotating in the AdS space. The main
observation providing this identification is that, for strings
rotating in the AdS space, the difference between the energy and the
spin of the configuration depends logarithmically on the spin. This
logarithmic dependence is characteristic of the anomalous dimensions
of the twist operators of the gauge theory. There is evidence about
that the logarithmic dependence does not acquire corrections if we
go from the perturbative to the strong coupling regime in the t
'Hooft coupling \cite{Frolov}.

  Although ${\cal N}=4$ super-Yang Mills cannot be considered as a realistic theory,
the AdS/CFT correspondence could be an useful tool in
realistic calculations. This is because the finite temperature
version of ${\cal N}=4$ SYM has certain analogies with realistic
elementary particle models, although the zero temperature version
has not (see for instance \cite{Raja}). Also, the purely
gravitational aspects of this correspondence are related to a wide
variety of problems in differential geometry (see for instance
\cite{Anderson}). Nevertheless, it is of interest is to generalize
this duality to other type of theories. This could be for instance
theories with less number of supercharges than 32
\cite{Malnu}-\cite{Witcon}, or to consider non conformal
field theories duals \cite{Yankie}, such as the
Klebanov-Strassler ones \cite{Klebstras}. For conformal field
theories, the AdS/CFT correspondence has been generalized to the
holographic principle \cite{Klebanov}, in which is stated that
\emph{any} $AdS$ vacuum of string or $M$ theory define a conformal
field theory. In the case of $AdS_5\times X_5$, being $X_5$ an
Einstein manifold, the central charge of the conformal field theory
is, in the large $N$ limit, inversely proportional to the volume of
$X_5$ \cite{Gubser2}. The holographic principle permits to consider gauge/string
duals with less number of supersymmetries, and have been generalized
for eleven dimensional backgrounds of the form $AdS_4\times X_7$,
which are duals to three dimensional superconformal field theories
\cite{Klebanov}.

  The holographic principle renewed the interest in constructing 5
and 7-dimensional Einstein manifolds and in particular those
admitting at least one conformal Killing spinor. The number of such
spinors will be related to the number of supersymmetries of the
conformal field theory. This leads to consider weak $G_2$ holonomy
spaces, Einstein-Sasaki spaces and tri-Sasaki ones. Several examples
have been constructed for instance in \cite{Bilal}-\cite{Mateos8}
and there have been certain success for finding gauge/gravity duals
corresponding to these backgrounds \cite{Mateos9}-\cite{Mateos}.

   A new step for finding gauge/gravity duals with less number of
supersymmetries was achieved in \cite{Lunin}, where it was
considered a three parameter deformations of ${\cal N}=4$ super-Yang
Mills  superpotential that preserves ${\cal N}=1$ supersymmetry
\cite{Leigh} (see also \cite{Dorey}). These deformations are called
$\beta$ deformations. The original superpotential of the theory
$W=Tr [[\Phi_1, \Phi_2], \Phi_3]$ is transformed as\be \lb{deaf} Tr(
\Phi_1 \Phi_2 \Phi_3 -\Phi_1 \Phi_3 \Phi_2 ) \to h \, Tr( e^{ i  \pi
\beta} \Phi_1\Phi_2 \Phi_3 - e^{- i \pi \beta} \Phi_1\Phi_3 \Phi_2 )
+ h' Tr( \Phi_1^3 +\Phi_2^3 +\Phi_3^3 ), \ee being $h, h' , \beta$
complex parameters, satisfying one condition by conformal
invariance. One election could be $h'=0$. Besides the $U(1)_R$
symmetry, there is a $U(1) \times U(1)$ global symmetry generated by
$$
U(1)_1:\;\;(\Phi_1,\Phi_2 ,\Phi_3 ) \to (\Phi_1,e^{i\varphi_1}\Phi_2
,e^{-i \varphi_1} \Phi_3 ),
$$
\be\lb{mmm} U(1)_2:\;\;(\Phi_1,\Phi_2 ,\Phi_3 ) \to (e^{-i
\varphi_2} \Phi_1,e^{i\varphi_2}\Phi_2 , \Phi_3 ), \ee which leaves
the superpotential and the supercharges invariant. Therefore there
is a two dimensional manifold of ${\cal N}=1$ CFT with a torus
symmetry. It was found in \cite{Lunin} that the $U(1)\times U(1)$
action is realized in gravity part as an isometry. The effect of the
$\gamma$-deformation of ${\cal N}=4$ super Yang-Mills induce in the
gravity dual the simple transformation \be \lb{tau} \tau\equiv B + i
\sqrt{g} \longrightarrow \tau \longrightarrow \tau' =\frac { \tau}{
1 + \gamma \tau}, \ee where $\sqrt{g}$ is the volume of the two
torus \cite{Lunin}. The transformation (\ref{tau}) indeed comes from
a known solution generating technique explained in \cite{Horowitz}.

  The transformations (\ref{tau}) are not the full $SL(2,R)$
transformations. Indeed (\ref{tau}) is the subgroup of $SL(2,R)$ for
which $\tau\to 0$ implies that $\tau'\to 0$. In fact, from
(\ref{tau}) it follows that $\tau_\gamma = \tau + o(\tau^2) $ for
small $\tau$. Transformations with these properties are the only
possible ones mapping a ten dimensional geometry which is non
singular to a new one also without singularities. The reason is that
the only points where a singularity can be introduced by performing
an $SL(2,R)$ transformation is where the two torus shrinks to zero
size. This shrink happens when $\tau' \to 0$ but for $\gamma$
transformations, this implies that $\tau\to 0$. Therefore, if the
original metric was non-singular, then the deformed metric is also
non singular \cite{Lunin}. The transformation (\ref{tau}) is the
result of doing a T-duality on one circle, a change of coordinates,
followed by another T-duality. This is another reason for which it
can be interpreted as a solution generating technique
\cite{Horowitz}. It has been applied recently in order to find
several deformed backgrounds in \cite{Mateos22}-\cite{Vazquez},
together with an analysis of their gauge field theory duals.
\\

\textit{Sketch of the present work}
\\

   In the present work we construct an infinite family of tri-Sasaki metrics
in seven dimensions and we find that all these metrics admit an
squashed version which is of weak $G_2$ holonomy. The idea behind
this construction is simple. Our starting point are the Swann
hyperKahler metrics \cite{Swann}, which are fibrations over
quaternion Kahler metrics $g_4$ of the form
$$
g_s=|u|^2g_4+|du+\omega u|^2,
$$
being $u$ certain quaternion coordinate and $\omega$ an imaginary
quaternion valued 1-form associated to the quaternion Kahler space.
Under the transformation $u\to\lambda u$ these metrics are scaled by
a factor $g_s\to\lambda^2 g_s$, thus are conical and define a family
tri-Sasaki metrics. We find the explicit expression for these
family, which is composed by fibrations over quaternion Kahler
spaces. The six dimensional space formed by the orbits of the Reeb
vector is indeed well known, it is the twistor space Z associated to
the quaternion Kahler base. The resulting reduced metric is the
natural Kahler-Einstein metric defined over Z \cite{Salomon}. This
result are presented in the proposition 1 of section 2.

   We show that if the quaternion Kahler base is CP(2), the
resulting tri-Sasaki metric is $N(1,1)_I$, which is known to admit
an squashed version $N(1,1)_{II}$ of weak $G_2$ holonomy
\cite{Donpa}. Guided by this result, we find that the squashed
version of any of the tri-Sasaki metrics that we are presenting are
of weak $G_2$ holonomy. They are indeed the weak $G_2$ holonomy
metrics defined by the exactly conical family of Bryant-Salamon
Spin(7) metrics \cite{Bryant}.

    We then manage to
find tri-Sasaki (and weak $G_2$ holonomy) examples which are locally
$T^3$ fibrations. These 7-dimensional Einstein metrics are fibered
over certian quaternion Kahler orbifolds, the quaternion Kahler
limit of the euclidean AdS-Kerr-Newman-Taub-Nut metrics. These
4-metrics itself corresponds to a Wick rotation of the Minkowski
Plebanski-Demianski metric \cite{Demianski} and were considered in
several contexts \cite{Demianski2}-\cite{Demianski6}. In some
limiting cases for the parameters, the euclidean versions tends to
$S^4$ or CP(2), which are smooth. We also consider the most general
quaternion Kahler $T^2$ fibrations, which were found in
\cite{Caldo}. The presence of orbifold singularities is of interest
in the context of AdS/CFT correspondence, because they lead to
situations generalizing those analyzed by Mandal et all in
\cite{Mandal}.

   We then lift these 7-metrics to supergravity solutions possessing
near horizon limit $AdS_4\times X_7$, being $X_7$ a tri-Sassaki or
weak $G_2$ holonomy space. We also construct solutions which does
not have this horizon limit. In the manifold limit $S^4$ of the quaternion
Kahler base we analyze
rotating membrane configurations and we reproduce the logarithmic
behaviour of E-S, thus these configurations are dual to the "twist"
operators of the dual conformal field theory.

     We also consider the effect of the
SL(2,R) deformation and construct
new supergravity solutions, the deformed ones. We reproduce the
logarithmic behaviour for the deformed background corresponding to
$S^4$.

    For completeness, we discuss another type of Kahler-Einstein
examples that are present in the literature and present some Calabi-Yau
metrics fibered over non symmetric Kahler spaces, by using the methods developed in
\cite{Bielawski}, \cite{Hwang}.

\section{Conical internal spaces}

A wide variety of supergravity backgrounds can be constructed as
fibrations over Ricci flat conical metrics. An $n$-dimensional
manifold $X_n$ develops a conical singularity if and only if it is
possible to find a coordinate system with a coordinate $r$ for which
the metric has the local form \be\lb{conica} g_n=dr^2+r^2 g_{n-1},
\ee being $g_{n-1}$ a metric tensor independent on $r$. The metric
$g_n$ will be singular at $r=0$, except if $g_{n-1}$ is the round
$n-1$ dimensional sphere. Any metric $g_n$ of the form
(\ref{conica}) is called a cone over $g_{n-1}$. There also exist
metrics taking the form (\ref{conica}) for large values of $r$, such
metrics are called asymptotically conical. If the conical metric
$g_n$ is Ricci-flat, that is, its Ricci tensor satisfies $R_{ij}=0$,
then $g_{n-1}$ is Einstein, which means that its Ricci tensor is
given by $R_{ab}\sim g_{ab}$. In this section we discuss the
geometrical properties of such cones and construct a large family.
The geometrical objects that enter in this construction are
Kahler-Einstein, Einstein-Sasaki, tri-Sasaki, hyperKahler, weak
$G_2$ and $Spin(7)$ holonomy spaces. We use the word space instead
of manifold in order to keep open the possibility of constructing
spaces admitting orbifold or other type of singularities. The reader
who is familiar with these concepts can jump to the the last two subsections, in
particular to the proposition 1.

\subsection{$Spin(7)$ and weak $G_2$ holonomy}

A weak $G_2$ holonomy space $X_7$ is a seven dimensional space with
a metric $g_7$ and admitting at least one conformal Killing spinor,
that is, an spinor satisfying $D_j \eta \sim \lambda \eta$
\cite{Gray}. Here $\lambda$ is a constant and $D_i$ is the covariant
derivative in spinor representation, which is defined by
$D_i=\partial_i+\omega_{iab}\gamma^{ab}$. The one form $\omega_{ab}$
is the spin connection on $X_7$ defined by the first Cartan equation
with zero torsion and $\gamma^{ab}$ is the antisymmetric product of
Dirac $\gamma^a$ matrices in seven dimensions. If the constant
$\lambda$ is zero, then the spinor will be covariantly constant and
it will be preserved after parallel transport along any closed
curve. This means that the holonomy will be in $G_2$, which is the
subgroup of $SO(7)$ which possess a one-dimensional invariant
subspace. The reduction of the holonomy to $G_2$ is equivalent to
the existence of a $G_2$ invariant three form
$$
\Phi=c_{abc}e^a\wedge e^b\wedge e^c
$$
which is covariantly constant, that is, $\nabla_X \Phi=0$ for any
vector field $X$. We denote as $e^a$ seven soldering forms for which
the metric is diagonal, i.e, $g_7=\delta_{ab}e^a\otimes e^b$ and
$c_{abc}$ are the octonion multiplication constants. This condition
holds if and only if $\Phi$ and its dual $\ast\Phi$ are closed.
Instead for a weak $G_2$ holonomy space the existence of a conformal
Killing vector is equivalent to the condition $d\Phi=\lambda
\ast\Phi$. Such spaces have generically $SO(7)$ holonomy and there
exist a frame for which the spin connection $\omega_{ab}$ satisfies
$$
\omega_{ab}=\frac{c_{abcd}}{2}\omega_{cd}-\lambda c_{abc}e^c,
$$
being $c_{abcd}$ the dual octonion constants. The last condition
implies that a weak $G_2$ holonomy manifold is always Einstein, i.e,
$R_{ij}=\lambda g_{ij}$. In the limit $\lambda\rightarrow 0$ the
space will be Ricci flat. This is the case for a $G_2$ holonomy
space or for any euclidean space admitting covariantly constant
Killing spinors.

By another side, spaces with holonomy in $Spin(7)\in SO(8)$ are
eight dimensional and also admitting one covariantly constant
Killing spinor, thus are Ricci flat. Similarly to its $G_2$
counterpart, they are characterized by a closed $Spin(7)$ invariant
four form
$$
\Phi_4=\widetilde{c}_{abcd}e^a\wedge e^b\wedge e^c\wedge e^d.
$$
We are in the middle dimension here and $\Phi_4=\ast\Phi_4$. Here
$\widetilde{c}_{abcd}$ are constants related to the octonion
constants and from its values it follows that $\Phi_4$ can be
reexpressed as \be\lb{esr} \Phi_4=e^8\wedge \Phi+\ast \Phi \ee being
$\Phi$ a $G_2$ invariant three-form constructed with the seven
remaining soldering forms. The expression (\ref{esr}) is the origin
of a correspondence between conical $Spin(7)$ holonomy metrics of
cohomogeneity one and weak $G_2$ holonomy ones. More precisely, any
of such $Spin(7)$ metrics is of the form
$$
g_8=dr^2+r^2g_7,
$$
being $g_7$ a metric of weak $G_2$ holonomy and conversely, for any
weak $G_2$ holonomy metric $g_7$ the cone $g_8$ will be of $Spin(7)$
holonomy. This can be seen as follows. Let us consider the choice of
the frame $\tilde e^a$ given by \be \lb{noli} \tilde e^8=dr, \qquad
\tilde e^a = -\frac{\lambda}{4}r e^a, \ee being $e^a$ seven
soldering forms corresponding to $g_7$. In principle there is an
$SO(8)$ freedom to choose our frame, but if the element $\tilde e^8$
is fixed as $\tilde e^8=dr$, then there it remains an $SO(7)$
freedom only. The first Cartan structure
$$
d\tilde e^a + \tilde\omega^{ab}\wedge\tilde e^b = 0,
$$
gives the decomposition \be\lb{soda} \omega^{ab} =
\tilde\omega^{ab},\qquad \tilde\omega^{8a} = {\lambda\over4} e^a,
\ee being $\omega_{ab}$ the spin connection for the seven
dimensional part. Let us assume that the form $\hat{\Phi}$
constructed with (\ref{noli}) is closed, this is what we mean about
cohomogeneity one. Then (\ref{noli}) and (\ref{esr}) gives
\ba\lb{camal} \widetilde{\Phi} = \left({\lambda r\over4}\right)^3\,
dr\wedge \Phi + \left({\lambda r\over4}\right)^4\, *\Phi \,, \cr
\label{w20} d \widetilde{\Phi} = - \left({\lambda
r\over4}\right)^3\, dr \wedge \bigl(d\Phi-\lambda
*\Phi\bigr) + \left({\lambda r\over4}\right)^4\, d *\Phi \,, \ea
where $\Phi$ and $*\Phi$ are the usual seven-dimensional 3- and
4-form constructed with the frame $e^a$. It is directly seen from
(\ref{camal}) that $Spin(7)$ holonomy condition, namely
$d\widetilde{\Phi}=0$ is equivalent to weak $G_2$ holonomy of the
seven-dimensional base space, that is, to the condition
$$
d\Phi-\lambda *\Phi=0,\;\;\;\;\;\;\;\;\;\;d *\Phi=0.
$$
The converse of this statement is also true. Indeed, equations
(\ref{noli}) are equivalent to
$$
\tilde\omega^{ab} = {1\over2}c_{abcd}\tilde \omega^{cd},
$$
which is the eight-dimensional self-duality condition implying the
reduction of the holonomy to a subgroup of $Spin(7)$. This is the
one to one correspondence we wanted to show \cite{Hitchin} (see also
\cite{Bilok}).

   For applications to marginal deformations of field theories it is
needed to focus on metrics with weak $G_2$ holonomy admitting $T^3$
actions. Examples are the Aloff-Wallach spaces $N(k,l)=SU(3)/U(1)$,
which possesses two different metrics $N(k,l)_I$ and $N(k,l)_{II}$.
Except for $N(1,1)_I$, which is tri-Sasaki, the remaining metrics
are of weak $G_2$ holonomy. For $N(1,1)$ we have the isometry group
$SU(3)\times SO(3)$ while for the other cases we have $SU(3)\times
U(1)$. Another example is the squashed seven sphere $SO(5)\times
SO(3)/SO(3)\times SO(3)$ with isometry group $SO(5)\times SO(3)$. If
the manifold is homogeneous, that is, if $X_7$ is of the form $G/H$
then it will be one of this type, see \cite{Castellani}-\cite{Peter}
for a detailed discussion. Our aim is to construct a more large
class of weak $G_2$ holonomy manifolds admitting a $T^3$ action, not
necessarily homogeneous.

\subsection{Einstein-Sassaki and Kahler-Einstein spaces}

     As in the previous subsection, let us consider an eight dimensional
space $X_8$ endowed with a metric $g_8$ and with holonomy in
$Spin(7)$. If $X_8$ possess two Killing spinors instead of one, then
the holonomy will be reduced further to $SU(4)\in Spin(7)$. In fact,
$SU(4)$ is the subgroup of $SO(8)$ with a two dimensional invariant
subspace. As is well known, any $2n$-dimensional metric with
holonomy $SU(n)$ is Calabi-Yau, an so is $g_8$. If in addition $g_8$
is conical, then the seven dimensional metric $g_7$ over which $g_8$
is fibered will be called Einstein-Sasaki. This metric will possess
two \emph{conformal} Killing spinors. If there is a third Killing
spinor, we have a further reduction of the holonomy to $Sp(2)\in
SU(4)$. Any $4n$ dimensional space with holonomy in $Sp(n)$ is
hyperKahler, in particular $g_8$. In this case $g_7$ will be called
tri-Sasaki because it admits three \emph{conformal} Killing spinors.
We can take this notions as definitions, for any value of $n$.
Clearly, any tri-Sasaki metric is Einstein-Sassaki and if we are in
seven dimensions they will be of weak $G_2$ holonomy. Also any
hyperKahler metric is Calabi-Yau, and in $d=8$ they will be
$Spin(7)$ metrics. The converse of these statements are obviously
non true.

    Any Calabi-Yau space is Ricci flat due to the presence of Killing
spinors, and is also Kahler. A Kahler structure over an space
$X_{2n}$ is defined by a doublet $(g_{2n}, I)$ composed by an even
dimensional metric $g_{2n}$, a $(1,1)$ antisymmetric tensor $I$ such
that $I^2=-I$ which is covariantly constant, that is $\nabla_X I=0$
being $\nabla$ the Levi-Civita connection and for which the metric
is quaternion hermitian (which means that
$g_{2n}(IX,IY)=-g_{2n}(X,Y)$ for any pair of vector fields $X, Y$ of
the tangent space at a given point). From the antisymmetry of $I$ it
follows that the $(2,0)$ tensor with components
$\Omega(X,Y)=g_{2n}(IX,Y)$ is a two form. The covariance of $I$
implies that $\Omega$ is closed and that $I$ is integrable, that is,
its Nijenhuis tensor vanish identically. This implies that $X_{2n}$
is a complex manifold. Sometimes the triplet $(g_{2n}, I, \Omega)$
is identified as the Kahler structure in the literature, but only if
the properties stated above are all satisfied.

   An Einstein-Sassaki space $X_{2n+1}$ is always odd dimensional and
can be constructed as an $R$ or $U(1)$-fibration over a
Kahler-Einstein metric. The local form of their metric is
\be\lb{lome} g_{2n+1}=(d\tau + A)^2 + g_{2n}, \ee being $\Omega=dA$
the Kahler form of the Kahler-Einstein metric $g_{2n}$. The metric
$g_{2n}$ is assumed to be $\tau$-independent. The vector
$\partial_{\tau}$ is Killing, and it is called the Reeb vector. If
the orbits of this vector are closed and the action is free, then
$X_{2n}$ is a manifold and the odd dimensional manifold $X_{2n+1}$
is regular. If the action has finite isotropy groups then $X_{2n}$
is an orbifold. In addition, the Einstein condition $R_{ij}\sim
g_{ij}$ for $g_{2n}$ has been shown to be equivalent to \cite{Besse}
\be\lb{hota} \rho=\Lambda \Omega \ee being
$\rho=-i\overline{\partial}\partial \log \; \det g$ the Ricci form
of the metric $g_{2n}$. The scalar curvature of $g_{2n}$ is
$2n\Lambda$.

\subsection{Quaternion Kahler and hyperKahler spaces}

   A quaternion Kahler space $M$ is an euclidean $4n$ dimensional
space with holonomy group $\Gamma$ included into the Lie group
$Sp(n)\times Sp(1)\subset SO(4n)$ \cite{Berger}-\cite{Ishihara}.
This affirmation is non trivial if $D>4$, but in $D=4$ there is the
well known isomorphism $Sp(1)\times Sp(1)\simeq SU(2)_L\times
SU(2)_R \simeq SO(4)$ and so to state that $\Gamma\subseteq
Sp(1)\times Sp(1)$ is equivalent to state that $\Gamma\subseteq
SO(4)$. The last condition is trivially satisfied for any oriented
space and gives almost no restrictions, therefore the definition of
quaternion Kahler spaces should be modified in $d=4$.

Here we do a brief description of these spaces, more details can be
found in the appendix and in the references therein. For any
quaternion exists three automorphism $J^i$ ($i=1$ ,$2$, $3$) of the
tangent space $TM_x$ at a given point $x$ with multiplication rule
$J^{i} \cdot J^{j} = -\delta_{ij} + \epsilon_{ijk}J^{k}$. The metric
$g_q$ is quaternion hermitian with respect to this automorphism,
that is \be\lb{hermoso} g_q(X,Y)=g(J^i X, J^i Y), \ee being $X$ and
$Y$ arbitrary vector fields. The reduction of the holonomy to
$Sp(n)\times Sp(1)$ implies that the $J^i$ satisfy the fundamental
relation \be\lb{rela2}
\nabla_{X}J^{i}=\epsilon_{ijk}J^{j}\omega_{-}^{k}, \ee being
$\nabla_{X}$ the Levi-Civita connection of $M$ and $\omega_{-}^{i}$
its $Sp(1)$ part. As a consequence of hermiticity of $g$, the tensor
$\overline{J}^{i}_{ab}=(J^{i})_{a}^{c}g_{cb}$ is antisymmetric, and
the associated 2-form
$$
\overline{J}^i=\overline{J}^{i}_{ab} e^a \wedge e^b
$$
satisfies \be\lb{basta}
d\overline{J}^i=\epsilon_{ijk}\overline{J}^{j}\wedge\omega_{-}^{k},
\ee being $d$ the usual exterior derivative.  Corresponding to the
$Sp(1)$ connection we can define the 2-form
$$
F^i=d\omega_{-}^i+\epsilon_{ijk}\omega_{-}^j \wedge \omega_{-}^k.
$$
For any quaternion Kahler manifold it follows that \be\lb{lamas}
R^i_{-}=2n\kappa \overline{J}^i, \ee \be\lb{rela} F^i=\kappa
\overline{J}^i, \ee being $\Lambda$ certain constant and $\kappa$
the scalar curvature. The tensor $R^a_{-}$ is the $Sp(1)$ part of
the curvature. The last two conditions implies that $g$ is Einstein
with non zero cosmological constant, i.e, $R_{ij}=3\kappa
(g_{q})_{ij}$ being $R_{ij}$ the Ricci tensor constructed from
$g_q$. The $(0,4)$ and $(2,2)$ tensors
$$
\Theta=\overline{J}^1 \wedge \overline{J}^1 + \overline{J}^2 \wedge
\overline{J}^2 + \overline{J}^3 \wedge \overline{J}^3,
$$
$$
\Xi= J^1 \otimes J^1 + J^2 \otimes J^2 + J^3 \otimes J^3
$$
are globally defined and covariantly constant with respect to the
usual Levi Civita connection for any of these spaces. This implies
in particular that any quaternion Kahler space is orientable.

In four dimensions  the Kahler triplet $\overline{J}_2$ and the one
forms $\omega^{a}_{-}$ are
$$
\omega^{a}_{-}=\omega^a_{0}- \epsilon_{abc}\omega^b_c,\qquad
\overline{J}_1=e^1\wedge e^2-e^3\wedge e^4,
$$
$$
\overline{J}_2=e^1\wedge e^3-e^4\wedge e^2\qquad
\overline{J}_3=e^1\wedge e^4-e^2\wedge e^3.
$$
In this dimension quaternion Kahler spaces are defined by the
conditions (\ref{rela}) and (\ref{lamas}). This definition is
equivalent to state that quaternion Kahler spaces are Einstein and
with self-dual Weyl tensor.

In the Ricci-flat limit $\kappa\to 0$ the holonomy of a quaternion
Kahler space is reduced to a subgroup of $Sp(n)$ and the resulting
spaces are hyperKahler. It follows from (\ref{rela}) and
(\ref{rela2}) that the almost complex structures $J_i$ are
covariantly constant in this case. Also, there exist a frame for
which $\omega_-^i$ goes to zero. In four dimensions this implies
that the spin connection corresponding to this frame is self-dual.

\subsection{An infinite tri-Sasaki family in detail}

   The results of this section are crucial for the purposes of the
present work. For this reason we will make the calculations in
detail. As we have stated, any hyperKahler conical metric $g_8$
define a tri-Sasaki metric by means of the formula
$g_8=dr^2+r^2g_7$. A well known family of conical hyperKahler
metrics are the Swann metrics \cite{Swann}, this are 4n dimensional
metrics but we will focus only in the case $d=8$. The metrics reads
\be\lb{Swann2} g_8=|u|^2 g_q + |du + u \omega_{-}|^2, \ee being
$g_q$ any 4-dimensional quaternion Kahler metric. In the expression
for the metric we have defined the quaternions
$$
u=u_0 + u_1 I + u_2 J + u_3 K ,\;\;\;\;\;\; \overline{u}= u_0 - u_1
I - u_2 J - u_3 K,
$$
and the quaternion one form
$$
\omega_{-}=\omega_{-}^1 I+\omega_{-}^2 J +\omega_{-}^3 K,
$$
constructed with the anti-self-dual spin connection. The
multiplication rule for the quaternions $I$, $J$ and $K$ is deduced
from
$$
I^2=J^2=K^2=-1, \qquad IJ=K, \qquad JI=-K
$$
The metric $g_q$ is assumed to be independent on the coordinates
$u_i$. We easily see that if we scale $u_0$, $u_1$,$u_2$,$u_3$ by
$t>0$ this scales the metric by a homothety $t$, which means that
the metrics (\ref{Swann2}) are conical. Therefore they define a
family of tri-Sasaki metrics, which we will find now. We first
obtain, by defining $\widetilde{u}_i=u_i/u$ that
$$
|du + u \omega_{-}|^2=(du_0-u_i\omega_{-}^i)^2 +(du_i+
u_0\omega_{-}^i + \frac{\epsilon_{ijk}}{2}u_k\omega_{-}^j)^2
$$
$$
=(\widetilde{u}_0 du+ u d\widetilde{u}_0-u
\widetilde{u}_i\omega_{-}^i)^2+(\widetilde{u}_i du+u
d\widetilde{u}_i+u \widetilde{u}_0\omega_{-}^i +
u\frac{\epsilon^{ijk}}{2}\widetilde{u}_j\omega_{-}^k)^2
$$
$$
=du^2+u^2(d\widetilde{u}_0-\widetilde{u}_i\omega_{-}^i)^2+u^2(d\widetilde{u}_i+
\widetilde{u}_0\omega_{-}^i +
\frac{\epsilon_{ijk}}{2}\widetilde{u}_j\omega_{-}^k)^2
$$
$$
+2u u_0 du (d\widetilde{u}_0-\widetilde{u}_i\omega_{-}^i)+2u u_i du
(d\widetilde{u}_i+ \widetilde{u}_0\omega_{-}^i +
\frac{\epsilon^{ijk}}{2}\widetilde{u}_j\omega_{-}^k).
$$
It is not difficult to see that the last two terms are equal to
$$
2u u_0 du (d\widetilde{u}_0-\widetilde{u}_i\omega_{-}^i)+2u u_i du
(d\widetilde{u}_i+ \widetilde{u}_0\omega_{-}^i +
\epsilon_{ijk}\widetilde{u}_i\widetilde{u}_k\omega_{-}^j)=\frac{d(\widetilde{u}_i^2)}{2}
+\frac{\epsilon_{ijk}}{2}\widetilde{u}_i\widetilde{u}_j\omega_{-}^k
$$
But the second term is product of a antisymmetric pseudotensor with
a symmetric expression, thus is zero, and the first term is zero due
to the constraint $\widetilde{u}_i^2=1$. Therefore this calculation
shows that \be\lb{va} |du + u
\omega_{-}|^2=du^2+u^2(d\widetilde{u}_0-\widetilde{u}_i\omega_{-}^i)^2+u^2(d\widetilde{u}_i+
\widetilde{u}_0\omega_{-}^i +
\frac{\epsilon_{ijk}}{2}\widetilde{u}_j\omega_{-}^k)^2. \ee By
introducing (\ref{va}) into (\ref{Swann2}) we find that $g_8$ is a
cone over the following metric \be\lb{trusa}
g_7=g_q+(d\widetilde{u}_0-\widetilde{u}_i\omega_{-}^i)^2+(d\widetilde{u}_i+
\widetilde{u}_0\omega_{-}^i +
\frac{\epsilon_{ijk}}{2}\widetilde{u}_j\omega_{-}^k)^2. \ee This is
the tri-Sasaki metric we were looking for. By expanding the squares
appearing in (\ref{trusa}) we find that
$$
g_7=g_q+(d\widetilde{u}_i)^2+(\omega_{-}^i)^2 +2\omega_{-}^1
(\widetilde{u}_0d\widetilde{u}_1-\widetilde{u}_1d\widetilde{u}_0
+\widetilde{u}_2d\widetilde{u}_3-\widetilde{u}_3d\widetilde{u}_2)
$$
\be\lb{roo} + 2\omega_{-}^2
(\widetilde{u}_0d\widetilde{u}_2-\widetilde{u}_2d\widetilde{u}_0
+\widetilde{u}_2d\widetilde{u}_1-\widetilde{u}_1d\widetilde{u}_3)
+2\omega_{-}^3
(\widetilde{u}_0d\widetilde{u}_3-\widetilde{u}_3d\widetilde{u}_0
+\widetilde{u}_1d\widetilde{u}_2-\widetilde{u}_2d\widetilde{u}_1).
\ee But the expression in parenthesis are a representation of the
$SU(2)$ Maurer-Cartan 1-forms, which are defined by
$$
\sigma_1=-(\widetilde{u}_0d\widetilde{u}_1-\widetilde{u}_1d\widetilde{u}_0
+\widetilde{u}_2d\widetilde{u}_3-\widetilde{u}_3d\widetilde{u}_2)
$$
$$
\sigma_{2}=-(
\widetilde{u}_0d\widetilde{u}_2-\widetilde{u}_2d\widetilde{u}_0
+\widetilde{u}_2d\widetilde{u}_1-\widetilde{u}_1d\widetilde{u}_3)
$$
$$
\sigma_{3}=-
(\widetilde{u}_0d\widetilde{u}_3-\widetilde{u}_3d\widetilde{u}_0
+\widetilde{u}_1d\widetilde{u}_2-\widetilde{u}_2d\widetilde{u}_1).
$$
Therefore the metric (\ref{trusa}) can be reexpressed in simple
fashion as \be\lb{trusa2} g_7=g_q+(\sigma_i-\omega_-^i)^2. \ee This
is one of the expressions that we will use along this work.

    Let us recall that there exist a coordinate system for which the
Maurer-Cartan forms are expressed as \be\lb{mcarta}
\sigma_1=\cos\varphi d\theta+\sin\varphi \sin\theta d\tau,\qquad
\sigma_2=-\sin\varphi d\theta+\cos\varphi \sin\theta d\tau,\qquad
\sigma_3=d\varphi+\cos\theta d\tau. \ee With the help of this
coordinates we will write (\ref{trusa2}) in more customary form for
tri-Sasaki spaces, namely \be\lb{custo} g_7=(d\tau+H)^2+g_6, \ee as
in (\ref{lome}). Here $g_6$ a Kahler-Einstein metric with Kahler
form $\overline{J}$ and $H$ a 1-form such that $dH=2\overline{J}$. A
lengthy algebraic calculation shows that the fiber metric is
$$
(\sigma_i-\omega_-^i)^2=(d\tau+\cos\theta
d\varphi-\sin\theta\sin\varphi\omega_-^1-
\cos\theta\sin\varphi\omega_-^2-\cos\theta\omega_-^3)^2
$$
$$
+(\sin\theta d\varphi-\cos\theta\sin\varphi\omega_-^1-
\cos\theta\cos\varphi\omega_-^2+\sin\theta\omega_-^3)^2
+(d\theta-\sin\varphi\omega_-^2 +\cos\varphi\omega_-^1)^2,
$$
from where we read that \be\lb{ache} H=\cos\theta
d\varphi-\sin\theta\sin\varphi\omega_-^1-
\cos\theta\sin\varphi\omega_-^2-\cos\theta\omega_-^3. \ee
The vector $\partial_{\tau}$ is the Reeb vector, and is a Killing vector.
The six dimensional metric  \be\lb{kalonste}
g_6= g_q+(d\theta-\sin\varphi\omega_-^2 +\cos\varphi\omega_-^1)^2,
\ee
$$
+(\sin\theta d\varphi-\cos\theta\sin\varphi\omega_-^1-
\cos\theta\cos\varphi\omega_-^2+\sin\theta\omega_-^3)^2,
$$
should be Kahler-Einstein. We will check that this is the case next.
\\

\textit{Another deduction of the tri-Sasaki metrics
(\ref{trusa2})}
\\

 We will prove now that the six dimensional space formed
by the orbits of the Reeb vector of the tri-Sasaki family
presented above is the twistor space Z associated to
the quaternion Kahler base. The resulting reduced
metric (\ref{kalonste}) is the natural Kahler-Einstein
metric defined over Z \cite{Salomon}. We need first to define what is Z.
Recall that for any quaternion Kahler space $M$, a linear
combination of the almost complex structures of the form
$J=\widetilde{v}_i J_i$ will be also an almost complex structure on
$M$. Here $\widetilde{v}^i$ denote three "scalar fields"
$\widetilde{v}^i=v^i/v$ being $v=\sqrt{v^iv^i}$. This fields are
assumed to be constant over $M$ and are obviously constrained by
$\widetilde{v}^i \widetilde{v}^i=1$. This means that the bundle of
almost complex structures over $M$ is parameterized by points on the
two sphere $S^2$. This bundle is what is known as the twistor space $Z$ of
$M$. The space $Z$ is endowed with the metric \be\lb{kahlo}
g_6=\theta_i \theta_i + g_q,\ee where
$\theta_i=d(\widetilde{v}^i)+\epsilon^{ijk}\omega_{-}^j
\widetilde{v}^k$. The constraint $\widetilde{v}^i \widetilde{v}^i=1$
implies that the metric (\ref{kahlo}) is six dimensional. It have
been shown that this metric together with the sympletic two form
\cite{Salomon}, \cite{Lolo} \be\lb{two}
\overline{J}=-\widetilde{u}_i\overline{J}_i
+\frac{\epsilon_{ijk}}{2}\widetilde{v}_i\theta_j\wedge \theta_k, \ee
constitute a \emph{Kahler} structure. The calculation of the Ricci
tensor of $g_6$ shows that it is also Einstein, therefore the space
$Z$ is \emph{Kahler-Einstein}. The expressions given below are
written for a quaternion Kahler metric normalized such that
$\kappa=1$, for other normalizations certain coefficients must be
included in (\ref{two}). By parameterizing the coordinates
$\widetilde{v}_i$ in the spherical form \be\lb{esferoso}
\widetilde{v}_1=\sin\theta \sin\varphi,\qquad
\widetilde{v}_2=\sin\theta \cos\varphi,\qquad
\widetilde{v}_3=\cos\theta, \ee we find that (\ref{kahlo}) is the
same as (\ref{kalonste}). The isometry group of the Kahler-Einstein
metrics is in general $SO(3)\times G$, being $G$ the isometry group
of the quaternion Kahler basis which also preserve the forms
$\omega_-^i$. The $SO(3)$ part is the one which preserve the
condition $\widetilde{v}_i \widetilde{v}_i=1$. Globally the isometry
group could be larger.

From the definition of Einstein-Sasaki geometry, it follows directly
that the seven dimensional metric \be\lb{owner}
g_7=(d\tau+H)^2+g_6=(d\tau+H)^2+\theta_i \theta_i + g_q, \ee will be
Einstein-Sassaki if $dH=2\overline{J}$, and we need to find an
explicit expression for such $H$. Our aim is to show that this form
is indeed (\ref{ache}). The expression (\ref{two}) needs to be
simplified as follows. We have that
$\theta_i=d(\widetilde{v}^i)+\epsilon^{ijk}\omega_{-}^j
\widetilde{v}^k$. Also by using the condition
$\widetilde{v}_i\widetilde{v}_i=1$ it is found that
$$
\widetilde{v}_i\theta_i=\widetilde{v}_i
d\widetilde{v}_i+\epsilon^{ijk}\widetilde{v}^i\omega_{-}^j
\widetilde{v}^k=\widetilde{v}_i d\widetilde{v}_i=
d(\widetilde{v}_i\widetilde{v}_i)=0.
$$
From the last equality it follows the orthogonality condition
$\widetilde{v}_i\theta_i=0$ which is equivalent to
$$
\theta_3=-\frac{(\widetilde{v}_1\theta_1+\widetilde{v}_2\theta_2)}{\widetilde{v}_3}.
$$
The last relation implies that
$$
\frac{\epsilon_{ijk}}{2}\widetilde{v}_i\theta_j\wedge
\theta_k=\frac{\theta_1\wedge \theta_2}{\widetilde{v}_3}
=\frac{d\widetilde{v}_1\wedge
d\widetilde{v}_2}{\widetilde{v}_3}-d\widetilde{v}_i\wedge
\omega_{-}^i+\frac{\epsilon^{ijk}}{2}\widetilde{v}_i\omega_{-}^j\wedge
\omega_{-}^k.
$$
By another side in a quaternion Kahler manifold with $\kappa=1$ we
always have
$$
\widetilde{J}_i=d\omega_{-}^i+\frac{\epsilon^{ijk}}{2}\omega_{-}^j\wedge
\omega_{-}^k.
$$
Inserting the last two expressions into (\ref{two}) gives a
remarkably simple expression for $\overline{J}$, namely
\be\lb{simple}
\overline{J}=-d(\widetilde{v}_i\omega_{-}^i)+\frac{d\widetilde{v}_1\wedge
d\widetilde{v}_2}{\widetilde{v}_3}. \ee By using (\ref{esferoso}) it
is obtained that
$$
\frac{d\widetilde{v}_1\wedge
d\widetilde{v}_2}{\widetilde{v}_3}=-d\varphi\wedge d\cos\theta.
$$
With the help of the last expression we find that (\ref{simple}) can
be rewritten as
$$
\overline{J}=-d(\widetilde{v}_i\omega_{-}^i)-d\varphi\wedge
d\cos\theta,
$$
from where it is obtained directly that the form $H$ such that
$dH=\overline{J}$ is \cite{yoga} \be\lb{simplon}
H=-\widetilde{v}_i\omega_{-}^i+\cos\theta d\varphi, \ee up to a
total differential term. By introducing (\ref{esferoso}) into
(\ref{simplon}) we find that $H$ is the same than (\ref{ache}), as
we wanted to show.

   It will be of importance for the purposes of the present work to
state these results in a concise proposition.
\\

{\bf Proposition}{ \it Let $g_q$ be a four dimensional Einstein
space with self-dual Weyl tensor, i.e, a quaternion Kahler space. We
assume the normalization $\kappa=1$ for $g_q$. Then the metrics
$$
g_6=g_q +(d\theta-\sin\varphi\omega_-^2 +\cos\varphi\omega_-^1)^2
$$
$$
+(\sin\theta d\varphi-\cos\theta\sin\varphi\omega_-^1-
\cos\theta\cos\varphi\omega_-^2+\sin\theta\omega_-^3)^2
$$
are Kahler-Einstein whilst
$$
g_7=(\sigma_i-\omega_{-}^i)^2+g_q, \qquad g_8=dr^2+r^2g_7
$$
are tri-Sasaki and hyperKahler respectively. Here $\omega_-^i$ is
the $Sp(1)$ part of the spin connection and $\sigma_i$ are the usual
Maurer-Cartan one forms over $SO(3)$. Moreover the "squashed" family
$$
g_7=(\sigma_i-\omega_{-}^i)^2+5g_q,
$$
is of weak $G_2$ holonomy.}
\\

We will consider the last sentence of this proposition in the next
section. In order to finish this section we would like to describe a
little more the Swann bundles. Under the transformation $u\to G u$
with $G: M \to SU(2)$ the $SU(2)$ instanton $\omega_{-}$ is gauge
transformed as $\omega_{-}\to G\omega_{-}G^{-1}+ GdG^{-1}$.
Therefore the form $du + \omega_{-}u$ is transformed as
$$
du + u \omega_{-}\rightarrow d(Gu) + (G\omega_{-}G^{-1}+ GdG^{-1})
Gu=G du+ (dG+G\omega_{-}-dG)u=G (du + u \omega_{-}),
$$
and it is seen that $du +  \omega_{-}u$ is a well defined
quaternion-valued one form over the chiral bundle. The metric
(\ref{Swann2}) is also well defined over this bundle. Associated to
the metric (\ref{Swann2}) there is a quaternion valued two form
\be\lb{quato} \widetilde{\overline{J}}=u\overline{J}\overline{u}+(du
+ u \omega_{-})\wedge \overline{(du + u \omega_{-})}, \ee and it can
be checked that the metric (\ref{Swann}) is hermitian with respect
to any of the components of (\ref{quato}). Also
$$
d\widetilde{\overline{J}}=du\wedge
(\overline{J}+d\omega_{-}-\omega_{-}\wedge
\omega_{-})\overline{u}+u\wedge
(\overline{J}+d\omega_{-}-\omega_{-}\wedge \omega_{-})d\overline{u}
$$
$$+u(d\overline{J}+\omega_{-}\wedge
d\omega_{-}-d\omega_{-}\wedge \omega_{-})\overline{u}.
$$
The first two terms of the last expression are zero due to
(\ref{rela}). Also by introducing (\ref{rela}) into the relation
(\ref{basta}) it is seen that
$$
d\overline{J}+\omega_{-}\wedge d\omega_{-}-d\omega_{-}\wedge
\omega_{-}=0,
$$
and therefore the third term is also zero. This means that the
metric (\ref{Swann2}) is hyperKahler with respect to the triplet
$\widetilde{\overline{J}}$. The Swann metrics have been considered
in several context in physics, as for instance in
\cite{Suan1}-\cite{Suan4}. It is an important tool also in
differential geometry because the quaternion Kahler quotient
construction correspond to hyperKahler quotients on the Swann
fibrations.

\subsection{A weak $G_2$ holonomy family by squashing}

   In \cite{Bryant} there were probably constructed
the first examples of $Spin(7)$ holonomy metrics. These examples are
fibered over four dimensional quaternion Kahler metrics defined over
manifold $M$.  This resembles the Swann metrics that we have
presented in (\ref{Swann2}), although the Bryant-Salamon were found
first. The anzatz for the $Spin(7)$ is \be\lb{Swann} g_8=g|u|^2
\overline{g} + f|du + u \omega_{-}|^2, \ee where $f$ and $g$ are two
unknown functions $f(r^2)$ and $g(r^2)$ which will be determined by
the requirement that the holonomy is in $Spin(7)$, i.e, the closure
of the associated 4-form $\Phi_4$. The analogy between the anzatz
(\ref{Swann}) and (\ref{Swann2}) is clear, in fact, if $f=g=1$ the
holonomy will be reduced to $Sp(2)$. A convenient (but not unique)
choice for $\Phi_4$ is the following \be\lb{expresate} \hat{\Phi}= 3
f g [\alpha\wedge \overline{\alpha}\wedge \overline{e}^t \wedge e +
\overline{e}^t \wedge e \wedge \alpha \wedge \overline{\alpha}]+ g^2
\overline{e}^t \wedge e \wedge\overline{e}^t\wedge e + f^2 \alpha
\wedge \overline{\alpha} \wedge \alpha \wedge \overline{\alpha} \ee
where $\alpha = du + u \omega_{-}$. After imposing the condition
$d\Phi_4=0$ to (\ref{Swann}) it is obtained a system of differential
equations for $f$ and $g$ with solution
$$
f=\frac{1}{(2\kappa r^2+c)^{2/5}},
$$
$$
g=(2\kappa r^2+c)^{3/5},
$$
and the corresponding metric \be\lb{sou} g_s=(2\kappa
r^2+c)^{3/5}\overline{g} + \frac{1}{(2\kappa
r^2+c)^{2/5}}|\alpha|^2. \ee Spaces defined by (\ref{sou}) are the
Bryant-Salamon $Spin(7)$ ones. The metrics (\ref{sou}) are non
compact (because $|u|$ is not bounded), and asymptotically conical.
They will be exactly conical only if $c=0$. This is better seen by
introducing spherical coordinates for $u$
$$
u_1=|u| \sin\theta\cos\varphi\cos\tau,
$$
$$
u_2=|u| \sin\theta\cos\varphi\sin\tau,
$$
$$
u_3=|u| \sin\theta\sin\varphi,
$$
$$
u_4=|u| \cos\theta,
$$
and defining the radial variable
$$
r^2=\frac{9}{20}(2\kappa |u|^2 + c)^{3/5}
$$
from which we obtain the spherical form of the metric \be \lb{spo} g
= {dr^2 \over \kappa (1- {c / r^{10/3}})} + {9 \over 100 \, \kappa}
r^2 \left(1 - {c \over r^{10/3}}\right) \left(\sigma^i -
\omega_{-}^i\right)^2 + {9 \over 20} \, r^2 \,\overline{g} \ee being
$\sigma^i$ the left-invariant one-forms on $SU(2)$
$$
\sigma_1=\cos\varphi d\theta + \sin\varphi \sin\theta d\tau
$$
$$
\sigma_2=-\sin\varphi d\theta + \cos\varphi \sin\theta d\tau
$$
$$
\sigma_3=d\varphi + \cos \theta d\tau.
$$
In this case it is clearly seen that (\ref{spo}) are of
cohomogeneity one and thus, by the results presented on the previous
section, they define a weak $G_2$ holonomy metric.

     Let us fix the normalization $\kappa=1$, as before. Then in the
limit $r>>c$ it is found the behavior \be\lb{cono2} g \approx dr^2 +
r^2\Omega, \ee being $\Omega$ a seven dimensional metric
asymptotically independent of the coordinate $r$, namely
\be\lb{brg2} \Omega= \left(\sigma^i - \omega_{-}^i\right)^2 + 5 g_q
\ee In particular the subfamilies of (\ref{spo}) with $c=0$ are
exactly conical and their angular part is (\ref{brg2}). The metrics (\ref{brg2})
are of weak $G_2$ holonomy and possesses an
$SO(3)$ isometry action associated with the $\sigma^i$. If also the
four dimensional quaternion Kahler metric has an isometry group $G$
that preserve the $\omega_{-}^i$, then the group is enlarged to
$SO(3)\times G$.

\subsection{An instructive test: the case $N(1,1)_I$ and $N(1,1)_{II}$}

    It is important to compare the weak $G_2$ holonomy metrics
(\ref{brg2}) and the tri-Sasaki metrics (\ref{trusa2}). The only
difference between the two metrics is a factor 5 in front of $g_q$
in (\ref{brg2}). Both metrics possess the same isometry group. At
first sight it sounds possible to absorb this factor 5 by a simple
rescale of the coordinates and therefore to conclude that both
metrics are the same. \emph{But this is not true}. We are fixing the
normalization $\kappa=1$ in both cases, thus this factor should be
absorbed only by an rescaling on the coordinates of the fiber. There
is no such rescaling. Therefore, due to this factor 5, both metrics
are different. This is what one expected, since they are metrics of
different type.

    We can give an instructive example to understand why this is so.
Let us consider the Fubini-Study metric on
CP(2). This metric is Kahler-Einstein and quaternion Kahler
simultaneously and there exists a coordinate system for it takes
the form \be\lb{fubi}
g_f=2d\mu^2+\frac{1}{2}\sin^2\mu\widetilde{\sigma}_3^2
+\frac{1}{2}\sin^2\mu\cos^2\mu(\widetilde{\sigma}_1^2+\widetilde{\sigma}_2^2).
\ee We have denoted the Maurer-Cartan one-forms of this expression
as $\widetilde{\sigma}_i$ in order to not confuse them with the
$\sigma_i$ appearing in (\ref{brg2}) and (\ref{trusa2}). The
anti-self-dual part of the spin connection is \be\lb{rolon}
\omega_-^1=-\cos\mu\widetilde{\sigma}_1, \qquad
\omega_-^2=\cos\mu\widetilde{\sigma}_2, \qquad
\omega_-^3=-\frac{1}{2}(1+\cos\mu)\widetilde{\sigma}_3. \ee The two
metrics that we obtain by use of (\ref{brg2}) and (\ref{trusa2}) are
\be\lb{tro} g_7=
2b\;d\mu^2+\frac{1}{2}\sin^2\mu\widetilde{\sigma}_3^2
+b\;\frac{1}{2}\sin^2\mu\cos^2\mu(\widetilde{\sigma}_1^2+b\;\widetilde{\sigma}_2^2)\ee
$$
+(\sigma_1+\cos\mu\widetilde{\sigma}_1)^2+(\sigma_2-\cos\mu\widetilde{\sigma}_2)^2+
(\sigma_3+\frac{1}{2}(1+\cos\mu)\widetilde{\sigma}_3)^2.
$$
If (\ref{brg2}) and (\ref{trusa2}) are correct, then $b=1$
corresponds to a tri-Sasaki metric and $b=5$ to a weak $G_2$
holonomy one. This is true. Locally this metrics are the same that
$N(1,1)_I$ and $N(1,1)_{II}$ given in \cite{Donpa}, which are known
to be tri-Sasaki and weak $G_2$. We see therefore that this number
five in front of the quaternion Kahler metric is relevant and change
topological properties of the metric (such as the number of
conformal Killing spinors).

\section{Examples of quaternion Kahler manifolds and orbifolds}

The tri-Sasaki and weak $G_2$ holonomy spaces presented
in proposition 1 are fibered over quaternion Kahler spaces in four
dimensions. Such spaces can be extended
to a wide variety of supergravity solutions. We are interested in
supergravity solutions
with three commuting Killing vectors.
In this case the SL(2,R) deformation technique described in
\cite{Lunin} can be applied, which in many cases correspond to marginal deformations
of the field theory duals. The Reeb vector is clearly one of the isometries,
and by inspection of the formulas of Proposition 1 it is seen that
three commuting Killing vectors will be obtained
if the quaternion Kahler base possess two commuting isometries which also preserve
the 1-forms $\omega_-^i$. We will refer to these spaces as toric quaternion Kahler spaces.
In this section we describe a large class of such spaces.

\subsection{Quaternion Kahler limit of AdS-Kerr-Newman-Taub-Nut}

   The spaces that we will present next are obtained
by a Wick rotation of the Plebanski and
Demianski solution \cite{Demianski} and have been
discussed in detail in \cite{Demianski2}-\cite{Demianski6}. After
taking certain scaling limit of the parameters, the distance element
becomes \be\lb{plebanski} g_q=\frac{x^2-y^2}{P}dx^2+
\frac{x^2-y^2}{Q}dy^2+\frac{P}{x^2-y^2}(d\alpha+y^2d\beta)^2+
\frac{Q}{x^2-y^2}(d\alpha+x^2d\beta)^2 \ee being $P(x)$ and $Q(y)$
polynomials of the form
\be\lb{dv}
P(x)=q-2 s x-t x^2- \kappa x^4,\qquad Q(y)=-P(y),
\ee
being $(q,s,t,\kappa)$ four parameters. These expressions can be
rewritten as
$$
P(x)=-\kappa(x-r_1)(x-r_2)(x-r_3)(x-r_4),\qquad Q(y)=-P(y),
$$
$$
r_1+r_2+r_3+r_4=0,
$$
the last condition comes from the fact that $P(x)$ contains no cubic
powers of $x$. The two commuting Killing vectors are
$\partial_{\alpha}$ and $\partial_{\beta}$.

  The metric (\ref{plebanski})
is invariant under the transformation $x\leftrightarrow y$. The
transformations $x\rightarrow -x$, $y\rightarrow -y$,
$r_i\rightarrow -r_i$ are also a symmetry of the metric. In addition
the symmetry $(x, y, \alpha, \beta)\rightarrow (\lambda x, \lambda
y, \frac{\alpha}{\lambda}, \frac{\beta}{\lambda^3})$,
$r_i\rightarrow \lambda r_i$ can be used in order to put one
parameter equal to one, so there are only three effective parameters
here. The domains of definition are determined by
$$
(x^2-y^2)P(x)\geq 0, \qquad (x^2-y^2)Q(y)\geq 0.
$$
The anti-self-dual part of the spin connection is
$$
\omega_-^1=\frac{\sqrt{PQ}}{y-x}d\beta, \qquad
\omega_-^3=\frac{1}{x-y}\bigg(\sqrt{\frac{Q}{P}}dx+\sqrt{\frac{P}{Q}}dy\bigg),
$$
\be\lb{party}
\omega_-^2=-\kappa(x-y)d\alpha+\frac{1}{x-y}\bigg(q-s(x+y)-t xy-
\kappa x^2 y^2\bigg)d\beta, \ee  (see for instance
\cite{Mahapatrah}). We will need (\ref{party}) in the following.

          The metrics (\ref{plebanski}) are the self-dual limit of the
AdS-Kerr-Newmann-Taub-Nut solutions, the last ones corresponds to the
polynomials
$$
P(x)=q-2 s x-t x^2-\kappa x^4,\qquad Q(y)=-q+2 s' x+t x^2+\kappa
x^4,
$$
and are always Einstein. But the self-duality condition holds if and only if $s'=s$,
as in (\ref{dv}). We
will concerned with this limit in the following, because is the one
which is quaternion Kahler. If we define the new coordinates
$$
y=\widetilde{r},\qquad x=a \cos\widetilde{\theta}+N,
$$
$$
\alpha=t+(\frac{N^2}{a}+a)\frac{\widetilde{\phi}}{\Xi},\qquad
\beta=-\frac{\widetilde{\phi}}{a\Xi},
$$
where we have introduced the parameters
$$
\Xi=1-\kappa a^2,\qquad q=-a^2+N^2(1-3\kappa a^2+3\kappa N^2),
$$
$$
s=N(1-\kappa a^2+4N^2), \qquad t=-1-\kappa a^2-6 \kappa N^2,
$$
then the functions $P$ and $Q$ are expressed as
$$
P=-a^2\sin^2\widetilde{\theta}[1-\kappa(4aN
\cos\widetilde{\theta}+a^2\cos^2\widetilde{\theta})],
$$
$$
Q=-\widetilde{r}^2-N^2+2s'\widetilde{r}+a^2+\kappa(\widetilde{r}^4-a^2\widetilde{r}^2-6\widetilde{r}^2N^2+3a^2N^2-3N^4),
$$
and the metric take the AdS-Kerr-Newman-Taub-Nut form
$$
g_q=\frac{\sin^2\widetilde{\theta}[1-\kappa(a^2\cos^2\widetilde{\theta}+4aN
\cos\widetilde{\theta})]}{R^2}[\;a
d\widetilde{t}-\frac{r^2-a^2-N^2}{\Xi}d\widetilde{\phi}\;]^2
$$
\be\lb{demi}
+\frac{\lambda^2}{R^2}[\;d\widetilde{t}+(\frac{a\sin^2\widetilde{\theta}}{\Xi}
-\frac{2N\cos\widetilde{\theta}}{\Xi})d\widetilde{\phi}\;]^2+\frac{R^2}{1-\kappa(a^2\cos^2\widetilde{\theta}+4aN
\cos\widetilde{\theta})}d\widetilde{\theta}^2+\frac{R^2}{\lambda^2}dr^2
, \ee being $R$ and $\lambda$ defined by
$$
R=\widetilde{r}^2-(a\cos\widetilde{\theta}+N)^2,
$$
$$
\lambda=\widetilde{r}^2+N^2-2s'\widetilde{r}-a^2-\kappa(\widetilde{r}^4-a^2\widetilde{r}^2-6\widetilde{r}^2N^2+3a^2N^2-3N^4).
$$
Notice that the self-dual limit corresponds to the choice
$s'=N(1-\kappa a^2+4aN^2)$ in all the expressions. The parameter
$\kappa$ is the scalar curvature of the metric and we fix
$\kappa=1$, as we did previously.

   These metrics have interesting limits. For $a=0$ and $N$
different from zero becomes the AdS Taub-Nut solution with local
metric \be\lb{ADSTN} g_q=V(\widetilde{r})(d\widetilde{t}-2N
\cos\widetilde{\theta}
d\widetilde{\phi})^2+\frac{d\widetilde{r}^2}{V(\widetilde{r})}+(\widetilde{r}^2-N^2)(d\widetilde{\theta}^2+\sin^2\widetilde{\theta}
d\widetilde{\phi}^2), \ee being $V(\widetilde{r})$ given by
$$
V(\widetilde{r})=\frac{\lambda}{R^2}=\frac{1}{\widetilde{r}^2-N^2}\bigg(\widetilde{r}^2+N^2
-(\widetilde{r}^4-6N^2\widetilde{r}^2-3N^4)-2s'\widetilde{r}\bigg).
$$
This metric has been considered in different contexts
\cite{Demianski2}-\cite{Demianski6}. The parameter $s'$ is a mass
parameter and $N$ is a nut charge. Both parameters are not
independent in the quaternion Kahler limit, in fact the self-duality
condition $s'=s$ relates them as $s'=N(1+4N^2)$. If the mass were
arbitrary then the metric will possess a "bolt", but in this case
the metric will possess a "nut", that is, a zero dimensional regular
fixed point set. The isometry group of (\ref{ADSTN}) is enhanced from
$U(1)\times U(1)$ to $SU(2)\times U(1)$ in this limit. The anti
self-dual part of the spin connection reads
$$
\omega_-^1=-\sqrt{(\widetilde{r}+N)V(\widetilde{r})}\sin
\widetilde{\theta}d\widetilde{\phi}, \qquad
\omega_-^3=\sqrt{(\widetilde{r}+N)V(\widetilde{r})}d\widetilde{\theta},
$$
\be\lb{rn}
\omega_-^2=(\widetilde{r}-N)d\widetilde{t}+g(\widetilde{r})\cos\widetilde{\theta}d\widetilde{\phi},
\ee being $g(\widetilde{r})$ defined by
$$
g(\widetilde{r})=\bigg(\frac{N^2(\widetilde{r}-N)+N(1+4N^2)+(1+6N^2)
\widetilde{r}- 2N \widetilde{r}^2}{\widetilde{r}-N}\bigg).
$$
By taking the further limit $N=0$, that is, but switching off the
mass and the charge, we obtain after introducing the new radius
$\widetilde{r}=\sin\widetilde{\rho}$ the following distance element
$$
g_q=\cos^2\widetilde{\rho}d\widetilde{t}^2+d\widetilde{\rho}^2+\sin^2\widetilde{\rho}(d\widetilde{\theta}^2+\sin^2\widetilde{\theta}
d\widetilde{\phi}^2).
$$
The anti-self-dual spin connection takes the simple form
$$
\omega_-^1=\cos\widetilde{\rho}\sin
\widetilde{\theta}d\widetilde{\phi}, \qquad
\omega_-^2=\sin\widetilde{\rho}d\widetilde{t}+\cos\widetilde{\theta}d\widetilde{\phi},\qquad
\omega_-^3=\cos\widetilde{\rho}d\widetilde{\theta},
$$
and it follows that we have obtained the metric of the sphere
$S^4=SO(5)/SO(4)$. If we would choose negative scalar curvature
instead, this limit would correspond to the non compact space
$SO(4,1)/SO(4)$. Both cases are maximally symmetric and for this
reason this is called the $AdS_4$ limit of the AdS-Taub-Nut
solution.

    The other 4-dimensional quaternion Kahler manifold is CP(2) with
the Fubbini-Study metric. This case is obtained by
defining the new coordinates $\widehat{r}=N(\widetilde{r}-N)$ and
$\widetilde{t}=2N\xi$ and taking the limit $N\to \infty$. The
result, after defining
$\widetilde{\rho}=\widehat{r}^2/4(1+\widehat{r}^2)$, is the metric
\be\lb{sepedos}
g_q=\frac{\widetilde{\rho}^2}{2(1+\widetilde{\rho}^2)^2}(d\xi-
\cos\widetilde{\theta}
d\widetilde{\phi})^2+\frac{2d\widetilde{\rho}^2}{(1+\widetilde{\rho}^2)^2}
+\frac{\widetilde{\rho}^2}{2(1+\widetilde{\rho}^2)^2}(d\widetilde{\theta}^2+\sin^2\widetilde{\theta}
d\widetilde{\phi}^2). \ee By noticing that $\sigma_3=d\xi-
\cos\widetilde{\theta} d\widetilde{\phi}$ and that
$\sigma_1^2+\sigma_2^2=d\widetilde{\theta}^2+\sin^2\widetilde{\theta}
d\widetilde{\phi}^2$ we recognize from (\ref{sepedos}) the Bianchi
IX form for the Fubbini-Study metric on CP(2)=$SU(3)/SU(2)$.

   Another possible limit of (\ref{demi}) is $N \to 0$, the result will be the
AdS-Kerr euclidean solution, namely
\be\lb{demi}
g_q=\frac{\sin^2\widetilde{\theta}(1-a^2\cos^2\widetilde{\theta})}{\widetilde{r}^2-a^2\cos^2\widetilde{\theta}}
(\;a
d\widetilde{t}-\frac{\widetilde{r}^2-a^2}{\Xi}d\widetilde{\phi}\;)^2\ee
$$
+\frac{\widetilde{r}^2-a^2\cos\widetilde{\theta}^2}{1-a^2\cos^2\widetilde{\theta}}
d\widetilde{\theta}^2+\frac{\widetilde{r}^2-a^2\cos\widetilde{\theta}^2}{(\widetilde{r}^2-a^2)(1-\widetilde{r}^2)}d\widetilde{r}^2
+\frac{(\widetilde{r}^2-a^2)(1-\widetilde{r}^2)}{\widetilde{r}^2-a^2\cos\widetilde{\theta}^2}(\;d\widetilde{t}+\frac{a\sin^2\widetilde{\theta}}{\Xi}
d\widetilde{\phi}\;)^2.$$ The anti-self-dual connection $\omega_-^i$
is in this case $$
\omega_-^1=-\frac{1}{\widetilde{r}-a\cos\widetilde{\theta}}\sqrt{(1-a^2\cos\widetilde{\theta}^2)
(\widetilde{r}^2-a^2)(1-\widetilde{r}^2)}
\frac{\sin\widetilde{\theta}}{\Xi}d\widetilde{\phi},
$$
$$
\omega_-^2=(\widetilde{r}-a\cos\widetilde{\theta})d\widetilde{t}+\frac{1}{\widetilde{r}-a\cos\widetilde{\theta}}
\frac{W(\widetilde{r}, \widetilde{\theta})}{\Xi}d\widetilde{\phi},
$$
\be\lb{rn2}
\omega_-^3=\frac{1}{\widetilde{r}-a\cos\widetilde{\theta}}\bigg(
\sqrt{\frac{(\widetilde{r}^2-a^)(1-\widetilde{r}^2)}{1-a^2\cos^2\widetilde{\theta}}}d\widetilde{\theta}
-\sqrt{\frac{1-a^2\cos^2\widetilde{\theta}}{(\widetilde{r}^2-a^2)(1-\widetilde{r}^2)}}a\sin{\widetilde{\theta}}d\widetilde{r}\bigg),
\ee where we have defined the function \be\lb{sh}
W(\widetilde{r},\widetilde{\theta})=[(\widetilde{r}-a\cos\widetilde{\theta})^2
-a+(1+a^2)\widetilde{r} \cos\widetilde{\theta}-a
\widetilde{r}^2\cos^2\widetilde{\theta}]. \ee The parameter $a$ is
usually called rotational parameter, although we have no the notion
of rotational black hole in euclidean signature. The mass parameter
$s$ and the nut charge are zero in this case.

\subsection{Toric quaternion Kahler spaces}

    We turn now to more general toric quaternion Kahler
orbifolds, following \cite{Caldo}. As we have mentioned, in four
dimensions quaternion Kahler spaces are the same than Einstein
spaces with self-dual Weyl tensor. The self-duality condition is
conformally invariant. This means that if a metric $g$ is self-dual,
then all the family $[g]$ of metrics obtained by $g$ by conformal
transformations is self-dual. The Einstein condition instead is not
invariant under conformal transformations.

  Let us focus first in the construction of self-dual families with
$U(1)\times U(1)$ isometry, the Einstein condition will be
considered afterwards. For any representative $g$ of a toric
conformal family $[g]$ there exist a coordinate system for which the
metric is expressed in the Gowdy form \be\lb{form} g = g_{ab}dx^a
dx^b + g_{\alpha\beta}dx^{\alpha} dx^{\beta}. \ee The latin and
greek indices takes values 1 and 2. Both $g_{ab}$ and
$g_{\alpha\beta}$ are supposed to be independent of the coordinates
$x^{\alpha}=(\alpha, \beta)$. The Killing vectors are then
$\partial_{\alpha}$ and $\partial_{\beta}$ and are commuting, so
there is a $U(1)\times U(1)$ action on the manifold. This
considerations are local and no assumption about the $U(1)$ actions
is made. \footnote{We are loosely speaking about toric conformal
families. If a metric $g$ has two commuting Killing vectors
$\partial_{\alpha}$ and $\partial_{\beta}$, then by a conformal
transformation $g\rightarrow \Omega^2 g$ depending on $(\alpha,
\beta)$ we will obtain a new metric which is not toric anymore.
Along this section the $\Omega$´s are supposed to be independent on
$(\alpha, \beta)$. In this case it is ensured that
$\partial_{\alpha}$ and $\partial_{\beta}$ will be Killing vectors
of every metric in $[g]$.}

Clearly, the part $g_{ab}dx^a dx^b$ in (\ref{form}) can be
interpreted as a two dimensional metric. By a theorem due to Gauss
it is known that every two dimensional metric is conformally flat.
Therefore the anzatz (\ref{form}) can be written as \be\lb{form2} g
= \Omega^2(d\rho^2+ d\eta^2) + g_{\alpha\beta}dx^{\alpha}
dx^{\beta}, \ee being now $g_{\alpha\beta}$ functions of the
coordinates $(\rho, \eta)$ and $\Omega$ a conformal factor with the
same coordinate dependence. Because self-duality is conformally invariant
we can make a conformal transformation to (\ref{form2}) and consider
the following metric \be\lb{form3} \overline{g} =\frac{(d\rho^2+
d\eta^2)}{\rho^2} + \overline{g}_{\alpha\beta}dx^{\alpha}
dx^{\beta}, \ee  without loosing generality.
The factor $\rho^2$ was introduced by convenience. It is natural to
express $\overline{g}_{\alpha\beta}dx^{\alpha} dx^{\beta}$ in terms
of certain functions $\overline{A}_i$ and $\overline{B}_i$ ($i=1,2$)
of $(\rho, \eta)$ as
\be\lb{form41}
\overline{g}_{\alpha\beta}dx^{\alpha} dx^{\beta}=(\overline{A}_0
d\theta - \overline{B}_0 d\varphi)^2 + ( \overline{A}_1 d\theta -
\overline{B}_1 d\varphi)^2.
\ee
But this is not the most simplifying form in order to impose the
self-duality condition. It is more convenient to write it as
\be\lb{ono} \overline{g}_{\alpha\beta}dx^{\alpha}
dx^{\beta}=\frac{(A_0 d\theta - B_0 d\varphi)^2 + ( A_1 d\theta -
B_1 d\varphi)^2}{(A_{0} B_{1}-A_{1}B_{0})^2}, \ee in terms of new
functions $A_i$ and $B_i$. The relation with the other functions
$\overline{A}_i$ and $\overline{B}_i$ is given by comparison of the
last two expressions. Although considering the anzatz (\ref{ono})
could seem non practical, the self-duality
condition became a linear differential system in terms of $A_i$ and
$B_i$.\footnote{An intuitive argument to see that an anzatz of the
form (\ref{form4}) could simplify the self-duality condition goes as
follows. It is known that for any group of four vector fields $e_i$,
the Ashtekar et all equations \cite{Ashtekar}
$$
[e_1,e_2]+[e_3, e_4]= 0, \;\;\;\; [e_1, e_3]+[e_4, e_2]= 0, \;\;\;\;
[e_1, e_4]+[e_2, e_3]= 0,
$$
define a self-dual metric $g=\delta_{ab}e^a \otimes e^b$ (here $e_i$
is the dual basis of $e^a$) called hypercomplex metric. The reader
can check that if we select the following soldering forms
$$
e^1=d\rho, \qquad e^2=d\eta,\qquad e^3=(\frac{A_0 d\theta - B_0
d\varphi}{A_{0} B_{1}-A_{1}B_{0}}),\qquad e^4=(\frac{A_1 d\theta -
B_1 d\varphi}{A_{0} B_{1}-A_{1}B_{0}})
$$
for the metric (\ref{form4}) then the inverse basis take the simple
form \be\lb{einan} e_1=\partial_{\rho},\;\;\; e_2=\partial_{\eta}
\ee \be\lb{einan2} e_3= B_1
\partial_{\theta} + A_1\partial_{\varphi},\;\;\; e_4= B_0
\partial_{\theta} + A_0 \partial_{\varphi},
\ee and the Ashtekar conditions became simply Cauchy-Riemann
equations
$$
(A_1)_{\rho}=(A_0)_{\eta}, \qquad (A_0)_{\rho}=-(A_1)_{\eta}
$$
$$
(B_1)_{\rho}=(B_0)_{\eta}, \qquad (B_0)_{\rho}=-(B_1)_{\eta}
$$
which are linear. If (\ref{form41}) is used instead, then the
resulting system will take a non linear "look".} Therefore it is
better to impose the self-duality condition to \be\lb{form4}
\overline{g} =\frac{(d\rho^2+ d\eta^2)}{\rho^2} + \frac{(A_0 d\theta
- B_0 d\varphi)^2 + ( A_1 d\theta - B_1 d\varphi)^2}{(A_{0}
B_{1}-A_{1}B_{0})^2}. \ee If also the commuting Killing vectors are
surface orthogonal, then the condition $W=\ast W$ give the linear
system \be\lb{joyce1} (A_0)_{\rho}+(A_1)_{\eta}=A_0/\rho, \ee
\be\lb{joyce2} (A_0)_{\eta}-(A_1)_{\rho}=0, \ee and the same
equations for $B_i$. This system was found by Joyce in \cite{Joyce}.
Surface orthogonality implies that the manifold $M$ corresponding to
(\ref{form4}) is of the form $M=N\times T^2$, being $T^2$ the two
dimensional torus. There exists other examples (as those in the
footnote) which are solution of the self-duality conditions but for
which this decomposition do not hold.

The \emph{linear} system (\ref{joyce1}) and (\ref{joyce2}) is enough
simple. It is easy to check that (\ref{joyce2}) implies the
existence of a potential function $G$ such that \be\lb{potencial1}
A_0=G_\rho;\;\;\;\;\;\; A_1=G_{\eta}. \ee Then (\ref{joyce1}) gives
that $G_{\rho\rho} + G_{\eta\eta}=G_{\rho}/\rho$. Inversely we
deduce from (\ref{joyce1}) the existence of another potential $V$
such that \be\lb{potencial1} A_0=-\rho V_\eta;\;\;\;\;\;\; A_1=\rho
V_{\rho}, \ee and (\ref{joyce2}) gives the Ward monopole equation
\cite{Ward} \be\lb{Wardy} V_{\eta\eta}+\rho^{-1}(\rho
V_{\rho})_{\rho}=0. \ee The relations \be\lb{shu} G_\rho=-\rho
V_{\eta};\;\;\;\;\;\;\ G_{\eta}=\rho V_{\rho}, \ee constitute a
Backlund transformation allowing to find a monopole $V$ starting
with a known potential $G$ or viceversa. The functions $B_i$ can be
also expressed in terms of another potential functions $G'$ and $V'$
satisfying the same equations than $V$ and $G$.

The next task is to extract the Einstein representatives of the
self-dual families defined in (\ref{form4}). This will be
automatically quaternion Kahler. In general, to impose the Einstein
condition directly to (\ref{form4}) could give no results. In order
to be general, a transformation
$\overline{g}\to\Omega^2\overline{g}$ should be made to the metrics
(\ref{form4}) and the Einstein condition $R_{ij}=\Lambda g_{ij}$
should be imposed to the transformed metrics. This requirement will
fix the factor $\Omega$ and will give further relations between
$A_i$ and $B_i$. The result obtained from this condition is that
\be\lb{wold} A_1 B_0 - A_0 B_1=\rho( A_0^2 + A_1^2) - G A_0, \ee and
that $\Omega^2=G$. From (\ref{wold}) it is obtained that $B_0 = \rho
A_1 + \xi_0$ and $B_1 = G - \rho A_0 + \xi_1$ with $A_1 \xi_0 = A_0
\xi_1$. The functions $\xi_i$ are determined by the requirement that
$B_i$ also satisfy the Joyce system (\ref{joyce1}) and
(\ref{joyce2}), the result is $\xi_0=-\eta A_0$ and $\xi_1=-\eta
A_1$. Then it is obtained that \be\lb{Caldon1}
A_0=G_\rho;\;\;\;\;\;\; A_1=G_{\eta} \ee \be\lb{Caldon2} B_0=\eta
G_{\rho}-\rho G_{\eta};\;\;\;\;\;\; B_{1}=\rho G_{\rho} + \eta
G_{\eta}-G. \ee By defining $G=\sqrt{\rho}F$ it follows from the
equation $G_{\rho\rho} + G_{\eta\eta}=G_{\rho}/\rho$ that F
satisfies
$$
F_{\rho\rho} + F_{\eta\eta}=\frac{3F}{4\rho^2}.
$$
The final expression of the metric is \cite{Pedersen}
$$
g_q=\frac{F^2-4\rho^2(F^2_{\rho}+F^2_{\eta})}{4F^2}\frac{d\rho^2+d\eta^2}{\rho^2}
$$
\be\lb{metric} +\frac{[(F-2\rho F_{\rho})u-2\rho
F_{\eta}v]^2+[-2\rho F_{\eta}u +(F+2\rho
F_{\rho})v]^2}{F^2[F^2-4\rho^2(F^2_{\rho}+F^2_{\eta})]}, \ee where
$u=\sqrt{\rho}d\alpha$, $v=(d\beta+\eta d\alpha)/\sqrt{\rho}$ and
$F(\rho, \eta)$ is a solution of the equation \be\lb{backly}
F_{\rho\rho}+F_{\eta\eta}=\frac{3F}{4\rho^2} \ee on some open subset
of the half-space $\rho>0$.

There exist a theorem that insure that a toric quaternion Kahler
manifold will always have surface orthogonal Killing vectors
\cite{Caldo}. Therefore the metrics (\ref{metric}) are the most
general toric quaternion Kahler ones. On the open set defined by
$F^2 >4\rho^2(F^2_{\rho}+F^2_{\eta})$ g has positive scalar
curvature, whereas $F^2 < 4\rho^2(F^2_{\rho}+F^2_{\eta})$ -g is
self-dual with negative scalar curvature. This are known as the
Calderbank-Pedersen metrics.

\subsection{The manifolds of the family: CP(2) and $S^4$}

    If we choose a potential $F$ independent on the coordinate $\eta$, then
it is obtained from (\ref{backly}) that it should be of the form
$F(\rho)=\rho^{3/2}-c\rho^{-1/2}$ being $c$ a constant. In this case
the metric (\ref{metric}) will have the explicit form
\be\lb{Kahler-eins}
g_K=\frac{\rho^2+c}{(\rho^2-c)^2}(d\rho^2+d\eta^2+\frac{d\alpha^2}{4})
+\frac{\rho^2}{(\rho^2-c)^2(\rho^2+c)}(d\beta+\eta d\alpha)^2, \ee
and the isometry group will be enlarged by the presence of a new
Killing vector $\partial_{\eta}+\beta\partial_{\alpha}$. The action
of the isometry group on the coordinates is given by
\be\lb{isometro} \alpha\rightarrow \alpha + a_1, \qquad
\eta\rightarrow \eta+a_2,\qquad \beta \rightarrow
\beta+a_3-a_2\alpha, \ee being $a_i$ constant parameters. It is
worthy to mention that the maximally symmetric quaternion Kahler
spaces possess $SO(5)$ or $SO(4,1)$ isometry group, therefore there
are can be at most two commuting isometries in a 4-dimensional
quaternion Kahler space. In other words, it is impossible that the
third isometry will commute with the others. Indeed the three
Killing vectors $T_i$ satisfy the three dimensional Heisenberg
algebra \be\lb{hai} [T_1, T_3]=T_2, \qquad [T_1, T_2]=[T_3, T_2]=0.
\ee The isometry group (\ref{isometro}) preserve the triplet
(\ref{threefor}).

If also $c=0$ then the potential function is $F=\rho^{3/2}$ and the
expression for the quaternion Kahler metric will be
\be\lb{Kahler-eins}
g_K=\frac{1}{\rho^2}(d\rho^2+d\eta^2+\frac{d\alpha^2}{4})
+\frac{1}{\rho^4}(d\beta+\eta d\alpha)^2. \ee Let us introduce the
two form
$$ \overline{J}'=\frac{d\alpha\wedge
d\eta}{2\rho^2}+d(\frac{1}{2\rho^2})\wedge (d\beta+\eta d\alpha),
$$
which preserved locally under the action (\ref{isometro}). The
$(1,1)$ tensor $J'$ defined through the relation $g_q(\cdot,
J'\cdot)=\overline{J}'$ is an almost complex structure defined over
the quaternion Kahler space $M$. The two form $\overline{J}'$ is
evidently closed, thus sympletic and it can be expressed
$\overline{J}'=dA$ being the one form $A$ given by \be\lb{formol}
A=-\frac{(d\beta+\eta d\alpha)}{2\rho^2}. \ee By introducing the
complex quantities \be\lb{elem} S=\rho^2+i(2\beta+\eta
\alpha)+C\overline{C},\qquad C=i\eta+\frac{1}{2}\alpha, \ee it
follows that the metric (\ref{Kahler-eins}) can be written as
\be\lb{Kahler-eins2} g_K=u\otimes \overline{u}+v\otimes \overline{v}
\ee where \be\lb{dero} u=\frac{1}{\rho}dC,\qquad
v=\frac{1}{2\rho^2}(dS + C d\overline{C}). \ee This is the classical
metric of the universal dilaton hypermultiplet \cite{Antoniadis} and
is known to be Kahler with Kahler potential \be\lb{kahl}
K=\log(S+\overline{S}-2C\overline{C}). \ee This means that space $M$
is not only sympletic, but Kahler. It is also quaternion-Kahler
(thus Einstein), therefore is \emph{Kahler-Einstein}. The form
(\ref{formol}) is expressed in the coordinates $(S, C, \overline{S},
\overline{C})$ as \be\lb{formol2}A=2(\frac{dS-d\overline{S}+C
d\overline{C}-\overline{C}dC}{S+\overline{S}-2C\overline{C}}). \ee
By going to the coordinate system defined by
$$
z_1=\frac{1-S}{1+S}, \qquad z_2=\frac{2C}{1+S}
$$
it is recognized that (\ref{dero}) is the Bargmann metric
\be\lb{Bargmann} g_q=\frac{dz_1 d\overline{z}_1+dz_2
d\overline{z}_2}{1-|z_1|^2-|z_2|^2}-\frac{(z_1 d\overline{z}_1+z_2
d\overline{z}_2)(\overline{z}_1 dz_1+
\overline{z}_2dz_2)}{(1-|z_1|^2-|z_2|^2)^2},\ee defined over the
unit open ball in $C^2$ given by $|z_1|^2+|z_2|^2<1$, with Kahler
potential $K=\log(1-|z_1|^2-|z_2|^2)$. This space is topologically
equivalent to the homogeneous symmetric space $SU(2,1)/SU(2)\times
SU(1)$. By going to spherical coordinates
$$
z_1=r\cos\frac{u}{2} \exp(i\frac{(v+w)}{2}),\qquad
z_2=r\sin\frac{u}{2} \exp(-i\frac{(v-w)}{2}),
$$
with $0<r<1$, $0<u<\pi$, $0<v<2\pi$ and $0<w<4\pi$, the Bargmann
metric take the Bianchi IX form \be\lb{Biancho}
g_q=\frac{dr^2}{(1-r^2)^2}+\frac{r^2\sigma_1^2}{(1-r^2)^2}+\frac{r^2(\sigma_2^2+\sigma_3^2)}{1-r^2}
\ee with manifest $SU(2)$ symmetry.

Let us note that the metrics (\ref{Kahler-eins}) can be extended
with no difficulty to an Einstein-Sasaki space in five dimensions by
use of (\ref{lome}), the local form of such metrics is \be\lb{esas2}
g_{es}=[d\tau  -\frac{(d\beta+\eta d\alpha)}{2\rho^2}]^2
+\frac{1}{\rho^2}(d\rho^2+d\eta^2+\frac{d\alpha^2}{4})
+\frac{1}{\rho^4}(d\beta+\eta d\alpha)^2,\ee and is seen that the
isometry group (\ref{isometro}) is an isometry of (\ref{esas2}). We
have three commuting Killing vectors, namely $\partial_{\alpha}$
$\partial_{\beta}$ and $\partial_{\tau}$, and so there is a $T^3$
action.

  For the Bargmann metric
$F^2<4\rho^2F_{\rho}^2$ and this means that $\kappa<0$. Thus the
construction of tri-Sasaki metrics presented in previous sections
can not be applied to this example. But by making the replacement
$z_i\to i z_i$ the Kahler potential of the Bargmann metric will be
converted into $K=\log(1+|z_1|^2+|z_2|^2)$. This is the potential
for the Fubbini-Study metric over CP(2), which is dual to the
Bergmann one. This metric is also Kahler-Einstein and quaternion
Kahler. Its metric tensor and the corresponding potential form $A$
can obtained from the formulas corresponding to the Bargmann metric
by making the replacement $z_i\to i z_i$. But different from the
Bargmann metric, this metric possess positive scalar curvature and
the construction presented in previous section can be applied. The
anti self-dual part of the spin connection of the Fubbini-Study
metric is given by
$$
\omega_-^1=-\frac{1}{2}\bigg(\frac{\overline{z}_2 dz_1+z_2
d\overline{z}_1}{(1+|z_1|^2)\sqrt{1+|z_1|^2+|z_2|^2}}\bigg), \qquad
\omega_-^2=\frac{i}{2}\bigg(\frac{\overline{z}_2 dz_1-z_2
d\overline{z}_1}{(1+|z_1|^2)\sqrt{1+|z_1|^2+|z_2|^2}}\bigg),
$$
\be\lb{fubbi} \omega_-^3=\frac{i}{2}\bigg(\frac{\overline{z}_1
dz_1-z_1 d\overline{z}_1+z_2 d\overline{z}_2-\overline{z}_2
dz_2}{(1+|z_1|^2)}\bigg). \ee From (\ref{trusa2}) it is obtained the
tri-Sasaki metric \be\lb{triso}
g_7=(\sigma_i-\omega_-^i)^2+\frac{dz_1 d\overline{z}_1+dz_2
d\overline{z}_2}{1+|z_1|^2+|z_2|^2}- \frac{(z_1 d\overline{z}_1+z_2
d\overline{z}_2)(\overline{z}_1 dz_1+
\overline{z}_2dz_2)}{(1+|z_1|^2+|z_2|^2)^2}, \ee which is fibered
over the Fubbini-Study metric. Here $\sigma_i$ the one Maurer-Cartan
forms in (\ref{mcarta}) and $\omega_-^i$ is given in (\ref{fubbi}).

  The same procedure can be applied to the sphere $S^4$, which, together with CP(2)
constitute the unique 4-dimensional quaternion Kahler spaces that
are manifolds. The corresponding Kahler-Einstein and tri-Sasaki
metrics are respectively
$$
g_6=\frac{1}{2}d\Omega_4^2+(d\theta-\sin\varphi A^2 +\cos\varphi
A^1)^2
$$
$$
+(\sin\theta d\varphi-\cos\theta\sin\varphi A^1-
\cos\theta\cos\varphi A^2+\sin\theta A^3)^2,
$$
$$
g_7=(A_i-\sigma^i)^2+\frac{1}{2}d\Omega_4^2,
$$
being $A_i$ the unit charge instantons on $S^4$ and $d\Omega_4^2$
the usual metric for the sphere.

\subsection{More general quaternion Kahler orbifolds}

  The space of metrics defined in (\ref{metric}) is very rich. They encode
many well known examples in the literature as well as new ones. We briefly
describe some of them, but a much more complete description can be found
in the original references \cite{Pedersen}-\cite{Pedersen2}.

  The Backglund transformation defined in (\ref{shu}) is a
correspondence between solutions $F$ of (\ref{backly}) and solutions
of the Ward monopole equation (\ref{Wardy}). The Ward monopole
equation describe hyperKahler metrics with two commuting Killing
vectors, which in cylindrical coordinates take the form \cite{Ward}
\begin{equation}
g=\frac{(dt+\rho V_{\rho }d\varphi )^{2}}{V_{\eta }}+V_{\eta }(d\rho
^{2}+d\eta ^{2}+\rho ^{2}d\varphi ^{2}).  \label{Wardfor}
\end{equation}%
Several of these metrics were consider recently in \cite{Calderbank}-\cite{Calderbank2}.
The commuting Killing vectors are $\partial_{t}$ and
$\partial_{\varphi}$. It is not difficult to see that these metrics
are hyperKahler. By defining the one form $ A=\rho V_{\rho }d\varphi
$ and $U=V_{\eta }$ the metrics (\ref{Wardfor}) take the
Gibbons-Hawking form \cite{Gibbonhak}
\begin{equation}
g=V^{-1}(dt+A)^{2}+Vdx_{i}dx_{j}\delta ^{ij},  \label{ashgib}
\end{equation}%
and it follows that $A$ and $V$ satisfy the linear system of
equations
\begin{equation}
\nabla V=\nabla \times A.  \label{Gibb-Hawk}
\end{equation}%
Any element of the family (\ref{Gibb-Hawk}) is hyper-K\"{a}hler with
respect to the hyper-K\"{a}hler triplet
\[
\overline{J}_{1}=(dt+A)\wedge dx-Vdy\wedge dz
\]%
\begin{equation}
\overline{J}_{2}=(dt+A)\wedge dy-Vdz\wedge dx  \label{transurop3}
\end{equation}%
\[
\overline{J}_{3}=(dt+A)\wedge dz-Vdx\wedge dy
\]%
which is actually $t$-independent. Therefore the Killing vector
$\partial_t$ is tri-holomorphic. Instead $\partial_{\varphi}$ is
not, that is, it do not preserve (\ref{transurop3}). It follows that
the Backglund transformation is a correspondence between toric
quaternion Kahler spaces and toric hyperKahler spaces with at least one
tri-holomorphic isometry.

  The elemental solution of the equation (\ref{Gibb-Hawk}) is given by
the single Wu-Yang monopole potential $A$ and the scalar field $V$
of the form
\begin{equation}
V=c+\frac{a}{r},\qquad A=\frac{a(ydx-xdy)}{r(r+z)}\qquad z>0,\qquad
\widetilde{A}'=\frac{a(ydx-xdy)}{r(r-z)}\qquad z\leq 0,
\label{scalar}
\end{equation}%
being $r$ the radius $r=\sqrt{x^{2}+y^{2}+z^{2}}$. The vector
potential $A$ is not globally defined in ${\bf R}^{3}$ due to the
presence of Dirac string singularities in the upper $z$ axis for
$A$ and in the lower $z$ axis for $%
\widetilde{A}'$. In the overlapping region both potentials differ
one to each other by a gauge transformation $\widetilde{A}'=A-2a\
d\arctan (y/x)$. Any array of Dirac monopoles will describe an
hyperKahler metric, but only if such monopoles are aligned along an
axis, then we will have axial symmetry as in the Ward case
(\ref{Wardfor}). The elementary Ward solutions are of the form
$$ U_{i}=a_{i}\log (\eta -\eta _{i}+\sqrt{(\eta -\eta _{i})^{2}+\rho
^{2}})
$$
and represent a monopole located in the position $(0,\eta _{i})$.
Any superposition of such elementary functions will give rise to a
toric hyperKahler metric. The Backglund transformation of the
solutions $U_i$ is given by \be\lb{elot}
F_{i}=\frac{\sqrt{a_i^2\rho^2+(a_i\eta-\eta_i)^2}}{\sqrt{\rho}}. \ee
Any superposition of these solutions, namely
$$
F=\sum_{k=1}^m
\frac{\sqrt{a_k^2\rho^2+(a_k\eta-b_k)^2}}{\sqrt{\rho}}.
$$
will give rise to a toric quaternion Kahler metric. There are also
elementary solution $F=\rho^{3/2}$ and $F=\rho^{-1/2}$, which are $\eta$-independent.

     For $m=2$ the solutions are called $2$-pole functions and are
given by
$$
F_1=\frac{1+\sqrt{\rho^2+\eta^2}}{\sqrt{\rho}};\;\;\;
F_2=\frac{\sqrt{(\rho)^2+(\eta+1)^2}}{\sqrt{\rho}}-
\frac{\sqrt{(\rho)^2+(\eta-1)^2}}{\sqrt{\rho}}.
$$
The first one gives rise to the spherical metric, while the second
one gives rise to the hyperbolic metric. This is seen by defining
the coordinates $(r_1, r_2)$ which are related to the hyperbolic
ones $(\rho, \eta)$ by the relation
$$
(r_1 + i r_2)^2=\frac{\eta-1+i\rho}{\eta+1+i\rho}.
$$
By writing the corresponding metric in terms of $(\rho, \eta)$ and
making the change to $(r_1, r_2)$ gives \cite{Pedersen} \be\lb{ou}
g=(1-r_1^2-r_2^2)^{-2}(dr_1^2+dr^2_2 + r_1^2d\theta_1^2 +
r_2^2d\theta_2^2), \ee which is the hyperbolic metric on the unit
ball on $R^4$. This is a conformally flat metric.

  Now, let us discuss the case of two monopoles on the $z$ axis.
Without loosing generality, it can be considered that the monopoles
are located in the positions $(0,0,\pm c)$. The potentials for this
configurations are
\[
V=\frac{1}{r_{+}}+\frac{1}{r_{-}},\qquad A=A_{+}+A_{-}=\left( \frac{z_{+}}{%
r_{+}}+\frac{z_{-}}{r_{-}}\right) d\arctan (y/x),\qquad r_{\pm }=\sqrt{%
x^{2}+y^{2}+(z\pm m)^{2}}.
\]%
This case corresponds to the Eguchi-Hanson instanton, whose metric,
in Cartesian coordinates, reads
\begin{equation}
g=\left( \frac{1}{r_{+}}+\frac{1}{r_{-}}\right) ^{-1}\left(
\;dt+\left( \frac{z_{+}}{r_{+}}+\frac{z_{-}}{r_{-}}\right)
\;d\arctan (y/x)\;\right) ^{2}+\left(
\;\frac{1}{r_{+}}+\frac{1}{r_{-}}\;\right)
(\;dx^{2}+dy^{2}+dz^{2}\;).  \label{eguchi}
\end{equation}%
In order to recognize the Eguchi-Hanson metric in its standard form
it is convenient to introduce a new parameter $a^{2}=8m,$ and the
elliptic coordinates defined by
\[
x=\frac{r^{2}}{8}\sqrt{1-(a/r)^{4}}\sin \varphi \cos \theta ,\quad y=\frac{%
r^{2}}{8}\sqrt{1-(a/r)^{4}}\sin \varphi \sin \theta ,\quad z=\frac{r^{2}}{8}%
\cos \varphi .
\]%
In this coordinate system it can be checked that
\[
r_{\pm }=\frac{r^{2}}{8}\left( 1\pm \left( a/r\right) ^{2}\cos
\varphi \right) ,\qquad z_{\pm }=\frac{r^{2}}{8}\left( \cos \varphi
\pm (a/r)^{2}\right) ,\qquad V=\frac{16}{r^{2}}\left(
1-(a/r)^{4}\cos ^{2}\varphi \right) ^{-1},
\]%
\[
A=2\;\left( 1-(a/r)^{4}\cos ^{2}\varphi \right) ^{-1}\;\left(
1-(a/r)^{4}\right) \;\cos \varphi \;d\theta ,
\]%
and, with the help of these expressions, it is found
\begin{equation}
g=\frac{r^{2}}{4}\left( \;1-(a/r)^{4}\;\right) \;(\;d\theta +\cos
\varphi
d\tau \;)^{2}+\left( \;1-(a/r)^{4}\;\right) ^{-1}\;dr^{2}+\frac{r^{2}}{4}%
\;(\;d\varphi ^{2}+\sin ^{2}\varphi d\tau \;)  \label{ego}
\end{equation}%
being $\tau =2t$. This is actually a more familiar expression for
the Eguchi-Hanson instanton, indeed \cite{Eguchi-Hanson}. Let us also note that the
Eguchi-Hanson metric corresponds to two monopoles in the z axis, but
if we choose $m^2=-1$ this will correspond to a to the potential for
an axially symmetric circle of charge. The corresponding metric is
called Eguchi-Hanson metric of the type I, and is always incomplete.

Lets go back to the Backglund transformed geometry corresponding to
the Eguchi-Hanson metric. The general "$3$-pole" solutions can be
written as
$$
F=\frac{a}{\sqrt{\rho}}+\frac{b+c/m}{2}\frac{\sqrt{\rho^2+(\eta+m)^2}}{\sqrt{\rho}}
+\frac{b-c/m}{2}\frac{\sqrt{\rho^2+(\eta-m)^2}}{\sqrt{\rho}}.
$$
By definition $-m^2=\pm 1$, which means that $m$ can be imaginary or
real. The corresponding solutions are denominated type I and type II
respectively, by analogy with the hyperKahler case. It is
interesting to note that for $c=0$ and $b=-1$ and defining the
coordinates $(t, \theta)$ by $\eta=(\cosh^2t-1)\cos\theta$ and
$\rho=2 \coth t \sinh^{-1} t\sin\theta$ the metric take the form
$$
g_q=4dt^2+\sinh^2 t(d\theta^2+\sin^2\theta d\varphi^2)+\frac{\sinh^2
2t}{2}(d\psi+\cos\theta d\varphi)^2.
$$
By making the further transformation $2t=\log(1+r)-\log(1-r)$ this
metric take the form (\ref{Biancho}) and therefore it is the
Bargmann metric. The Bargmann metric can also be obtained with the
function $F=\rho^{3/2}$ and this means that different solutions of
the equation (\ref{backly}) can give rise to the same metric. In
fact, the Bargmann metric can be recovered for  $c=0$ and $b=1$ and
also for $c=1$ and $b=0$. There exist certain freedom in the choice
of $F$ that leaves the metric invariant. This freedom allows in
particular to set $a=1$. If $c=0$ and $b$ is arbitrary then the
metric reduce to the Pedersen metric \cite{Pedor} (see also \cite{Caldo}) \be\lb{Bianchu}
g_q=\frac{(w^2+q^2)}{(w-sq r^2)^2}\bigg(\frac{ wr^2+s}{1+q
r^2}dr^2+\frac{r^2}{4}\bigg(\frac{1+q r^4}{w r^2 + s}\sigma_1^2+(w
r^2+s)(\sigma_2^2+\sigma_3^2)\bigg)\bigg), \ee being $w$, $q$ and
$s$ three parameters. We should be aware that these metrics
possesses only one parameter up to an homothety, the other two can
be selected to one by a suitable rescaling. This is in accordance
with that, by construction, the only parameter is $b$. The advantage
of this notation is that several limits are better understood. The
scalar curvature of this metrics is $-48wq/(w^2+q^2)$ and we see
that in the limit $w=0$ or $q=0$ the metric will be hyperKahler. In
the first case the metric reduce to the Taub-Nut one \cite{Taub-Nut}, in the second
corresponds to the I and II Eguchi-Hanson metrics.

  It is natural to introduce the cylindrical coordinate system
$$
\rho= \sqrt{R^2 \pm 1} \cos\theta, \;\;\;\; \eta=R \sin\theta,
$$
where $\theta$ takes values in the interval $(-\pi/2, \pi/2)$. In
these coordinates \be\lb{unade} \sqrt{\rho}F=1+b R+c \sin\theta, \ee
\be\lb{dosde}
\rho^{-1}[\frac{1}{4}F^2-\rho^2(F_{\rho}^2+F_{\eta}^2)]= \frac{b(R
\mp b) + c(\sin\theta + c)}{R^2 \pm \sin^2\theta}. \ee The zeroes of
$F$ are the conformal infinite of the metric, while the zeroes of
$\frac{1}{4}F^2-\rho^2(F_{\rho}^2+F_{\eta}^2)$ are singularities
separating the domains of positive and negative scalar curvature.
For the type II metrics the coordinate $R$ can take values between 1
and $\infty$. For $b=1$ and $c=0$ it is obtained the Fubbini-Study
metric on CP(2), while for $b=-1$ and $c=0$ or $b=0$ and $c=\pm 1$
it is obtained the Bargmann metric. By introducing the vector
$(b,c)$ we have that along the lines joining these four points the
metrics will be bi-axial Bianchi. Along the lines joining $(1, 0)$
with the other points there will live Bianchi IX metrics whereas on
the lines between $(-1, 0)$, $(0, 1)$ and $(0, -1)$ the metric is
Bianchi VIII. For the type I metrics the value of $R$ is non
restricted. The zeroes of $F$ defining the conformal infinite are
$R_{\infty}=-(1+c\sin\theta)/b$. The zeroes of
$\frac{1}{4}F^2-\rho^2(F_{\rho}^2+F_{\eta}^2)$ are
$R_{\pm}=(b^2+c^2+c\sin\theta)/b$. The case $b=0$ and correspond to
Bianchi VIII metrics. If $c=0$ we obtain the Pedersen metrics. The
conformal infinite is $R_{\infty}=1/b$ and $R_{\pm}=b$ \cite{Pedersen}.

There have  been found certain quaternion Kahler deformations of the
Taub-Nut gravitational instantons and other examples in
\cite{Valent}, and the relation between these metrics and the
$m$-pole solutions has been worked out in that reference. Also we
would like to remark that the orbifolds (\ref{plebanski}) can be
represented in the form (\ref{metric}) but the coordinate change is
rather complicated and we will not describe it here, see
\cite{Bryanthermi}-\cite{Aposgaucho}. Higher multi-instanton
solutions, or $m$-pole solutions are a linear combinations of the
form
$$
F=\sum_k^m\frac{\sqrt{a_k^2\rho^2+(a_k\eta-b_k)^2}}{\sqrt{\rho}}
$$
for some real parameters $(a_k, b_k)$ for $1<k<m$. But there is an
$SL(2,R)$ action that leaves the metric invariant up to an overall
factor and therefore, as a vector space, $(a_k, b_k)$ is $2m-dim
SL(2,R)-1=2(m-2)$-dimensional. The $m$-pole solutions arise as
quaternion-Kahler quotients of $HP^{m-1}$ by an $(m-2)$ dimensional
subtorus of a maximal torus $T^m$ in $Sp(m)$, all these metrics are
therefore defined on a compact orbifolds \cite{Galicki2}.
Applications of them to the universal hypermultiplet have been
found, for instance in \cite{Ketov}-\cite{Antoniadis}. Other
applications has been considered in \cite{yo}-\cite{Angelova}.

\section{Explicit tri-Sassaki and weak $G_2$ metrics and supergravity solutions}

\subsection{Tri-Sassaki and weak $G_2$ over AdS-Kerr and AdS-Taub-Nut}

  We are now in position to construct compact
tri-Sasaki and weak $G_2$ holonomy metrics. The main ingredient in
this construction is the proposition 1, applied to limiting cases of
the euclidean Plebanski-Demianski solution (\ref{demi}) or to the
toric metrics (\ref{metric}). But before to start is convenient to
mention that there exist in the literature Einstein-Sasaki spaces
fibered over the so called "orthotoric" Kahler-Einstein spaces
\cite{Mateos4}-\cite{orto}. As was show in \cite{orto} this spaces
can be obtained by taking certain scaling limit of the euclidean
Plebanski-Demianski metrics (\ref{demi}). In particular, there were
found several toric Einstein-Sasaki metrics defined over $S^2\times
S^3$. Nevertheless, those spaces are 5-dimensional and are
fibrations over 4-dimensional Kahler-Einstein spaces. Instead, we
are presenting a 7-dimensional family which is fibered over
4-dimensional quaternion Kahler spaces. Thus, in principle, our
examples bears no relation to those found in
\cite{Mateos4}-\cite{orto}.

    After this comment, we turn our attention now to the construction of tri-Sasaki
(and weak $G_2$) metrics implicit in proposition 1.
\\

\textit{The AdS-Taub-Nut case}
\\

It is direct, by using proposition 1 and the lifting formula
(\ref{brg2}), to work out tri-Sassaki and weak $G_2$ holonomy
metrics fibered over the AdS-Taub-Nut metrics (\ref{ADSTN}), the
result is
$$
g_7=(\sqrt{(\widetilde{r}+N)V(\widetilde{r})}\sin
\widetilde{\theta}d\widetilde{\phi}+\sigma_1)^2+
\bigg((\widetilde{r}-N)d\widetilde{t}+g(\widetilde{r})\cos\widetilde{\theta}d\widetilde{\phi}-\sigma_2\bigg)^2
$$
$$
+(\sqrt{(\widetilde{r}+N)V(\widetilde{r})}d\widetilde{\theta}-\sigma_3)^2+b\;\bigg(\;V(\widetilde{r})(d\widetilde{t}-2N
\cos\widetilde{\theta}
d\widetilde{\phi})^2+\frac{d\widetilde{r}^2}{V(\widetilde{r})}+(\widetilde{r}^2-N^2)(d\widetilde{\theta}^2+\sin^2\widetilde{\theta}
d\widetilde{\phi}^2)\bigg).
$$
Although the base quaternion Kahler space possess $SU(2)\times U(1)$
isometry, this group does not preserve the fibers, so the isometry
group is $SU(2)'\times U(1)^2$, being the $SU(2)'$ group related to
the Maurer-Cartan forms of the fiber metric and $U(1)^2$ generated
by $\partial_{\widetilde{t}}$ and $\partial_{\widetilde{\phi}}$. Let
us notice that we have a third commuting Killing vector, which is
the Reeb vector $\partial_{\tau}$, which is present in the
expression for the Maurer-Cartan forms $\sigma_i$. Therefore we have
a $T^3$ subgroup of isometries. By taking into account the explicit
form of the $\sigma_i$´s given in (\ref{mcarta}) we obtain the
following metric components
$$
g_{\widetilde{t}\widetilde{t}}=(\widetilde{r}-N)^2+ b
V(\widetilde{r}), \qquad g_{\widetilde{\phi}\widetilde{\phi}}=4b N^2
V(\widetilde{r})\cos^2\widetilde{\theta}
+(\widetilde{r}^2-N^2)\sin^2\widetilde{\theta}
$$
$$
g_{\widetilde{\theta}\widetilde{\theta}}=b(\widetilde{r}^2-N^2)+(\widetilde{r}+N)V(\widetilde{r}),
\qquad g_{\widetilde{r}\widetilde{r}}=\frac{b}{V(\widetilde{r})},
\qquad g_{\tau\tau}= g_{\varphi\varphi}=g_{\theta\theta}=1
$$
$$
g_{\widetilde{t}\widetilde{\phi}}=- 2N b
V(\widetilde{r})\cos\widetilde{\theta}+(\widetilde{r}-N)g(\widetilde{r})\cos\widetilde{\theta},
\qquad g_{\widetilde{t}\tau}=-(\widetilde{r}-N)\cos\varphi
\sin\theta
$$
\be\lb{azere}
g_{\widetilde{\phi}\tau}=\sqrt{(\widetilde{r}+N)V(\widetilde{r})}\sin\widetilde{\theta}
\sin\theta \sin\varphi+g(\widetilde{r})\cos\widetilde{\theta}
\sin\theta \cos\varphi \ee
$$
g_{\widetilde{\phi}\theta}=\sqrt{(\widetilde{r}+N)V(\widetilde{r})}\sin\widetilde{\theta}
\cos\varphi+g(\widetilde{r})\cos\widetilde{\theta} \sin\varphi
$$
$$
g_{\widetilde{\theta}\tau}=-\sqrt{(\widetilde{r}+N)V(\widetilde{r})}\cos\theta,
\qquad
g_{\widetilde{\theta}\varphi}=-\sqrt{(\widetilde{r}+N)V(\widetilde{r})}
$$
$$
g_{\widetilde{t}\theta}=-(\widetilde{r}-N)\sin\varphi
\sin\theta,\qquad g_{\tau\varphi}=\cos\theta,
$$
the remaining components are all zero. The parameter $b$ take the
values 1 or 5, $b=1$ corresponds to an Einstein-Sassaki metric,
while $b=5$ corresponds to a weak $G_2$ holonomy metric.
\\

\textit{The AdS-Kerr-Newmann case}
\\

For the rotating case, that is, for the AdS-Kerr-Newman metrics
(\ref{demi}) we obtain the metrics \be\lb{demil}
g_q=\bigg(\frac{\sqrt{f(\widetilde{\theta})
c(\widetilde{r})d(\widetilde{r})}}{e(\widetilde{r},
\widetilde{\theta})}
\frac{\sin\widetilde{\theta}}{\Xi}d\widetilde{\phi}-\sigma_1\bigg)^2+\bigg(e(\widetilde{r},
\widetilde{\theta})d\widetilde{t}+ \frac{W(\widetilde{r},
\widetilde{\theta})}{\Xi e(\widetilde{r},
\widetilde{\theta})}d\widetilde{\phi} -\sigma_2\bigg)^2 \ee
$$
+\bigg(
\sqrt{\frac{c(\widetilde{r})d(\widetilde{r})}{f(\widetilde{\theta})}}\frac{d\widetilde{\theta}}{e(\widetilde{r},
\widetilde{\theta})}
-\sqrt{\frac{f(\widetilde{\theta})}{c(\widetilde{r})d(\widetilde{r})}}\frac{a\sin{\widetilde{\theta}}}{e(\widetilde{r},
\widetilde{\theta})}d\widetilde{r} -\sigma_3\bigg)^2+\frac{b
\;f(\widetilde{\theta}) \sin^2\widetilde{\theta}
}{\widetilde{r}^2-a^2\cos^2\widetilde{\theta}} (\;a
d\widetilde{t}-\frac{c(\widetilde{r})}{\Xi}d\widetilde{\phi}\;)^2
$$
$$
+ \frac{b\;
c(\widetilde{r})d(\widetilde{r})}{\widetilde{r}^2-a^2\cos\widetilde{\theta}^2}(\;d\widetilde{t}
+\frac{a\sin^2\widetilde{\theta}}{\Xi}
d\widetilde{\phi}\;)^2+\frac{\widetilde{r}^2-a^2\cos\widetilde{\theta}^2}{f(\widetilde{\theta})}
b\;d\widetilde{\theta}^2+\frac{\widetilde{r}^2-a^2\cos\widetilde{\theta}^2}{c(\widetilde{r})d(\widetilde{r})}b
\;d\widetilde{r}^2 ,
$$
where we have introduced the functions
$$
f(\widetilde{\theta})=1-a^2\cos^2\widetilde{\theta},\qquad
c(\widetilde{r})=\widetilde{r}^2-a^2, \qquad
d(\widetilde{r})=1-\widetilde{r}^2\qquad e(\widetilde{r},
\widetilde{\theta})=\widetilde{r}-a\cos\widetilde{\theta}.
$$
The local isometry is $SU(2)\times U(1)^2$ and as before, the
vectors $\partial_{\widetilde{t}}$,  $\partial_{\widetilde{\phi}}$.
and $\partial_{\tau}$ generate a $T^3$ isometry subgroup. From
expression (\ref{demil}) we read the following components
$$
g_{\widetilde{t}\widetilde{t}}=\frac{b\;
c(\widetilde{r})d(\widetilde{r})}{\widetilde{r}^2-a^2\cos\widetilde{\theta}^2}
+a^2\frac{b \;f(\widetilde{\theta}) \sin^2\widetilde{\theta}
}{\widetilde{r}^2-a^2\cos^2\widetilde{\theta}} +e^2(\widetilde{r},
\widetilde{\theta})
$$
$$
g_{\widetilde{\phi}\widetilde{\phi}}=\frac{b\;
c(\widetilde{r})d(\widetilde{r})}{\widetilde{r}^2-a^2\cos\widetilde{\theta}^2}\frac{a^2\sin^4\widetilde{\theta}}{\Xi^2}
+\frac{b \;f(\widetilde{\theta}) \sin^2\widetilde{\theta}
}{\widetilde{r}^2-a^2\cos^2\widetilde{\theta}}\frac{c^2(\widetilde{r})}{\Xi^2}
+\frac{W^2(\widetilde{r}, \widetilde{\theta})}{\Xi^2
e^2(\widetilde{r}, \widetilde{\theta})} +\frac{f(\widetilde{\theta})
c(\widetilde{r})d(\widetilde{r})}{e^2(\widetilde{r},
\widetilde{\theta})} \frac{\sin^2\widetilde{\theta}}{\Xi^2}
$$
$$
g_{\widetilde{\theta}\widetilde{\theta}}=\frac{\widetilde{r}^2-a^2\cos\widetilde{\theta}^2}{f(\widetilde{\theta})}
b+\frac{1}{e^2(\widetilde{r},
\widetilde{\theta})}\frac{c(\widetilde{r})d(\widetilde{r})}{f(\widetilde{\theta})},
\qquad
g_{\widetilde{r}\widetilde{r}}=\frac{\widetilde{r}^2-a^2\cos\widetilde{\theta}^2}{c(\widetilde{r})d(\widetilde{r})}b
+\frac{f(\widetilde{\theta})}{c(\widetilde{r})d(\widetilde{r})}\frac{a^2\sin^2{\widetilde{\theta}}}{e^2(\widetilde{r},
\widetilde{\theta})},
$$
$$
g_{\widetilde{r}\widetilde{\theta}}=-\frac{a\sin{\widetilde{\theta}}}{e^2(\widetilde{r},
\widetilde{\theta})}, \qquad
g_{\widetilde{r}\varphi}=\sqrt{\frac{f(\widetilde{\theta})}{c(\widetilde{r})d(\widetilde{r})}}
\frac{a\sin{\widetilde{\theta}}}{e(\widetilde{r},
\widetilde{\theta})} \qquad
g_{\widetilde{r}\tau}=\sqrt{\frac{f(\widetilde{\theta})\cos\theta}{c(\widetilde{r})d(\widetilde{r})}}\frac{a\sin{\widetilde{\theta}}\cos\theta}{e(\widetilde{r},
\widetilde{\theta})}
$$
$$
g_{\widetilde{t}\widetilde{\phi}}= \frac{b\;
c(\widetilde{r})d(\widetilde{r})}{\widetilde{r}^2-a^2\cos\widetilde{\theta}^2}\frac{a\sin^2\widetilde{\theta}}{\Xi}
+a\frac{b \;f(\widetilde{\theta}) \sin^2\widetilde{\theta}
}{\widetilde{r}^2-a^2\cos^2\widetilde{\theta}}
\frac{c(\widetilde{r})}{\Xi} +\frac{W(\widetilde{r},
\widetilde{\theta})}{\Xi},
$$
$$
g_{\tau\tau}= g_{\varphi\varphi}=g_{\theta\theta}=1, \qquad
g_{\widetilde{t}\tau}=-e(\widetilde{r},\widetilde{\theta})\cos\varphi
\sin\theta
$$
\be\lb{azere2}
g_{\widetilde{\phi}\tau}=-\frac{\sqrt{f(\widetilde{\theta})
c(\widetilde{r})d(\widetilde{r})}}{e(\widetilde{r},
\widetilde{\theta})} \frac{\sin\widetilde{\theta}}{\Xi} \sin\theta
\sin\varphi-\frac{W(\widetilde{r}, \widetilde{\theta})}{\Xi
e(\widetilde{r}, \widetilde{\theta})} \sin\theta \cos\varphi \ee
$$
g_{\widetilde{\phi}\theta}=\frac{W(\widetilde{r},
\widetilde{\theta})}{\Xi e(\widetilde{r}, \widetilde{\theta})}
\sin\varphi+\frac{\sqrt{f(\widetilde{\theta})
c(\widetilde{r})d(\widetilde{r})}}{e(\widetilde{r},
\widetilde{\theta})} \frac{\sin\widetilde{\theta}}{\Xi}\cos\varphi
$$
$$
g_{\widetilde{\theta}\tau}=
-\sqrt{\frac{c(\widetilde{r})d(\widetilde{r})}{f(\widetilde{\theta})}}\frac{\cos\theta}{e(\widetilde{r},
\widetilde{\theta})}, \qquad
g_{\widetilde{\theta}\varphi}=-\sqrt{\frac{c(\widetilde{r})d(\widetilde{r})}{f(\widetilde{\theta})}}\frac{1}{e(\widetilde{r},
\widetilde{\theta})}
$$
$$
g_{\widetilde{t}\theta}=-e(\widetilde{r},
\widetilde{\theta})\sin\varphi,\qquad g_{\tau\varphi}=\cos\theta
$$
and the other components are zero. In the limit $a=0$ the base metric reduce to
$S^4$, the resulting tri-Sasaki metrics is
$$
g_7=(\sin\widetilde{\rho}d\widetilde{t}+\cos\widetilde{\theta}d\widetilde{\phi}+\sin\varphi
d\theta -\cos\varphi \sin\theta d\tau)^2 +
(\cos\widetilde{\rho}d\widetilde{\theta}-d\varphi - \cos \theta
d\tau)^2$$ \be\lb{chacha} +(\cos\widetilde{\rho}\sin
\widetilde{\theta}d\widetilde{\phi}-\cos\varphi d\theta -\sin\varphi
\sin\theta
d\tau)^2+b\cos^2\widetilde{\rho}d\widetilde{t}^2+bd\widetilde{\rho}^2+b\sin^2\widetilde{\rho}(d\widetilde{\theta}^2+\sin^2\widetilde{\theta}
d\widetilde{\phi}^2) . \ee  By another side, the tri-Sasaki and weak $G_2$ metrics
fibered over CP(2) are those corresponding to the spaces $N(1,1)_I$ and $N(1,1)_{II}$
and were given in (\ref{tro}), so we will not discuss them again.

\subsection{An infinite family of weak G2 and tri-Sassaki orbifolds}

Let us now turn our attention to the construction of the tri-Sasaki
and the Kahler-Einstein metrics fibered over the toric quaternion
Kahler metrics (\ref{backly}). The anti-self-dual part of the spin
connection $\omega_-^i$ possess a remarkable simple form in terms of
the potential function $F$ \cite{Pedersen} \be\lb{threefor}
\omega_-^1= \frac{1}{F}[-\rho F_{\eta}\frac{d\rho}{\rho}
+(\frac{1}{2}F+\rho F_{\rho})\frac{d\eta}{\rho}],\quad
\omega_-^2=\frac{u}{F},\qquad \omega_-^3=\frac{v}{F}. \ee The
Kahler-Einstein metric defined over the twistor space Z of
(\ref{metric}) is obtained directly from (\ref{kahlo}) and is given
by
$$ g_6=\frac{F^2-4\rho^2(F^2_{\rho}+F^2_{\eta})}{4F^2}\frac{d\rho^2+d\eta^2}{\rho^2} $$
\be\lb{warrior} +\frac{[(F-2\alpha F_{\rho})u-2\rho
F_{\eta}v]^2+[-2\rho F_{\eta}u +(F+2\rho
F_{\rho})v]^2}{F^2[F^2-4\rho^2(F^2_{\rho}+F^2_{\eta})]} \ee
$$
 +(d\theta-\sin\varphi\omega_-^2
+\cos\varphi\omega_-^1)^2+(\sin\theta
d\varphi-\cos\theta\sin\varphi\omega_-^1-
\cos\theta\cos\varphi\omega_-^2+\sin\theta\omega_-^3)^2
$$
The Kahler form for (\ref{warrior}) is $\overline{J}=dH$ where
\be\lb{ache2} H= \frac{\sin\theta \sin\varphi}{2\rho F}\bigg(-2\rho
F_{\eta}d\rho +(F+2\rho F_{\rho})d\eta\bigg)+ \frac{\sin\theta
\cos\varphi u}{F} +\frac{\cos\theta v}{F}-\cos\theta d\varphi, \ee
and with the help of this expressions we obtain the tri-Sasaki
metrics
$$
g_7=(d\tau+H)^2+g_6,
$$
being $H$ defined in (\ref{ache2}). Both expressions for the
tri-Sasaki and the Kahler-Einstein metrics are completely determined
in terms of a single eigenfunction $F$ of the hyperbolic laplacian.
Indeed the components of $g_7$ are given explicitly by
$$
(g_7)_{\rho\rho}=(g_q)_{\rho\rho}+\frac{F_{\eta}^2}{F^2},\qquad
(g_7)_{\rho\eta}=\frac{F_{\eta}}{F^2}(F+2\rho F_{\rho}) \qquad
(g_7)_{\eta\eta}=(g_q)_{\eta\eta}+\frac{1}{\rho^2
F^2}(\frac{F}{2}+\rho F_{\rho})^2,\qquad
$$
$$
(g_7)_{\alpha\alpha}=(g_q)_{\alpha\alpha}+\frac{(1+\rho^2)}{\rho
F^2},\qquad
(g_7)_{\alpha\beta}=(g_q)_{\alpha\beta}+\frac{2\eta}{\rho
F^2},\qquad (g_7)_{\beta\beta}=(g_q)_{\beta\beta}+\frac{1}{\rho F^2}
$$
$$
(g_7)_{\rho\theta}=\frac{2F_{\eta}}{F}\cos\theta,\qquad
(g_7)_{\rho\tau}=\frac{2 F_{\eta}}{F}\sin\theta \sin\varphi
$$
\be\lb{sasa} (g_7)_{\eta\theta}=-\frac{1}{\rho F}(F+2\rho
F_{\rho})\cos\theta,\qquad (g_7)_{\eta\tau}=-\frac{1}{\rho
F}(F+2\rho F_{\rho})\sin\theta\sin\varphi \ee
$$
(g_7)_{\alpha\theta}=\frac{2\sqrt{\rho}}{F}\sin \varphi,\qquad
(g_7)_{\alpha\varphi}=-\frac{2\eta}{\sqrt{\rho}F},\qquad
(g_7)_{\alpha\tau}=-\frac{2\sqrt{\rho}}{F}\sin\theta
\cos\varphi+\frac{2\eta}{\sqrt{\rho}F}\cos\theta,
$$
$$
(g_7)_{\beta\varphi}=-\frac{2}{\sqrt{\rho}F},\qquad
(g_7)_{\beta\tau}=-\frac{2}{\sqrt{\rho}F}\cos\theta
$$
$$
(g_7)_{\theta\theta}=(g_7)_{\tau\tau}=(g_7)_{\varphi\varphi}=1,\qquad
(g_7)_{\tau\varphi}=\cos\theta
$$
and the remaining components are zero.

   The tri-Sasaki metric possess an $SU(2)$ isometry group
associated with the $\sigma^i$ and a $T^2$ isometry of the
quaternion Kahler base. Therefore the isometry group is at least
$SO(3)\times T^2$. The Killing vectors are
$$
K_{1}=\partial_{\alpha},\qquad K_{2}=\partial_{\beta}
$$
\be\lb{kilo} K_{3}=\partial_{\tau},\qquad
K_{4}=\cos\tau\partial_{\varphi}-\coth\varphi\sin\tau\partial_{\tau}
+\frac{\sin\tau}{\sin\varphi}\partial_{\theta} \ee
$$
K_{5}=-\sin\tau\partial_{\varphi}-\coth\varphi\cos\tau\partial_{\tau}
+\frac{\cos\tau}{\sin\varphi}
\partial_{\theta}
$$
with commutation rule
$$
[K_{1}, K_{i}]=[K_{2}, K_{i}]=0,\qquad i=1,..,5 \qquad [K_{i},
K_{j}]=\epsilon_{ijk} K_{k},\qquad i,j,k=3,4,5.
$$
Both the tri-Sasaki metric and the Kahler-Einstein one possesses
three commuting Killing vectors. For the Kahler Einstein metric the
vectors are $\partial_{\theta}$, $\partial_{\alpha}$ and
$\partial_{\beta}$, for the tri-Sasaki metric they are
$\partial_{\tau}$, $\partial_{\alpha}$ and $\partial_{\beta}$.

By making the replacement $g_q\to 5 g_q$ in the formulas above we
obtain a family of weak $G_2$ holonomy metric. Locally the isometry
group will be the same than for them than for the
tri-Sasaki ones that we have presented.

\subsection{Supergravity solutions fibered over Einstein spaces}

Let us describe how to construct supergravity backgrounds fibered
over conical Ricci-flat metrics and their role in the AdS/CFT
correspondence. Consider an stack of $N$ parallel Dp branes. The
general form of such background is
$$
g_{10}=H_p^{-1/2}(r)g_{1,p}+ H_p^{1/2}(r)(dr^2 + r^2 g_{8-p}),
$$
\be\lb{generol} e^{2\phi-2\phi_{\infty}}=H_p^{\frac{p-3}{2}}, \qquad
A_{p+1}=-\frac{1}{2}(H_p^{-1}-1) dx^0\wedge ...\wedge dx^p, \ee
where
$$
H_p(r)=1+2^{5-p}\pi^{\frac{5-p}{2}}g_s
N_c\Gamma(\frac{7-p}{2})\frac{\alpha'^{\frac{7-p}{2}}}{r^{7-p}}.
$$
The metric $g_{8-p}$ is Einstein and is assumed to be independent on
$r$ and also independent on the Minkowski coordinates $(t,x,y,z)$.
If the Dp branes are flat, the light open spectrum is $U(N_c)$ super
Yang Mills in $p+1$ dimensions. We have that
$g_{YM}=2\pi^{p-2}g_s\alpha'^{\frac{p-3}{2}}$ being
$g_s=e^{2\phi_{\infty}}$. The field theory limit is obtained by
taking $\alpha\to 0$ such that $g_{YM}$ is fixed. For $p<3$ the ten
dimensional Newton constant goes to zero and the theory is decoupled
from the bulk. Instead for $p>3$ the constant $g_s$ goes to infinite
and a dual description is convenient in order to analyze the
decoupling problem. In order to have finite energy configurations in
the field theory limit one should consider the near horizon limit in
the IIB background. Such limit is obtained by taking $r\to 0$ and
$\alpha'\to 0$ such that the quantity with energy units
$U=r/\alpha'$ is fixed. For any $p$ the resulting metric will be
$$
g_{IIB}=\alpha'[(\frac{d_p g_{YM}^2 N}{U^{7-p}})^{-1/2}g_{1,p}+
(\frac{d_p g_{YM}^2 N}{U^{7-p}})^{1/2}(dU^2+U^2 d\Omega_{8-p})]
$$
$$
e^{\phi}=(2\pi)^{2-p}g_{YM}^2(\frac{d_p g_{YM}^2
N}{U^{7-p}})^{\frac{3-p}{4}}, \qquad d_p=2^{7-2p}
\pi^{\frac{9-3p}{2}}\Gamma(\frac{7-p}{2}).
$$
The Yang-Mills coupling constant $g_{YM}$ is not dimensionless for
any $p$, but the effective constant $g_{eff}^2\sim g_{YM}^2NU^{p-3}$
is. The low energy description of super Yang-Mills can be trusted
for $g_{eff}^2<<1$ which means that $U>>(g_{YM}^2N)^{1/(3-p)}$ for
$p<3$ and $U<<(g_{YM}^2N)^{1/(3-p)}$ for $p>3$. In the ultraviolet
limit $U\to\infty$ the string coupling vanish for $p<3$ and the
theory is UV free. In the other case a dual description is needed,
which is in accordance with the fact that the corresponding SYM
theories are not renormalizable and new degrees of freedom appears
at short distances.

The situation is different for $p=3$, in which the AdS/CFT
correspondence fully applies. The type IIB supergravity solutions of
the form
$$
g_{10}=H^{-1/2}(-dt^2+dx^2+dy^2+dz^2)+ H^{1/2}(dr^2 + r^2 g_{5}),
$$
\be\lb{genero} e^{2\phi}=e^{2\phi_{\infty}}, \qquad
F_5=(1+\ast)dH_3^{-1}\wedge dx^0\wedge dx^1\wedge dx^2\wedge dx^3,
\ee being $H(r)$ an harmonic function over the Ricci flat metric
given by
$$
H(r)=1+\frac{L^4}{r^4},\qquad L^4=4\pi g_8 N \alpha^2.
$$
Such solutions represent an stack of N parallel D3 branes separated
by some distance called $r$. Then it follows that the six
dimensional metric in (\ref{genero}) possess a conical singularity,
except for the round five sphere. It is also Ricci-flat by
construction and therefore $g_5$ is Einstein.  If the Ricci flat
metric is indeed flat then the theory living in the D3 brane
decouples from the bulk and the branes come close together. The
resulting theory is $N=4$ super Yang Mills. By another side the near
horizon limit of the geometry of (\ref{genero}) is $AdS_5\times
X_5$, being $X_5$ the Einstein space over which $g_5$ is defined. In
the context of the AdS/CFT correspondence the gauge field theory
living on the D3 brane at the conical singularity is identified as
the dual of type IIB string theory on $AdS_5\times X_5$. The open
and closed string massive modes decouple by taking the limit
$\alpha'\to 0$ and the Planck length $l_p=g_s^{1/2}\alpha'$ goes to
zero because $g_s$ is given in terms of the dilaton, which is
constant. The $AdS$ factor reflects that the dual field theory is
conformally invariant. The number of supersymmetries of the gauge
theory is related to the number of independent Killing vectors,
which depends on the holonomy of the cone. Instead for $p$ is
different from 3 the curvature in the near horizon limit is $R\sim
1/g_{eff}$ which is U dependent, thus no AdS factor appears. This
reflects that $U(N_c)$ super Yang Mills theory is not conformal
invariant. The same happen for non flat branes.

There are also of interest eleven dimensional supergravity solutions
over a manifold with local form $M_3\times X_8$, being the manifold
$X_8$ Ricci-flat and developing a conical singularity. The generic
supergravity solution in consideration is of the form
\be\lb{genero2} g_{11}=H^{-2/3}(-dt^2+dx^2+dy^2)+ H^{1/3}(dr^2+r^2
g_{7}), \ee
$$
F=\pm dx\wedge dy\wedge dt\wedge dH^{-1}
$$
where
$$
H(r)=1+\frac{2^5\pi^2N l_p^6}{r^6}.
$$
This solution describe $N$ M2 branes. The near horizon limit of this
geometry is obtained taking the 11 dimensional Planck length $l_p\to
0$ and keeping fixed $U=r^2/l_p^3$. The resulting background is
$AdS_4\times X_7$, being $X_7$ is an Einstein manifold with
cosmological constant $\Lambda=5$, and the radius of $AdS_4$ is
$2R_{AdS}=l_p(2^5\pi^2 N)^{1/6}$. Such solutions have the generic
form \be\lb{solu} g_{11}=g_{AdS}+g_7, \qquad F_4\sim \omega_{AdS},
\ee being $\omega_{AdS}$ the volume form of $AdS_4$. If $X_7$ is the
round sphere the radius will be the same than the $AdS$ part. This
is the flat case and it is conjectured that the dual theory is the
2+1 dimensional $N=8$ superconformal field theory living on the $M2$
brane. The isometry group $SO(7)$ of the sphere reflects the fact
that a $N=8$ SCFT is invariant under $SO(7)$ subgroup instead of
$SO(8)$.  The quantization of the flux of the tensor $F$ implies
that the constant $\alpha$ is quantized in units of $l_{11}^6$,
being $l_{11}$ the Planck length in eleven dimensions.  This
backgrounds are in general associated to three dimensional conformal
field theories arising as the infrared limit of the world volume
theory of N coincident M2 branes located a the singularity of
$M_3\times X_8$. Also in this case, the number of supersymmetries of
the field theory is determined by the holonomy of $X_8$. In the case
of $Spin(7)$, $SU(4)$ or $Sp(2)$ holonomies we have $N=1,2,3$
supersymmetries, respectively. This implies that the 7-dimensional
cone will be of weak $G_2$ holonomy (if the eight dimensional metric
is of cohomogenity one, see below), tri-Sassaki or a
Sassaki-Einstein, respectively. If $g_8$ is flat, then we have the
maximal number of supersymmetries, namely eight.

\subsection{Supergravity backgrounds over tri-Sassaki and weak $G_2$}

Let us construct supergravity backgrounds
corresponding to the Einstein 7-metrics (\ref{sasa}) or (\ref{azere}),
(\ref{azere2}) and (\ref{chacha}) . The generic 11-dimensional
supergravity solution is $$ g_{11}=H^{-2/3}(-dt^2+dx^2+dy^2)+
H^{1/3}dr^2+r^2 H^{1/3}\bigg(
(g_7)_{\alpha\alpha}d\alpha^2+(g_7)_{\alpha\beta}d\alpha\otimes
d\beta $$ $$+(g_7)_{\alpha\phi}d\alpha\otimes
d\phi+(g_7)_{\beta\beta}d\beta^2 + (g_7)_{\beta\phi}d\beta \otimes
d\phi+ (g_7)_{\phi\phi}d\phi^2 +Q_{\alpha}d\alpha +
Q_{\beta}d\beta+Q_{\phi}d\phi+ \widetilde{H}\bigg),
$$
 \be\lb{genero2}
F=\pm dx\wedge dy\wedge dt\wedge dH^{-1} \ee being $H$ an harmonic
function over the hyperKahler cone. In particular if $H=H(r)$ we
have
\be\lb{drco}
H(r)=1+\frac{2^5\pi^2N l_p^6}{r^6}.
\ee
In the expression for the metric we have introduced the 1-forms $Q$
and the symmetric tensor $\widetilde{H}$ given by
$$
\widetilde{H}=d\theta^2+d\varphi^2+(g_7)_{\theta\eta}d\theta\otimes
d\eta +(g_7)_{\rho\rho}d\rho^2+ (g_7)_{\rho\eta}d\rho\otimes d\eta
+(g_7)_{\eta\eta}d\eta^2,
$$
$$
Q_{\alpha}=(g_7)_{\alpha\theta}d\theta+
(g_7)_{\alpha\varphi}d\varphi, \qquad Q_{\beta}=
(g_7)_{\beta\varphi}d\varphi, \qquad
Q_{\phi}=(g_7)_{\phi\varphi}d\varphi+ (g_7)_{\phi\eta}d\eta +
(g_7)_{\phi\rho}d\rho.
$$
This supergravity solution describe $N$ M2 branes. The near horizon
limit of this geometry is obtained taking the 11 dimensional Planck
length $l_p\to 0$ and keeping fixed $U=r^2/l_p^3$. The resulting
background is $AdS_4\times X_7$, being $X_7$ is an Einstein manifold
with cosmological constant $\Lambda=5$, and the radius of $AdS_4$ is
$2R_{AdS}=l_p(2^5\pi^2 N)^{1/6}$. Such solutions have the generic
form \be\lb{solu} g_{11}=g_{AdS}+g_7, \qquad F_4\sim \omega_{AdS},
\ee being $g_7$ an Einstein metric over $X_7$ and $\omega_{AdS}$ the
volume form of $AdS_4$.
\\

\textit{Non AdS backgrounds and harmonic functions}
\\

   Non $AdS_4$ backgrounds are also of interest because they
are related to non conformal field theories. Therefore it is of
interest to find harmonic functions with are functions not only of
the radius $r$, but also of other coordinates of the internal space.

   We will now give here a simple way to construct non trivial harmonic
functions. Let us notice that all the 4-dimensional quaternion
Kahler orbifolds that we have constructed have two commuting Killing
vectors which also preserve the one forms $\omega_-^i$. This vector
also preserve the Kahler triplet
$d\overline{J}=d\omega_-+\omega_-\wedge \omega_-$. Consequently they
preserve the hyperKahler triplet (\ref{quato}) for the corresponding
Swann fibration. Such vectors are therefore Killing and
tri-holomorphic (thus tri-hamiltonian). For any eight dimensional
hyperKahler metric with two commuting Killing vectors there exist a
coordinate system in which takes the form \cite{Poon}
\be\lb{gengibbhawk} g_8=U_{ij}dx^i\cdot dx^j+
U^{ij}(dt_i+A_i)(dt_j+A_j), \ee being $(U_{ij}, A_i)$ solutions of
the generalized monopole equation
$$
F_{x_{\mu}^i
x_{\nu}^j}=\epsilon_{\mu\nu\lambda}\nabla_{x_{\lambda}^i}U_j,
$$
\be\lb{genmonop}
\nabla_{x_{\lambda}^i}U_j=\nabla_{x_{\lambda}^j}U_i, \ee
$$
U_i=(U_{i1}, U_{i2}),
$$
The coordinates $(x^1_i, x^2_i)$ with $i=1, 2, 3$ are the momentum
maps of the tri-holomorphic vector fields $\partial/\partial \theta$
and $\partial/\partial \varphi$, but we do not need to go in further
details. In the momentum map system the 11-dimensional supergravity
solution reads \be\lb{11sugra2}
g_{11}=H^{-2/3}g_{2,1}+H^{1/3}[U_{ij}dx^i\cdot dx^j+
U^{ij}(dt_i+A_i)(dt_j+A_j)], \ee \be\lb{11sugraF}
F=\pm\omega(E^{2,1})\wedge dH^{-1}, \ee and the harmonic condition
on $H$ is expressed as
$$
U^{ij}\partial_i \cdot \partial_j H=0.
$$
All the Swann hyperKahler cones that we have presented are toric,
and therefore they can be expressed as
$$
g_8=dr^2+r^2g_7=U_{ij}dx^i\cdot dx^j+ U^{ij}(dt_i+A_i)(dt_j+A_j)
$$
Let us recall that, as a consequence of (\ref{genmonop}), we have
that $\partial_i\cdot\partial_j U_{ij}=0$, which implies that
$U^{ij}\partial_i\cdot\partial_j U_{ij}=0$. This means that any
entry $U_{ij}$ is an harmonic function over the hyperKahler cone.
The matrix $U^{ij}$ is determined by the relation
$U^{ij}=g_8(\partial_i,
\partial_j)$, and the inverse matrix $U_{ij}$ will give us three
independent non trivial harmonic functions for the internal space in
consideration.

   As an example we can consider the cone $g_8=dr^2+r^2g_7$ being
$g_7$ the tri-Sasaki metric corresponding to the AdS-Taub-Nut
solution (\ref{azere}). For this cone we have that
$$
U^{\widetilde{t}\widetilde{t}}=r^2(\widetilde{r}-N)^2+  r^2
V(\widetilde{r}), \qquad U^{\widetilde{\phi}\widetilde{\phi}}=4
N^2r^2 V(\widetilde{r})\cos^2\widetilde{\theta}
+(\widetilde{r}^2-N^2)r^2\sin^2\widetilde{\theta}
$$
\be\lb{noche} U^{\widetilde{t}\widetilde{\phi}}=- 2Nr^2
V(\widetilde{r})\cos\widetilde{\theta}
+r^2(\widetilde{r}-N)g(\widetilde{r})\cos\widetilde{\theta}. \ee By
defining $\Delta=U^{\widetilde{t}\widetilde{t}}
U^{\widetilde{\phi}\widetilde{\phi}}-(U^{\widetilde{t}\widetilde{\phi}})^2$
we obtain the following harmonic functions \be\lb{dia}
U_{\widetilde{t}\widetilde{t}}=\frac{U^{\widetilde{\phi}\widetilde{\phi}}}{\Delta},
\qquad U_{\widetilde{\phi}\widetilde{\phi}}=
-\frac{U^{\widetilde{t}\widetilde{t}}}{\Delta}, \qquad
U_{\widetilde{t}\widetilde{\phi}}=\frac{U^{\widetilde{t}\widetilde{\phi}}}{\Delta}.
\ee In the $S^4$ manifold limit we obtain more simple expressions,
namely
$$
U_{\widetilde{\phi}\widetilde{\phi}}=-\frac{1}{r^2}\bigg(\frac{1}{\sin^2\widetilde{\theta}
+\sin^2\widetilde{\rho}\cos^2\widetilde{\theta}
-\sin^4\widetilde{\rho}\cos^2\widetilde{\theta}}\bigg),
$$
\be\lb{arm2}
U_{\widetilde{\phi}\widetilde{t}}=\frac{1}{r^2}\bigg(\frac{\sin^2\widetilde{\theta}
+\sin^2\widetilde{\rho}\cos^2\widetilde{\theta}}{\sin^2\widetilde{\theta}
+\sin^2\widetilde{\rho}\cos^2\widetilde{\theta}
-\sin^4\widetilde{\rho}\cos^2\widetilde{\theta}}\bigg) \ee
$$
U_{\widetilde{t}\widetilde{t}}=\frac{1}{r^2}\bigg(\frac{\sin^2\widetilde{\rho}\cos\widetilde{\theta}}{\sin^2\widetilde{\theta}
+\sin^2\widetilde{\rho}\cos^2\widetilde{\theta}
-\sin^4\widetilde{\rho}\cos^2\widetilde{\theta}}\bigg)
$$
For the AdS-Kerr-Newman case (\ref{azere2}) we have
$$
U^{\widetilde{t}\widetilde{t}}=r^2\frac{
c(\widetilde{r})d(\widetilde{r})}{\widetilde{r}^2-a^2\cos\widetilde{\theta}^2}
+a^2r^2\frac{f(\widetilde{\theta}) \sin^2\widetilde{\theta}
}{\widetilde{r}^2-a^2\cos^2\widetilde{\theta}} +r^2
e^2(\widetilde{r}, \widetilde{\theta}),
$$
$$
U^{\widetilde{t}\widetilde{\phi}}=r^2 \frac{
c(\widetilde{r})d(\widetilde{r})}{\widetilde{r}^2-a^2\cos\widetilde{\theta}^2}\frac{a\sin^2\widetilde{\theta}}{\Xi}
+ar^2\frac{f(\widetilde{\theta}) \sin^2\widetilde{\theta}
}{\widetilde{r}^2-a^2\cos^2\widetilde{\theta}}
\frac{c(\widetilde{r})}{\Xi} +r^2\frac{W(\widetilde{r},
\widetilde{\theta})}{\Xi},
$$
$$
U^{\widetilde{\phi}\widetilde{\phi}}=r^2\frac{
c(\widetilde{r})d(\widetilde{r})}{\widetilde{r}^2-a^2\cos\widetilde{\theta}^2}\frac{a^2\sin^4\widetilde{\theta}}{\Xi^2}
+r^2\frac{f(\widetilde{\theta}) \sin^2\widetilde{\theta}
}{\widetilde{r}^2-a^2\cos^2\widetilde{\theta}}\frac{c^2(\widetilde{r})}{\Xi^2}
+r^2\frac{W^2(\widetilde{r}, \widetilde{\theta})}{\Xi^2
e^2(\widetilde{r}, \widetilde{\theta})}
+r^2\frac{f(\widetilde{\theta})
c(\widetilde{r})d(\widetilde{r})}{e^2(\widetilde{r},
\widetilde{\theta})} \frac{\sin^2\widetilde{\theta}}{\Xi^2}
$$
and again, the three functions $U^{ij}/\Delta$ are harmonic
functions over the internal hyperkahler space. Notice that
$\Delta\sim r^4$ and therefore all these harmonic functions depends
on $r$ as $1/r^2$. Finally, for the Swann metrics fibered over the
toric orbifolds (\ref{metric}) we find \be\lb{harmono} U_{\alpha\alpha}=
\frac{F}{r^2}(\frac{1}{2}F+\rho F_{\rho}) \qquad  U_{\alpha\beta}=\frac{\rho
F_{\eta}F}{r^2}, \qquad  U_{\beta\beta}=\frac{F}{r^2}(\frac{1}{2}F-\rho
F_{\rho}), \ee are also harmonic functions over the cone, depending
on a solution $F$ of a linear differential equation. All these
harmonic functions provide non $AdS_4$ horizon limits.

\section{Gamma deformations of supergravity backgrounds}

\subsection{Deformations of 11-supergravity solutions}

  Let us describe in more detail the $SL(2,R)$ solution
generating technique sketched in the introduction. This technique
was applied in order to find the dual of marginal deformed field
theories in \cite{Lunin}. One usually starts with a solution of the
eleven dimensional supergravity with $U(1)\times U(1)\times U(1)$
isometry. Any of such solutions can be written in the generic form
\be\lb{as} g_{11}=\Delta^{1/3}M_{ab}D\alpha_a
D\alpha_b+\Delta^{-1/6}\widetilde{g}_{\mu\nu}dx^{\mu}dx^{\nu}, \ee
$$
C_3=C D\alpha_1\wedge D\alpha_2\wedge D\alpha_3+C_{1(ab)} \wedge
D\alpha_a\wedge D\alpha_b+ C_{2(a)}\wedge D\alpha_a+C_{(3)},
$$
with the indices a,b=1,2,3 are associated to three coordinates
$\alpha_1$, $\alpha_2$ and $\alpha_3$. The metric and the field
$C_3$ does not dependent on these coordinates and the greek indices
$\mu$, $\nu$ run over the remaining eight dimensional coordinates.
We have introduced the covariant derivative
$D\alpha_i=d\alpha_i+A_i$, being $A_i$ a triplet of
$\alpha_i$-independent one forms. The expression (\ref{as}) possess
a manifest $SL(3,R)$ symmetry for which the coordinates $\alpha_i$
of (\ref{as}) and the tensor fields $M$ and $A_i$ have the following
transformation law \be\lb{sl3} \left(\begin{array}{c}
  \alpha_1 \\
  \alpha_2 \\
  \alpha_3
\end{array}\right)'=(\Lambda^{T})^{-1}\left(\begin{array}{c}
  \alpha_1 \\
  \alpha_2 \\
  \alpha_3
\end{array}\right),
\ee  \be\lb{sl2} M'=\Lambda M \Lambda^T \qquad
\left(\begin{array}{c}
  A_1 \\
  A_2 \\
  A_3
\end{array}\right)'=(\Lambda^{T})^{-1}\left(\begin{array}{c}
  A_1 \\
  A_2 \\
  A_3
\end{array}\right).
\ee
$$
\left(\begin{array}{c} {C}_{23\mu}\\{C}_{31\mu}\\{ C}_{12\mu}
\end{array}\right)\rightarrow
(\Lambda^T)^{-1} \left(\begin{array}{c} {C}_{23\mu}\\{C}_{31\mu}\\{
C}_{12\mu}
\end{array}\right),\qquad
\left(\begin{array}{c} {C}_{1\mu\nu}\\{ C}_{2\mu\nu}\\C_{3\mu\nu}
\end{array}\right)\rightarrow \Lambda
\left(\begin{array}{c} {  C}_{1\mu\nu}\\{
C}_{2\mu\nu}\\{C}_{3\mu\nu}
\end{array}\right).
$$
The full isometry group of 11-dimensional supergravity compactified
on a three torus is $SL(3,R)\times SL(2,R)$. The $SL(3,R)$ group
leaves the background (\ref{as}) unaltered. Following \cite{Lunin}
and \cite{Cremer} we will deform these $T^3$ invariant backgrounds
by an element of $SL(2,R)$. These deformation is a solution
generating technique which does not leave the background unchanged,
but gives new supergravity backgrounds. We find convenient to define
a complex parameter $\tau=C+i \Delta^{1/2}$ which, under the
$SL(2,R)$ action is transformed as \be\lb{taum} \tau\longrightarrow
\frac{a \tau + b}{c \tau +d}; \qquad \Lambda=\left(\begin{array}{cc}
  a & b \\
  c & d
\end{array}\right)\in SL(2,R).
\ee The eight dimensional metric $g_{\mu\nu}$ and the tensor $C_2$
are invariant under this action. The tensor $C_{(1)ab\mu}$ and
$A_{a\mu}$ form a doublet in similar way that the RR and NSNS two
form fields do in IIB supergravity, their transformation law is
\be\lb{doblete} B^a=\left(\begin{array}{c}
  2 A_a \\
  -\epsilon^{abc}C_{(1)bc}
\end{array}\right),\qquad B_a\longrightarrow \Lambda^{-T}B_a.
\ee The field strength $C_3$ also form a doublet with its magnetic
dual with transformation law \be\lb{doblete}
H=\left(\begin{array}{c}
  F_4 \\
  \Delta^{-1/2}\ast_{8} F_{4} + C_{(0)}F_{4}
\end{array}\right),
\qquad H\longrightarrow \Lambda^{-T}H, \ee being the Hodge operation
taken with respect to the eight dimensional metric $\widetilde{g}$.
As we discussed in the introduction, this transformation deform the
original metric (\ref{as}) and the deformed metric will be regular
only with elements of the form \cite{Lunin} \be\lb{gamo}
\Lambda=\left(\begin{array}{cc}
  1 & 0 \\
  \gamma & 1
\end{array}\right)\in SL(2,R),
\ee which constitute a subgroup called $\gamma$-transformations. We
will be concerned with such transformations in the following.

   If the fields $C$, $C_1$ and $C_2$ are zero, it follows that $A_i$
and $\widetilde{g}_{\mu\nu}$ are unchanged by a
$\gamma$-transformation and $C_1$ and $C_2$ remains zero. The
deformation then give the new fields \be\lb{inge}
\Delta'=G^2\Delta,\qquad C'=-\gamma G \Delta,\qquad
G=\frac{1}{1+\gamma^2 \Delta}. \ee By inspection of the
transformation rule (\ref{doblete}) it follows that \be\lb{gener}
F_4'=F_4-\gamma\Delta^{-1/2}\ast_{8} F_{4}-\gamma d(G\Delta
D\alpha_1\wedge D\alpha_2 \wedge D\alpha_3). \ee The
$\gamma$-deformed eleven dimensional metric results \cite{Cremer}
\be\lb{ede} g_{11}=G^{-1/3}(G\Delta^{1/3}M_{ab}D\alpha_a
D\alpha_b+\Delta^{-1/6}\widetilde{g}_{\mu\nu}dx^{\mu}dx^{\nu}). \ee
Note that if the initial four form $F_4$ was zero, then from the
last term in (\ref{gener}) a non trivial flux is obtained in the
deformed background.

\subsection{A family of deformed backgrounds}

  As we
have already mentioned, any eight-dimensional Ricci flat metric
$g_8$ can be extended to an eleven dimensional supergravity solution
of the form
$$
g_{11}=H^{-2/3}(-dt^2+dx^2+dy^2)+ H^{1/3}g_8
$$
$$
C^3=\pm H^{-1}dx\wedge dy\wedge dt, \qquad F^4=\pm dx\wedge dy\wedge
dt\wedge dH^{-1},
$$
being $H$ an harmonic function over $g_8$. If the metric $g_8$ is a
cone then we can reexpress the eleven dimensional metric as
$$
g_{11}=H^{-2/3}(-dt^2+dx^2+dy^2)+ H^{1/3}(dr^2+r^2g_7)
$$
being $g_7$ an Einstein metric. If $g_8$ is hyperKahler then $g_7$
will be tri-Sasaki, if $g_8$ is of $Spin(7)$ holonomy with
cohomogeneity one, then $g_7$ will be of weak $G_2$ holonomy. We
have constructed a whole family of tri-Sasaki metrics in
(\ref{azere}), (\ref{azere2}) and (\ref{chacha}). In addition, the
replacement $g_q\to 5 g_q$ in all these expressions give a family of
metrics with weak $G_2$ holonomy. All
these metrics possess three commuting Killing vectors, namely
$\partial_{\tau}$, $\partial_{\alpha}$ and $\partial_{\beta}$.
Therefore they are suitable to apply the $SL(2,R)$ solution
generating technique described previously. The corresponding
11-dimensional supergravity background with three commuting
isometries is
$$
g_{11}=H^{-2/3}(-dt^2+dx^2+dy^2)+
H^{1/3}dr^2+H^{1/3}r^2\bigg((g_7)_{\widetilde{t}\widetilde{t}}d\widetilde{t}^2+
(g_7)_{\widetilde{t}\widetilde{\phi}}d\widetilde{t}\otimes
d\widetilde{\phi}
$$
$$
+(g_7)_{\widetilde{t}\tau}d\widetilde{t}\otimes d\tau
+(g_7)_{\widetilde{\phi}\widetilde{\phi}}d\widetilde{\phi}^2+
(g_7)_{\widetilde{\phi}\tau}d\widetilde{\phi} \otimes d\tau+
(g_7)_{\tau\tau}d\tau^2 +Q_{\widetilde{t}}d\widetilde{t} +
Q_{\widetilde{\phi}}d\widetilde{\phi}+Q_{\tau}d\tau+
\widetilde{H}\bigg), $$ \be\lb{solon} C^3=\pm H^{-1}dx\wedge
dy\wedge dt, \qquad F^4=\pm dx\wedge dy\wedge dt\wedge dH^{-1}, \ee
where we have defined \be\lb{dicoro}
Q_{\widetilde{t}}=(g_7)_{\widetilde{t}\theta}d\theta, \qquad
Q_{\widetilde{\phi}}= (g_7)_{\widetilde{\phi}\theta}d\theta, \qquad
Q_{\tau}=(g_7)_{\tau\varphi}d\varphi+(g_7)_{\tau\widetilde{\theta}}d\widetilde{\theta}
+ (g_7)_{\tau\widetilde{r}}d\widetilde{r}, \ee
$$
\widetilde{H}=d\theta^2+d\varphi^2
+(g_7)_{\widetilde{\theta}\varphi}d\widetilde{\theta}\otimes
d\varphi+(g_7)_{\widetilde{r}\widetilde{r}}d\widetilde{r}^2+
(g_7)_{\widetilde{r}\widetilde{\theta}}d\widetilde{r}\otimes
d\widetilde{\theta}
+(g_7)_{\widetilde{r}\varphi}d\widetilde{r}\otimes
d\varphi+g_{\widetilde{\theta}\widetilde{\theta}}d\widetilde{\theta}^2
$$
Under the replacement $g_q\to 5g_q$ we will obtain a new
supergravity solution which is fibered over a weak $G_2$ holonomy
space. The solution generating technique applies to both cases
exactly in the same manner, but the dual field theories will possess
different number of supercharges. Let us note that if the harmonic
function is selected to be
$$
H(r)=1+\frac{2^5\pi^2N l_p^6}{r^6}.
$$
then the near horizon limit will be
$$
g_{11}=g_{AdS}+ g_7.
$$
But we can consider backgrounds with other horizon limits by
considering harmonic functions such as those constructed in
(\ref{noche})-(\ref{harmono}).

  Our task now is to
find a local coordinate system for which (\ref{solon}) takes the
manifest $T^3$ symmetric form \be\lb{as}
g_{11}=\Delta^{1/3}M_{ab}D\phi_a
D\phi_b+\Delta^{-1/6}\widetilde{g}_{\mu\nu}dx^{\mu}dx^{\nu} \ee with
the indices a,b=1,2,3 are associated to the isometries
$\phi_1=\widetilde{t}$, $\phi_2=\widetilde{\phi}$ and $\phi_3=\tau$
and the greek indices $\mu$, $\nu$ running over the remaining eight
dimensional coordinates. We need to introduce the following
quantities
$$
\left(\begin{array}{c}
  A_1 \\
  A_2 \\
  A_3
\end{array}\right)=(g_7)^{ab}\left(\begin{array}{c}
  Q_{\widetilde{t}} \\
  Q_{\widetilde{\phi}} \\
  Q_{\tau}
\end{array}\right)
$$
\be\lb{conto} h=\widetilde{H}-r^2H^{1/3}(g_7)_{ab}A_a A_b \ee With
the help of these quantities it is not difficult to check that the
metric takes the form
$$
g_{11}=g_8+H^{1/3}r^2 (g_7)_{ab}(d\phi_a+A_a)(d\phi_b+A_b)
$$
being $g_8$ given by \be\lb{don}
g_8=H^{-2/3}(-dt^2+dx^2+dy^2)+H^{1/3}(dr^2+r^2 h). \ee
 By further defining
$$
\Delta=[\det\Omega^{ab}],\qquad \widetilde{g}=\Delta^{1/6}g_8, $$
\be\lb{conto} D\phi_a=d\phi_a+A_a,\qquad
\Omega^{ab}=r^2H^{1/3}(g_7)_{ab} \qquad
M_{ab}=\frac{\Omega^{ab}}{\det (\Omega^{ab})^{1/3}}. \ee the metric
take the desired form with manifest $SL(3,R)$ symmetry \be\lb{deseo}
g_{11}=\Delta^{1/3}M_{ab}D\phi_a
D\phi_b+\Delta^{-1/6}\widetilde{g}_{\mu\nu}dx^{\mu}dx^{\nu},\ee By
using these quantities together with formulas
(\ref{inge})-(\ref{ede}) we obtain the following the deformed
backgrounds.

 For the orbifolds (\ref{sasa}) we have that
 $$
Q_{\alpha}=(g_7)_{\alpha\theta}d\theta+
(g_7)_{\alpha\varphi}d\varphi, \qquad Q_{\beta}=
(g_7)_{\beta\varphi}d\varphi, \qquad
Q_{\phi}=(g_7)_{\phi\varphi}d\varphi+ (g_7)_{\phi\eta}d\eta +
(g_7)_{\phi\rho}d\rho,
$$
$$
\widetilde{H}=d\theta^2+d\varphi^2+(g_7)_{\theta\eta}d\theta\otimes
d\eta +(g_7)_{\rho\rho}d\rho^2+ (g_7)_{\rho\eta}d\rho\otimes d\eta
+(g_7)_{\eta\eta}d\eta^2.
$$
The explicit form of the matrix $\Omega_{ab}$ turns out to be
\be\lb{lus} \Omega_{ab}=r^2H^{1/3}\left(\begin{array}{ccc}
                        1 & -\frac{2\sqrt{\rho}}{F}\sin\theta
\cos\varphi+\frac{2\eta}{\sqrt{\rho}F}\cos\theta & -\frac{2}{\sqrt{\rho}F}\cos\theta \\
                        -\frac{2\sqrt{\rho}}{F}\sin\theta
\cos\varphi+\frac{2\eta}{\sqrt{\rho}F}\cos\theta &
(g_q)_{\alpha\alpha}+\frac{(1+\rho^2)}{\rho F^2} &
(g_q)_{\alpha\beta}+\frac{2\eta}{\rho
F^2} \\
                        -\frac{2}{\sqrt{\rho}F}\cos\theta & (g_q)_{\alpha\beta}+\frac{2\eta}{\rho
F^2} & (g_q)_{\beta\beta}+\frac{1}{\rho F^2}
                      \end{array}\right),
\ee being $g_q$ the quaternion Kahler metric defined in
(\ref{metric}). All these quantities are expressed in terms of a
single eigenfunction $F$ of the laplacian operator in the hyperbolic
two dimensional space. By using the formulas (\ref{inge}),
(\ref{gener}) and (\ref{ede}) we obtain directly the deformed
backgrounds, as before.

\subsection{Explicit formulas for the spherical case}

It will be instructive to repeat this procedure to the background
(\ref{chachal}) fibered over $S^4$. We need to define the relevant
quantities first. From the definition $\Omega^{ab}=g_{ab}$ being $a,
b=1,2,3$ and $\phi_1=\widetilde{t}$, $\phi_2=\widetilde{\phi}$ and
$\phi_3=\tau$, we find the following toric fiber metric
$$
\Omega^{\widetilde{t}\widetilde{t}}=\Omega^{\widetilde{\phi}\widetilde{\phi}}=\Omega^{\tau\tau}=1,
\qquad
\Omega^{\widetilde{t}\widetilde{\phi}}=\sin\widetilde{\rho}\cos\widetilde{\theta},
$$
\be\lb{con}
\Omega^{\widetilde{t}\tau}=-\sin\widetilde{\rho}\sin\theta
\cos\varphi, \qquad \Omega^{\widetilde{\phi}\tau}=-\sin\theta(
\sin\varphi\cos\widetilde{\rho}\sin\widetilde{\theta}
+\cos\varphi\cos\widetilde{\theta}). \ee The determinant
$\Delta=\det \Omega^{ab}$ of this matrix is $$
\Delta=-\sin\widetilde{\rho}\sin\theta
\cos\varphi[\;\sin\widetilde{\rho}\cos\widetilde{\theta}\sin\theta(\;
\sin\varphi\cos\widetilde{\rho}\sin\widetilde{\theta}
+\cos\varphi\cos\widetilde{\theta}\;)+\sin\widetilde{\rho}\sin\theta
\cos\varphi\;]
$$
\be\lb{dotor} +[\;1-\sin^2\theta(
\sin\varphi\cos\widetilde{\rho}\sin\widetilde{\theta}
+\cos\varphi\cos\widetilde{\theta})^2\;]-\sin\widetilde{\rho}\cos\widetilde{\theta}
(\sin\widetilde{\rho}\cos\widetilde{\theta}+\sin\widetilde{\rho}\sin\theta
\cos\varphi). \ee From (\ref{con}) and (\ref{dotor}) we define the
matrix $M_{ab}=\Omega^{ab}/\Delta^{1/3}$ with unit determinant.
Also, from the definition $\Omega_{ab}=g^{ab}$ of the inverse matrix
we obtain that
$$
\Omega_{\widetilde{t}\widetilde{t}}=\frac{1}{\Delta}\bigg(1-\sin^2\theta(
\sin\varphi\cos\widetilde{\rho}\sin\widetilde{\theta}
+\cos\varphi\cos\widetilde{\theta})^2\bigg),
$$
$$
\Omega_{\widetilde{\phi}\widetilde{\phi}}=\frac{1}{\Delta}\bigg(1+\sin^2\widetilde{\rho}\sin^2\theta
\cos^2\varphi\bigg),\qquad
\Omega_{\tau\tau}=\frac{1}{\Delta}\bigg(1-\sin^2\widetilde{\rho}\cos^2\widetilde{\theta}\bigg),
$$
\be\lb{edto}
\Omega_{\widetilde{t}\widetilde{\phi}}=\frac{1}{\Delta}\bigg(\sin\widetilde{\rho}\cos\widetilde{\theta}-
\sin^2\theta\sin\widetilde{\rho}\cos\varphi(
\sin\varphi\cos\widetilde{\rho}\sin\widetilde{\theta}
+\cos\varphi\cos\widetilde{\theta})\bigg), \ee
$$
\Omega_{\widetilde{t}\tau}=\frac{1}{\Delta}\bigg(\sin\theta\sin\widetilde{\rho}\cos\varphi
-\sin\widetilde{\rho}\cos\widetilde{\theta}\sin\theta(
\sin\varphi\cos\widetilde{\rho}\sin\widetilde{\theta}
+\cos\varphi\cos\widetilde{\theta})\bigg),
$$
$$
\Omega_{\widetilde{\phi}\tau}=\frac{1}{\Delta}
\bigg(\sin\theta\sin^2\widetilde{\rho}\cos\widetilde{\theta}\cos\varphi-\sin\theta(
\sin\varphi\cos\widetilde{\rho}\sin\widetilde{\theta}
+\cos\varphi\cos\widetilde{\theta})\bigg).
$$
The one forms $Q_i$ are
$$
Q_{\widetilde{t}}=\sin\widetilde{\rho}\sin\varphi d\theta, \qquad
Q_{\widetilde{\phi}}=(\cos\widetilde{\theta}\sin\varphi
-\cos\widetilde{\rho}\sin\widetilde{\theta}\cos\varphi)d\theta
$$
\be\lb{comon} Q_{\tau}=-\cos\theta
(d\varphi+\cos\widetilde{\rho}d\widetilde{\theta}) \ee With the help
of (\ref{edto}) and (\ref{comon}) we define the one forms $A_a$ and
the covariant derivative $D_a$ by
\be\lb{foron}A_a=\Omega_{ab}Q_b,\qquad D\phi_a=d\phi_a+A_a.\ee  The
metric $\widetilde{H}$ defined in (\ref{dicoro}) is \be\lb{cont}
\widetilde{H}=d\theta^2+d\varphi^2+d\widetilde{\rho}^2+d\widetilde{\theta}^2-
2\cos\widetilde{\rho}d\widetilde{\theta}\otimes d\varphi, \ee and
therefore the metric in (\ref{conto}) reads \be\lb{cuntu}
h=\widetilde{H}-(g_7)_{ab}A_a A_b. \ee With the help of formulas
(\ref{as})-(\ref{deseo}) we obtain the deformed background, the
result is \be\lb{dofo}
g_d=(1+\gamma^2\Delta)^{1/3}\bigg(g_{11}-\frac{\gamma^2
\Delta}{1+\gamma^2\Delta}\Omega_{ab}Q_a Q_b\bigg)\ee
$$
C= - \frac{k \sin u}{3} \sinh^3 \rho dt \wedge du\wedge
dv-\frac{\gamma \Delta}{1+\gamma^2\Delta}D\widetilde{t}\wedge
D\widetilde{\phi}\wedge D\tau,
$$
being $g_{11}$ the undeformed metric (\ref{chachal}). Notice that
(\ref{dofo}) is explicit because all the quantities are defined by
(\ref{dotor}), (\ref{edto}) and (\ref{comon}). The procedure is
completed.

  We will consider IIB reductions of these backgrounds and their
deformations in the appendix.

\subsection{Rotating supermembrane solutions}

   We have presented an infinite family of eleven dimensional supergravity
backgrounds possessing at least three commuting Killing vectors.
These backgrounds are supposed to be dual to three dimensional
conformal field theories arising as the infrared limit of the world
volume theory of N coincident M2 branes located a the singularity of
$M_3\times X_8$. Because the eight dimensional geometry is
hyperKahler we expect $N=3$ supersymmetry in the superconformal
field theory. But is difficult to guess which is the explicit form
of the dual field theory and we are no attempting to obtain an
explicit lagrangian here. Nevertheless it has been
suggested that the correspondence between semiclassical strings with
high angular momentum and long operators can be generalized to
membranes \cite{Nunezo}. In the string case the configurations have
energy proportional to the ´T Hooft scale and thus are dual to
operators with large dimensions \cite{Gubser}. For a rotating string
in $AdS_5$ the difference between the energy and the spin depends
logarithmically on the spin, therefore the dual operators should
possess dimensions with the same dependence. This operators are
twist operators that are responsible for violations of Bjorken
scaling, and it has been shown that no corrections to the
logarithmic behaviour appears in the strong coupling limit
\cite{Frolov}. This correspondence have been generalized to
membranes, in which the relation between the spin, the J-charges and
the energy should be related to the anomalous dimensions of certain
operators of the conformal field theory \cite{Nunezo}.

   Therefore it is of interest to consider rotating membrane
configurations over the supergravity backgrounds that we have
constructed. Recall that the supermembrane action is given by
\be\lb{sumem} S=-\frac{1}{2\pi^2
l_{11}^3}\int\bigg(\frac{(-\gamma)^{1/2}}{2}[\gamma_{ij}\frac{\partial
X_{\mu}}{\partial \sigma_i} \frac{\partial X_{\nu}}{\partial
\sigma_j} G_{\mu\nu}-1]+\epsilon_{ijk}\frac{\partial
X_{\nu}}{\partial \sigma_i}\frac{\partial X_{\nu}}{\partial
\sigma_j}\frac{\partial X_{\nu}}{\partial \sigma_k}
C_{\mu\nu\rho}\bigg)d^3\sigma \ee where $\sigma_i=(\tau, \sigma,
\lambda)$ are the world volume coordinates, $\gamma_{ij}$ the
worldvolume metric, $X_{\mu}$ are the then target space coordinates
and $G_{\mu\nu}$ the target metric. We have also the 3-form
$C_{\mu\nu\rho}$ and the corresponding field strength is $H=dC$. The
equations of motion derived from (\ref{sumem}) is \be\lb{moushon}
\gamma_{ij}=\partial_i X_{\mu}\partial_j X_{\nu} G_{\mu\nu} \ee
$$
\partial_i\bigg( (-\gamma)^{1/2} \gamma^{ij} \partial_j X^{\rho}\bigg) = -
(-\gamma)^{1/2}\gamma^{ij} \partial_i X^{\mu} \partial_j X^{\nu}
\Gamma^{\rho}_{\mu\nu}(X)-\epsilon^{ijk} \partial_i X^{\mu}
\partial_j X^{\nu} \partial_k X^{\sigma} H^{\rho}_{\mu\nu\sigma}(X)
$$
The three diffeomorphisms of the action can be fixed by a gauge
described by the constraints \be\lb{const} \gamma_{0\alpha} =
\partial_0 X^{\mu} \partial_{\alpha} X^{\nu} G_{\mu\nu}(X) = 0 , \ee
$$
\gamma_{0 0} + L^2 \det \;[ \gamma_{\alpha\beta} \;] = \partial_0
X^{\mu} \partial_0 X^{\nu} G_{\mu\nu}(X) + L^2 \det \;[
\partial_{\alpha} X^{\mu} \partial_{\beta} X^{\nu} G_{\mu\nu}(X) \;] = 0 ,
$$
being $L$ a constant fixed by the equations of motion. By imposing
the constraints (\ref{const}) to (\ref{sumem}) we obtain the
following action in gauge fixed form \be\lb{gauge} S = \frac{1}{2
(2\pi)^2 L l_{11}^3} \int  \bigg( \partial_0 X^{\mu} \partial_0
X^{\nu} G_{\mu\nu}(X) - L^2 \det\;[ \partial_{\alpha} X^{\mu}
\partial_{\beta} X^{\nu} G_{\mu\nu}(X)
\;] \ee
$$
+ 2 L \epsilon^{ijk} \partial_i X^{\mu} \partial_j X^{\nu}
\partial_k X^{\rho} C_{\mu\nu\rho}(X) \bigg) d^3\sigma.
$$
In Poincare coordinates the $AdS_4$ space is
parameterized as
$$
g_{AdS}=- \cosh^2\rho dt^2 + d\rho^2 + \sinh^2\rho (du^2 + \sin^2u
dv^2).
$$
and the eleven dimensional background becomes
$$
\frac{1}{l_{11}^2} g_{11}^2=- \cosh^2\rho dt^2 + d\rho^2 +
\sinh^2\rho (du^2 + \sin^2u dv^2) + r_{12} g_7 ,
$$
$$
F_4 = k \cosh\rho \sinh^2\rho \sin u dt \wedge d\rho \wedge du
\wedge dv,\;\;\; C= - \frac{k \sin u}{3} \sinh^3 \rho dt \wedge
du\wedge dv ,
$$
being $r_{12}$ the relative radius between the $AdS_4$ and the
internal space and $k$ is a constant determined by the equation of
motions.

   We will study now the case when the metric on $X_7$ is
the tri-Sasaki metric fibered over the sphere $S^4$, the metric is
given in (\ref{chacha}). Our eleven
dimensional background is
$$
g_{11}=-\cosh^2\rho dt^2+d\rho^2+\sinh^2\rho(du^2+\sin^2u dv^2)
$$
\be\lb{chachal} +(\cos\widetilde{\rho}\sin
\widetilde{\theta}d\widetilde{\phi}-\cos\varphi d\theta -\sin\varphi
\sin\theta
d\tau)^2+(\cos\widetilde{\rho}d\widetilde{\theta}-d\varphi - \cos
\theta d\tau)^2 \ee
$$
+(\sin\widetilde{\rho}d\widetilde{t}+\cos\widetilde{\theta}d\widetilde{\phi}+\sin\varphi
d\theta -\cos\varphi \sin\theta
d\tau)^2+\cos^2\widetilde{\rho}d\widetilde{t}^2+d\widetilde{\rho}^2
+\sin^2\widetilde{\rho}(d\widetilde{\theta}^2+\sin^2\widetilde{\theta}
d\widetilde{\phi}^2). $$ The first configuration is one rotating in the
AdS background and for which the
third direction is wrapped in the Reeb direction
$$
\rho=\rho(\sigma), \qquad t=k\widetilde{\tau}, \qquad
u=\frac{\pi}{2}, \qquad v=\omega \widetilde{\tau},\qquad
\theta=\frac{\pi}{2}, \qquad \varphi=\frac{\pi}{2}, $$ \be\lb{ruta}
\tau=\lambda \delta \qquad \widetilde{\rho}=\frac{\pi}{2}, \qquad
\widetilde{\theta}=\frac{\pi}{2}, \qquad
\widetilde{\phi}=\nu_{\phi}\widetilde{\tau}, \qquad
\widetilde{t}=\nu_t \widetilde{\tau}. \ee The second configuration
we will analyze is one in which the membrane rotates in the internal
space (R-charge) and the third direction is wrapped in AdS, namely
$$
\rho=\rho(\sigma), \qquad t=k\widetilde{\tau}, \qquad
u=\frac{\pi}{2}, \qquad v=\lambda \delta,\qquad
\theta=\frac{\pi}{2}, \qquad \varphi=\frac{\pi}{2}, $$
$$
\tau=\nu \widetilde{\tau} \qquad \widetilde{\rho}=\frac{\pi}{2},
\qquad \widetilde{\theta}=\frac{\pi}{2}, \qquad
\widetilde{\phi}=\nu_{\phi}\widetilde{\tau}, \qquad
\widetilde{t}=\nu_t \widetilde{\tau}.
$$
For the first configuration by selecting $L=1$ we have that
$$
g_{\mu\nu}\partial_{\sigma}X^{\mu}\partial_{\widetilde{\tau}}X^{\nu}=
g_{\mu\nu}\partial_{\delta}X^{\mu}\partial_{\widetilde{\tau}}X^{\nu}=
g_{\mu\nu}\partial_{\sigma}X^{\mu}\partial_{\delta}X^{\nu}=0,
$$
$$
g_{\mu\nu}\partial_{\widetilde{\tau}}X^{\mu}\partial_{\widetilde{\tau}}X^{\nu}=
-\kappa^2\cosh^2\rho+\omega^2\sinh^2\rho+\nu_{\phi}^2+\nu_t^2.
$$
$$
g_{\mu\nu}\partial_{\sigma}X^{\mu}\partial_{\sigma}X^{\nu}=(\frac{d\rho}{d\sigma})^2,
$$
and the equations of motion gives the further relation
$$
\frac{d\rho}{d\sigma}=\sqrt{-\kappa^2\cosh^2\rho+\omega^2\sinh^2\rho+\nu_{\phi}^2+\nu_t^2}.
$$
Inserting this equation into the action give us \be\lb{inti}
S=-P\int
d\widetilde{\tau}\int^{\rho_0}_0\sqrt{-\kappa^2\cosh^2\rho+\omega^2\sinh^2\rho+\nu_{\phi}^2+\nu_t^2}
\ee We have that
$$
E=-\frac{\delta S}{\delta
k}\sim\int^{\rho_0}_0\frac{\cosh^2\rho}{\sqrt{-\kappa^2\cosh^2\rho+\omega^2\sinh^2\rho+\nu_{\phi}^2+\nu_t^2}},\qquad
$$
$$
S=\frac{\delta S}{\delta
\omega}\sim\int^{\rho_0}_0\frac{\sinh^2\rho}{\sqrt{-\kappa^2\cosh^2\rho+\omega^2\sinh^2\rho+\nu_{\phi}^2+\nu_t^2}}
$$
But the integral (\ref{inti}) is one of those appearing in
\cite{Gubser}, and we obtain from here that $E-S\sim\log S$, which
is what we wanted to show.

For the other configuration we have that \be\lb{inti2} S=-P\int
d\widetilde{\tau}\int^{\rho_0}_0\sinh\rho\sqrt{-\kappa^2\cosh^2\rho+\nu^2+\nu_{\phi}^2+\nu_t^2}
\ee In this case we have no place for spin, in this case we have
energy and R-charge angular momentum $J$
$$
E=-\frac{\delta S}{\delta k},\qquad J=\frac{\delta S}{\delta \nu}.
$$
From (\ref{inti2}) it is obtained that for long membranes, we have
$E=J+$...

    Let us consider now the rotating
configuration (\ref{ruta}) in the deformed background (\ref{dofo}).
For this configuration we
have that $\Delta=1$, $\Omega^{ab}=1$ and $Q_i$ and
$D\widetilde{t}\wedge D\widetilde{\phi}\wedge D\tau$ are zero. This
means  that the effective metric that the membrane  see rotating
over (\ref{dofo}) or (\ref{chachal}) is essentially the same. Thus
the logarithmic behaviour of the difference $E-S$ is reproduced for
the deformed background. We find this interesting because, while the
undeformed background is a direct product of $AdS_4$ with a seven
space, the deformation is not.

\section{Kahler-Einstein over Kahler-Einstein and other examples}

Till the moment have found an explicit expression for
Kahler-Einstein metrics defined over the twistor space Z of any four
dimensional quaternion Kahler space.  We have also have found the
corresponding Einstein-Sassaki metrics and we have checked, in
accordance with \cite{Swann}, that the eight-dimensional cone over
them is hyperKahler. It is indeed a Swann metric. Thus such
Einstein-Sasaki metrics admit three conformal Killing vectors and
are tri-Sasaki. This is different than other Kahler-Einstein spaces
appearing in the literature, for which the Einstein-Sasaki metrics
admits only two conformal Killing vectors. In this section we review
some Kahler-Einstein metrics which are fibered over Kahler metrics
of lower dimensions  \cite{Pedpoon}. In general they correspond to
Einstein-Sasaki metrics which are not tri-Sasaki. We consider them
construction of importance, because they encode several known spaces
appearing in the literature.

\subsection{The Pedersen-Poon hamiltonian approach}

   We consider a Kahler space $M$ with metric $g$, a Kahler form $\Omega$ and a complex
structure $J$. We assume the presence of $n$ Killing vectors
$(X_1,..,X_N)$ for which ${\cal L}_{X_i}\Omega=0$ which means that
the generalized torus $T^n$ act through holomorphic isometries over
$M$. An holomorphic isometry is also hamiltonian, that is, ${\cal
L}_{X_i}J=0$. The Killing vectors are linearly independent in a
dense open set of $M$, and are isotropic, that is, $\Omega(X,X)=0$.
This implies that $JX$ is orthogonal to every component of $X$. From
the relation
$$
{\cal L}_{X_i} \Omega=i_{X_i} d\Omega+ d(i_{X_i} \Omega)=d(i_{X_i}
\Omega)=0,
$$
being $i_{X_i}$ the contraction of the vector field $X_i$ with the
two form $\Omega$, it follows the existence of N functions $z_i$,
called momentum maps, defined through the relations \be\lb{moma}
dz_i=i_{X_i}\Omega. \ee The manifold $M$ can be viewed as a torus
bundle over a real manifold of dimension $2m-N$, being $m$ the
complex dimension of $M$. By denoting the N fiber coordinates as
$(t_1,..,t_N)$ it follows that the metric take the form \be\lb{tori}
g=h+w_{ij}dz^i dz^j+(w^{-1})_{ij}(dt_i+\theta_i)(dt_j+\theta_j), \ee
in the momentum map system, being $\theta_i$ certain 1-forms defined
over the base space $h$ of the bundle. The matrix $w_{ij}$ is
symmetric and positively definite.

     The manifold obtained by quotient of $M$ by the torus $T^N$ is
described by the coordinates $z_i$ and other complex coordinates
$\xi_{\nu}$ with $\nu=1,..,m-N$. The metric $h$ is $2(m-N)$
dimensional, but depend on the coordinates $z_i$ as evolution
parameters. In other words $h$ is the metric on the quotient space
of each level set of the momentum maps. Both the matrix $u$ and the
base metric $h$ are in principle $z_i$ dependent and $t_i$
independent. The metric $h$ is Kahler \cite{Galicki} and therefore
complex, and can be written in complex form \be\lb{cmp}
h=h_{ab}d\xi^a d\overline{\xi}^b. \ee From the definition of $\Omega$ it follows
directly that \be\lb{der} g(J X_i, X_j)=\Omega(X_i, X_j)=dz_i(X_j),
\ee and therefore
$$
-J(dt_i+\theta_i)=w_{ij} dz_j.
$$
This implies that
$$
i (dt_i+\theta_i)+w_{ij} dz_j,
$$
are $(0,1)$ type forms. The metric (\ref{tori}) can be expressed in complex form
as \be\lb{tori2} g=h_{ab}d\xi^a d\overline{\xi}^b+(w^{-1})_{ij}i
[w_{ij} dz_j+i(dt_i+\theta_i)][w_{ij} dz_j-i(dt_i+\theta_i)], \ee
and the corresponding Kahler form is \be\lb{decon}
\Omega=\Omega_h+dz_{i}\wedge (dt_{i}+\theta_i). \ee From the fact
that $\Omega$ is closed, it is obtained a differential system
involving $u$, the Kahler metric $h_{ab}$ and $\theta_i$. The
resulting equation is \be\lb{unto}
d\theta_k=\frac{i}{2}\frac{\partial h_{ab}}{\partial z_k}d\xi^a
\wedge d\overline{\xi}^b+i\frac{\partial w_{kl}}{\partial \xi^k}dz^j
\wedge d\xi^k-i\frac{\partial w_{kl}}{\partial \overline{\xi}^k}dz^j
\wedge d\overline{\xi}^k, \ee and the integrability condition
$d(d\theta_k)=0$ is equivalent to the equation \be\lb{int}
\frac{\partial^2 h_{ab}}{\partial z_{i}\partial z_{j}}
+\frac{\partial^2 w_{ij}}{\partial \xi^a
\partial \overline{\xi}^b}=0.
\ee The constructed metric is Kahler. It will be also Einstein if
\be\lb{hota} \rho=\Lambda \Omega \ee being
$\rho=Ric(J\cdot,\cdot)=-i\overline{\partial}\partial \log \; \det
g$ the Ricci form of the metric $g$, and the scalar curvature will
be $2m\Lambda$. The resulting system for Kahler-Einstein metrics was
worked out in \cite{Pedpoon}. By defining the function $u$ by
\be\lb{barry} u=\log \det h -\log \det w, \ee and using that in a
Kahler manifold $-i\overline{\partial}\partial=dJd$ it follows that
the system (\ref{hota}) is equivalent to \be\lb{frida} d(Jdu)=A
\Omega, \ee where we have defined $A=-2\Lambda$. By taking into
account the expression of $\Omega$ (\ref{decon}) it is obtained from the last condition
the following differential system
$$
4\frac{\partial^2 u}{\partial \xi_{\lambda}\partial
\overline{\xi}_{\mu}} +\frac{\partial u}{\partial
z_k}(w^{-1})_{kl}\frac{\partial h_{\lambda \mu}}{\partial z_l}=A
h_{\lambda\mu},
$$
$$
\frac{\partial(\frac{\partial u}{\partial
z_k}(w^{-1})_{kl})}{\partial \xi_{\lambda}}=0,\qquad
\frac{\partial(\frac{\partial u}{\partial
z_k}(w^{-1})_{kl})}{\partial z_{i}}=A \delta_{il}.
$$
The last equation implies that
$$
\frac{\partial u}{\partial z_k}(w^{-1})_{kl}=Az_l +B.
$$
From all this discussion it follows that our toric Kahler-Einstein
metrics are described by the system
$$
 4\frac{\partial^2 u}{\partial \xi_{\lambda}\partial
\overline{\xi}_{\mu}} +(Az_l+B_l)\frac{\partial h_{\lambda
\mu}}{\partial z_l}=A h_{\lambda\mu}
$$
\be\lb{pefo} \frac{\partial u}{\partial z_k}(w^{-1})_{kl}=Az_l+B_l,
\ee
$$ \frac{\partial^2 h_{ab}}{\partial z_{i}\partial z_{j}}
+\frac{\partial^2 w_{ij}}{\partial \xi^a
\partial \overline{\xi}^b}=0.
$$
These equations describe metrics with commuting Killing vectors. But
in order to have a free torus action the coordinates $t_k$ should be
periodically identified. This is achieved if the closed form
\be\lb{integral} d\theta_k=\frac{i}{2}\frac{\partial
h_{ab}}{\partial z_k}d\xi^a \wedge d\overline{\xi}^b+i\frac{\partial
w_{kl}}{\partial \xi^k}dz^j \wedge d\xi^k-i\frac{\partial
w_{kl}}{\partial \overline{\xi}^k}dz^j \wedge d\overline{\xi}^k \ee
is an integral form for any $k$. In this case there will not be
singularities if the coordinate $t_k$ is periodic.

     In the $N=1$ case, that is, when there is only one $U(1)$ holomorphic isometry,
the system (\ref{pefo}) is reduced to
$$
 4\frac{\partial^2 u}{\partial \xi_{\lambda}\partial
\overline{\xi}_{\mu}} +(Az+B_l)\frac{\partial h_{\lambda
\mu}}{\partial z}=A h_{\lambda\mu}
$$
\be\lb{pef} \frac{\partial u}{\partial z_k}(w^{-1})=Az+B, \ee
$$ \frac{\partial^2 h_{ab}}{\partial z^2}
+\frac{\partial^2 w}{\partial \xi^a
\partial \overline{\xi}^b}=0.
$$
Following \cite{Pedpoon} it is known that the system can be
simplified by imposing that the Kahler quotient metrics obtained
from each set of levels are homothetic, that is, $h=f(z)q$ being the
metric $q$ independent on the $z$ coordinate. A further
simplification is obtained if the length function $w$ is just a
function of $z$. In this case it follows from (\ref{pef}) that
$f(z)=Cz+D$ being $C$ and $D$ constants, and that
$$
-\frac{\partial^2 \log \det q}{\partial\xi_{\lambda}\partial
\xi_{\nu}}=k q_{\lambda\nu},
$$
being the constant $k$ defined by $4k=BC-AD$. This mean that $q$ is
also a Kahler-Einstein metric with scalar curvature $4nk$. The class
$F_k$ reduce in this case to the  Chern form of the Kahler-Einstein
base, which takes values $2\pi Z$ for any Kahler-Einstein metric.
Thus the metric that we are presenting are defined on a circle
bundle. The length function $w$ is obtained from the second
(\ref{pefo}). The result shows that there is no lost of generality
in selecting $C=1$ and $D=0$, the solution is given by \be\lb{pp}
w=\frac{z^n}{p z^{n+2}+q z^{n+1}+s} \ee being $p=A/n+2$, $q=B/n+1$
and $s$ another constant. The local form of this subfamily of
metrics is \be\lb{novo} g'=z q + w dz^2 + \frac{(d\tau + A)^2}{w}.
\ee being $A$ given by
$$
dA=\Omega_q
$$
The Kahler form of the new metric is simply \be\lb{niuka}
\Omega'=z\Omega_q+dz\wedge (d\tau + A), \ee and can also be
expressed as \be\lb{niuka2} \Omega'=d(zA)+dz\wedge d\tau=dA',\qquad
A'=zA+z d\tau. \ee The coordinate $z$ plays a role of a momentum map
of the isometry $\partial_{\tau}$.

\subsection{Complete metrics}

   Let us consider a $2n$ dimensional Kahler-Einstein metric
with sectional curvature normalized to one. This condition together
with $B=4\kappa$ fix the value $B=n+1$. The metric (\ref{novo})
takes in this case the following form \be\lb{kafu}
g_6=\frac{dr^2}{V}+\frac{r^2}{4}V(dt+A)^2+\frac{r^2}{4}g_{fs}, \ee
being $V$ given by \be\lb{pote}
V=1-(\frac{a}{r})^{2n+2}-\frac{\Lambda}{2(n+2)}r^2. \ee There is an
apparent singularity at the zeroes $r_0$ of $V$. Nevertheless such
singularities can be removed for certain values of the parameters of
the metric \cite{Pedpoon}-\cite{Donpage}. If $\Lambda>0$ then the
metric will be complete if and only if $a=0$ and the base space is
$CP(n)$ with its canonical metric. In this case the total space will
be $CP(n+1)$ with the Fubbini-Study metric \cite{Donpage}. If
instead $\Lambda<0$ there exist another complete metrics for certain
values of the parameters \cite{Pedpoon}. This is seen as follows.
Let us consider the fiber metric
$$
g_f=\frac{dr^2}{V}+\frac{r^2}{4}V dt^2,
$$
and let us introduce the radial coordinate $R^2=r^2 V$. The fiber
metric have apparent singularities at the zeroes of $V$ and the
coordinate $R$ tends to zero near the singularities. By
differentiating (\ref{pote}) it is obtained that
\be\lb{magda}
\frac{dV}{dr}=\frac{2}{r}U, \qquad
U=(n+1)(\frac{a}{r})^{2n+2}-\frac{\Lambda}{2(n+2)}r^2,
\ee
and in terms of these quantities the fiber can be reexpressed as
\be\lb{lab} g_f=(1+\frac{R^2
r^2}{U})^{-2}\frac{dR^2}{U^2}+\frac{R^2}{4}dt^2. \ee
In a singularity point $r_0$ we have that $V(r_0)=0$. Let us also suppose that
$U(r_0)=p\in Z$. In
this specific case the fiber metric (\ref{lab}) near the singularity takes the form
$$
g_f\simeq \frac{1}{p^2}(dR^2+R^2d\tau^2),
$$
being $2p\tau=t$. This means that the fiber metric extends smoothly
across the singularity $R=0$. The question now is to find values of
the parameters $\Lambda$ and $a$ such that the conditions
$U(r_0)=p\in Z$ and $V(r_0)=0$ are realized. By using the expressions
(\ref{magda}) for $U$ and $V$ it is found that these conditions reduce
to an algebraic equation for $\Lambda$ and $a$ with solution
 \be\lb{poro} \Lambda=\frac{2(n+1-p)}{r_0^2}, \qquad
a^{2n+2}=(r_0)^{2n+2}\frac{p+1}{n+2}, \ee which gives a further relation
\be\lb{smu}
(a^2\Lambda)^{n+1}=(2n+2-2p)^{n+1}\frac{p+1}{n+2}.
\ee
Thus, the metric extends smoothly across the singularity only if the parameters
$a$ and $\Lambda$ are related by (\ref{smu}).
We see from (\ref{poro}) that if $\Lambda<0$
then $p\geq n+2$. Also
$$
\frac{dV}{dr}=\frac{2n+2}{r^{2n+3}}a^{2n+2}-\frac{\Lambda}{n+2}r>0
$$
for $r>0$. This means that the metric is non singular for $r>r_0$.
In particular if the Fubbini-Study metric is used as the base space,
then the desingularization procedure presented above corresponds to
the desingularization $O(-p)\to C^{n+1}/Z_p$, being $O(-(n+1))$ the
canonical bundle of $P^n$  \cite{Pedpoon}.

   Also, the case $\Lambda<0$ corresponds to the parameters $p\geq 0$ and $s\leq 0$ in
(\ref{pp}). The fiber metric $g_f$ is two dimensional and by Gauss
theorem, it is conformally flat. This means that there exist a
coordinate system $(\rho, \tau)$ such that \be\lb{l1} p^2 g_f=p^2 w
dz^2+\frac{4d\tau^2}{w}=\Omega^2(\rho)(d\rho^2+\rho^2d\tau^2), \ee
being $\Omega^2$ a conformal factor.
From this equality we get the relation
$$
\frac{d\rho}{dz}=p\frac{w^{1/2}}{\Omega}, \qquad
\frac{4}{w}=\rho^2\Omega^2
$$
By differentiating the second one we get that
$$
4\frac{dw^{-1}}{dz}=2\rho^2\Omega \frac{d\Omega}{dz}+2\rho\Omega^2
\frac{d\rho}{dz}=2\Omega \rho^2 \frac{d\Omega}{dz}+2p\Omega^2 \rho
\frac{w^{1/2}}{\Omega}.
$$
By introducing the first relation one obtains that
$$
4\frac{dw^{-1}}{dz}=\frac{2}{w\Omega}\frac{d\Omega}{dz}+p \qquad
\Longleftrightarrow \qquad \frac{d\log(w\Omega^2)}{dz}=-pw
$$
From the last equation we obtain that
$$
\Omega^2 w=C \exp\{-p\int_0^z w dz\}
$$
and this, together with the second (\ref{l1}) yields
$$
\rho^2=\frac{4}{w\Omega^2}=4C^{-1} \exp\{p\int_0^z w dz\}.
$$
From (\ref{pp}) it is obtained that $$ w^{-1}=p z^{2}+q
+\frac{s}{z^{n}}
$$
and using that for $\Lambda\leq 0$ we have $p\geq 0$ and $s\leq 0$
it follows that for any positive constant $C_0$ there exists a value
$z_0
> \sqrt{r_0}$ such that any $z>z_0$ we have that
$$
w^{-1}\geq \frac{p}{n+2}z^2+C_0^2,
$$
from where it follows that
$$
\int_{\sqrt{r_0}}^z
w\leq\int_{\sqrt{r_0}}^{z_0}w+\frac{A_0}{C_0^2}(\arctan
\frac{z}{A_0}-\arctan \frac{z_0}{A_0}), \qquad
A_0=C_0\sqrt{\frac{n+2}{A}}.
$$
From the last inequality it is seen that for $z\to\infty$ the
function $\rho$ approaches to a constant. In other words $\rho$ is a
bounded function and hence the fiber metrics are defined on an open
disk. This result is independent on the choice of the
base space and is one of the key ingredient to prove that the open
disk bundle of $O(-p)\to P^n$ admit a complete Kahler-Einstein
metric with negative scalar curvature with $SU(n+1)\times U(1)$
invariant \cite{Pedpoon}.

\subsection{The Calabi-Yau limit}

    It is of interest to consider the
Ricci-flat limit of the metrics defined by (\ref{pote}) and
(\ref{kafu}). The resulting metric will be Ricci-flat Kahler, thus
Calabi-Yau, and its holonomy will be included in $SU(3)$. But we
have already mentioned that these metrics are complete only if the
parameters are constrained by (\ref{poro}). In the Ricci-flat limit
this condition is not satisfied and thus we do not have criteria to
know if the result will be a complete metric, except in the case
$a=0$ for which $V=1$.  We see that it deserve the attention to
study the Ricci flat limit $A=-2\Lambda=0$ of equations (\ref{pefo})
directly, instead of taking the limit to known solutions.

    In references \cite{Bielawski}-\cite{Hwang} there have been made
certain advances in constructing complete Calabi-Yau metrics, which
we describe here briefly. The Ricci flat limit of the system
(\ref{pefo}) is
$$
4\frac{\partial^2 u}{\partial \xi_{\lambda}\partial
\overline{\xi}_{\mu}} +B\frac{\partial h_{\lambda \mu}}{\partial
z}=0
$$
\be\lb{pefo2} \frac{\partial u}{\partial z}= w, \ee $$
4\frac{\partial^2 h_{ab}}{\partial z^2} +\frac{\partial^2
w}{\partial \xi^a
\partial \overline{\xi}^b}=0.
$$
From (\ref{barry}) it is obtained that $u=\log w^{-1}\det h$. From
these equality together with the second equation (\ref{pefo2}) it is
deduced that \be\lb{gila} w^{-1}=\frac{\int_0^z \det h}{\det h}. \ee
In addition, by multiplying the first of (\ref{pefo2}) by
$d\xi_{\lambda}\wedge \xi_{\mu}$ and summing over the repeated
indices gives \be\lb{jjd} \frac{d}{dz}\Omega_h(z)=-i\partial
\overline{\partial}u. \ee Combining formula (\ref{jjd}) with the
definition of $u$ gives \be\lb{jjd2}
\frac{d}{dz}\Omega_h(z)=\rho(h)-i\partial \overline{\partial}\log w
\ee and from the last formula together with (\ref{gila}) we obtain
\be\lb{jamiro} \frac{d}{dz}\Omega_h(z)=-i\partial
\overline{\partial}\int_0^z \det h. \ee If we were able to find a
triplet $(h(z), \Omega_h(z), w)$ solving these equations then we
will construct a Calabi-Yau metric in six dimensions with local form
\be\lb{coloyo} g=h+w dz^2 +w^{-1}(d\tau+A)^2, \ee being $A$ a 1-form
obtained from (\ref{unto}), which in our case reduces to
$$
dA=\frac{i}{2}\partial_z h_{ab}d\xi^a \wedge
d\overline{\xi}^b+i\frac{\partial w}{\partial \xi^k}dz \wedge
d\xi^k-i\frac{\partial w}{\partial \overline{\xi}^k}dz \wedge
d\overline{\xi}^k.
$$
Here $(\xi^k, \overline{\xi}^k)$ are complex coordinates for $h(z)$.

    A simple solution can be found starting
with a four dimensional Kahler-Einstein metric $g_4$ with Kahler
form $K$ defined over a manifold $X_4$, as in the previous
subsection. Let us consider the two form \be\lb{tutu}
\Omega_h(z)=\overline{J}+z\rho(\overline{J}), \ee being
$\overline{J}$ the Kahler form for $g_4$. Because $g_4$ is
Kahler-Einstein we have that $\rho(\overline{J})=\Lambda
\overline{J}$. This implies that
$$
\Omega_h(z)=(1+\Lambda z)\overline{J}.
$$
We also have that
$$
\Omega_h(z)\wedge \Omega_h(z)=P(z)\overline{J}\wedge \overline{J}
$$
being $P(z)=(1+\Lambda z)^2$ and therefore
$\rho(\overline{J})=\rho(\Omega_h)$. By introducing (\ref{tutu}) in
(\ref{jjd2}) and using $\rho(\overline{J})=\rho(\Omega_h)$ we see
that $\Omega_h(z)$ is a solution of (\ref{jjd2}). The corresponding metric $h(z)$
is simply a z-dependent dilatation of $g_4$, namely
\be\lb{alfa}
h(z)=(1+\Lambda z)g_4,
\ee
and from (\ref{gila}) we see that
\be\lb{ville} w^{-1}=\frac{\int_0^z (1+ \Lambda x)^2dx}{(1+ \Lambda
z)^2}=\frac{(1+\Lambda z)^3-1}{3\Lambda(1+\Lambda z)^2}.
\ee
By defining
$r^2=1+\Lambda z$ the metric take the following asymptotically
conical form \be\lb{lametra}
g_6=\frac{r^2}{9}(1-\frac{1}{r^6})(d\tau+A)^2+\frac{
r^6}{(r^6-1)}dr^2+\frac{\Lambda}{3} r^2 g_4 \ee with
$dA=\overline{J}$. This metric possess holonomy in $SU(3)$ and
depends on $\Lambda$ and the other parameters of the basis $g_4$. An
important result given in \cite{Bielawski} is that if $\Lambda>0$
then the metric (\ref{lametra}) is complete over the canonical
bundle $K_X$ of $X_4$. In addition it is clear that (\ref{lametra})
is asymptotically conical, that is, for large values of $r$ it tends
to a cone of the form $dr^2+r^2g_5$ being $g_5$ the Einstein-Sassaki
metric given by
$$
g_5=\frac{1}{9}(dt+A)^2+\frac{\Lambda}{3} g_4.
$$
Nevertheless this metric is Calabi-Yau for any value of $r$, not
only asymptotically, and thus (\ref{lametra}) provides a deformation
of such cones without spoiling the Calabi-Yau condition.

  In order to illustrate this construction, we can extend the
Kahler-Einstein metrics obtained in \cite{Mateos8}-\cite{Mateos4} to
an asymptotically conical Calabi-Yau metric. The result is
$$g_6=\frac{r^6}{(r^6-1)}dr^2+
\frac{r^2}{9}(1-\frac{1}{r^6})(d\tau-\cos\theta d\varphi+y
(d\beta+c\cos\theta d\varphi))^2+2r^2
\frac{(1-cy)}{6}(d\theta^2+\sin^2\theta d\varphi^2)
$$
\be\lb{ascon}
+\frac{2r^2dy^2}{w(y)q(y)}+\frac{2w(y)q(y)}{36}(d\beta+c\cos\theta
d\varphi)^2 \ee where we have defined
$$
w(y)=\frac{2(a-y^2)}{1-cy}, \qquad q(y)=\frac{a-3y^2+2cy^3}{a-y^2}.
$$
If we select $c=0$ and $a=3$ the metric will be asymptotically a
cone over $T^{1,1}$. If we select instead $c=a=1$ then the five
dimensional metric will be $S^5$.

  There exist another solutions $(h(z), \Omega_h(z), w)$
of the system (\ref{gila})-(\ref{jjd2}) that can be found starting
with a Kahler manifold $M$ with metric $g_4$ which is not Einstein,
but possess constant eigenvalues of the Ricci curvature. Let us
consider first the case in which the metric possess two different
eigenvalues $\lambda_1$ and $\lambda_2$ with multiplicity two. This
case has been studied recently in \cite{Aposdrago}. We will show
that (\ref{tutu}) still represents a solution although in this case
$\rho(\overline{J})\neq\Lambda \overline{J}$. For any Kahler
manifold $M$ with constant eigenvalues of the Ricci curvature the
Ricci form $\rho$ and the Kahler form $\overline{J}$ will be
generically
$$
\overline{J}=\overline{J}_1+\overline{J}_2,\qquad
\rho=\lambda_1\overline{J}_1+\lambda_2\overline{J}_2.
$$

In addition we always have that
$$
\rho\wedge \overline{J}=s\omega
$$
being $\omega$ the volume form of $M$ and $s$ the scalar curvature,
which in our case is $s=2\lambda_1+2\lambda_2$. We also have that
$$
\overline{J}_1\wedge\overline{J}_1=\overline{J}_2\wedge
\overline{J}_2=0,\qquad \overline{J}_1\wedge \overline{J}_2=2\omega.
$$
By using this relations it is direct to check that
$$
\Omega_h(z)\wedge\Omega_h(z)=P(z)\overline{J}\wedge \overline{J},
$$
being $P(z)=(1+\lambda_1 z)(1+\lambda_2 z)$. Therefore we have again
that $\rho(\overline{J})=\rho(\Omega_h)$ and thus (\ref{tutu}) is a
solution of (\ref{jjd2}).  The function $w$ is
given by (\ref{gila}), the result is
\be\lb{iu}
w^{-1}=\frac{\int_0^z P(x)dx}{P(z)}=\frac{1+\lambda_1
z}{2\lambda_1}+\frac{1+\lambda_2
z}{2\lambda_2}-\frac{1}{(1+\lambda_1 z)(1+\lambda_2
z)}(\frac{1}{2\lambda_1}+\frac{1}{2\lambda_2}).
\ee
The metric $h(z)$ is the metric for which
$\Omega_h(z)$ is the Kahler form. The procedure in order to find it is as follows.
One need to find a basis of soldering forms $\widetilde{e}^i(z)$ such that
$$
\Omega_h(z)=\widetilde{e}^1(z)\wedge \widetilde{e}^2(z)
+\widetilde{e}^3(z)\wedge \widetilde{e}^4(z)
$$
The metric $h(z)$ will be given by $h(z)=\widetilde{e}^i(z)\otimes \widetilde{e}^i(z)$.

  An special case of Kahler spaces with two degenerate eigenvalues is given as follows.
The two forms $\overline{J}_i$ are characterized by
$$
\overline{J}_1=\overline{J}(\pi_{1} X, \pi_{1} Y), \qquad \qquad
\overline{J}_2=\overline{J}-\overline{J}_1
$$
being $\pi_{1}$ the projection from $TM$ to $E_{1}$, being $E_{1}$
the corresponding $J$ invariant subspace associated to the
eigenvalue $\lambda_1$. The closure of $\overline{J}$ and $\rho$
implies that $\overline{J}_i$ are also closed. The almost complex
structure $\widetilde{J}$ defined by $\widetilde{J}|_{E_1}=J|_{E_1}$
and $\widetilde{J}|_{E_2}=-J|_{E_2}$ commutes with $J$ and the
corresponding two form
$$
\overline{\widetilde{J}}=\overline{J}_1-\overline{J}_2,
$$
is sympletic and possess opposite orientation with respect to the
one defined by $\overline{J}$. This means that
$$
\overline{\widetilde{J}}\wedge
\overline{\widetilde{J}}=-\overline{J}\wedge \overline{J}=2\omega.
$$
It has been shown that the sympletic two form $\widetilde{J}$ is
integrable if and only if the base space $M$ is a direct product of
two Kahler-Einstein spaces \cite{Aposdrago}. In this case $(g_4,
\widetilde{J})$ will be a Kahler structure with orientation opposite
to $(g_4, J)$, and $\overline{J}_1$ and $\overline{J}_2$ will be the
Kahler forms for such metrics. As an example we can consider the product of the
two dimensional Fubbini-Study metric $g_{fs}$ with the Bergmann one $g_{b}$. We normalize
the curvature as $\lambda=\pm 1$. With the corresponding Kahler forms $\overline{J}_{fs}$
and $\overline{J}_b$ we consider the two form
$$
\Omega_{h(z)}=(1+z)\overline{J}_1+(1-z)\overline{J}_2,
$$
which by construction is a solution of (\ref{gila}). The
corresponding metric $h(z)$ is given by
$$
h(z)=(1+z)g_{fs}
+(1-z)g_b.
$$
From (\ref{iu}) we see that $w=1/z$ in this case, and the Calabi-Yau metric reads
\be\lb{au}
g_6=(1+z)g_{fs}
+(1-z)g_b+\frac{1
}{z}dz^2+z(d\tau+A)^2, \ee being $A$ given by
$dA=\overline{J}_1-\overline{J}_2$. Observe that in general
$dA=\rho(\overline{J})$.

       Another class of Kahler manifolds with constant Ricci eigenvalues
are homogeneous Kahler manifolds, for which the holomorphic
isometries acts transitively. There also exist non homogeneous
Kahler metrics in the literature with constant eigenvalues of the
Ricci curvature. An example is the family  \be\lb{waid} g=e^u x
(dx^2+dy^2)+x dz^2+\frac{1}{x}(dt+ydz)^2, \ee which possess this
property if $u$ is a function satisfies \be\lb{lubil}
u_{xx}+u_{yy}=s x e^u. \ee The constant $s$ is the scalar curvature
of the metric. The family (\ref{waid}) is Kahler and in general non
homogeneous, except for certain subcases. For instance by selecting
$u=3\log x$ it is obtained the Kahler metric \be\lb{dragon}
g=\frac{dx^2}{x^2}+\frac{dy^2}{x^2}+x dz^2+\frac{(dt + y
dz)^2}{x^2}, \ee with Kahler form
$$
\overline{J}=-dz^2 \wedge dy+dy\wedge d(\frac{1}{x}).
$$
This metric possess two different eigenvalues of the curvature
tensor. If we make the variable change \be\lb{varchan}
u_1=\frac{x^2+y^2-1}{2x},\qquad v_1=-\frac{y}{x},\qquad u_2=t,\qquad
v_2=z \ee then the metric takes the form \be\lb{dragon2}
g=(-u_1+\sqrt{u_1^2+v_1^2+1})du_2^2+(u_1+\sqrt{u_1^2+v_1^2+1})dv_2^2-2v_1
du_2dv_2 \ee
$$
\frac{1}{(u_1^2+v_1^2+1)}[(1+v_1^2)du_1^2+(1+u_1^2)dv_1^2-2u_1v_1du_1
dv_1].
$$
It has been shown that this metric is homogeneous and non symmetric
in \cite{Aposgaucho} and the Ricci eigenvalues are $(0, 0,
-\frac{3}{8}, -\frac{3}{8})$. But the metrics (\ref{waid}) are non
homogeneous in general.

  It has been shown that in general, the resulting
Calabi-Yau metric will be complete if the Ricci eigenvalues are all
positive \cite{Bielawski}. This is not the case for
for many of the examples that we have constructed so we can not decide
whether or not the resulting Calabi-Yau metrics are complete. It is then of interest
to classify which solutions of the equation (\ref{waid}) give rise to metrics
with positive eigenvalues. Nevertheless this could be a hard task, due to the non
linear nature of (\ref{waid}).

\section{Discussion}

  Along this paper we considered an infinite family of tri-Sasaki
7-metrics and its squashed version, which are of weak $G_2$
holonomy. We have found in particular, a large class of examples
with $T^3$ isometry. We constructed several new supergravity
backgrounds and their deformed by use of the Maldacena-Lunin
prescription. This should correspond to a marginal deformation in
the dual theory. We have found in certain manifold limit a rotating
configuration reproducing the logarithmic behaviour of the
difference between the spin and the energy. We have found the same
behaviour for the deformed background, although this is not a direct
product of $AdS_4$ with a seven dimensional space.

    We want to emphasize that there is an underlying linear structure describing
all the backgrounds presented along this work. This is given by the (\ref{backly}) and in
fact, all the spaces that we have presented here are completely determined
in terms of solutions of this equation. It will be nice to make a more
deep analysis of the dual conformal
theories of these backgrounds. Notice that the complete examples
that we have presented are defined in terms of certain twistors.
This is in part, a consequence of the underlying linear structure.
It will be interesting
to understand how these twistors are realized in the dual conformal
field theory. Perhaps the methods presented in \cite{Mateos20} could be useful
for this purpose. Another interesting task is to figure out the pp-wave limit of these
backgrounds and to study rotating configurations in the orbifold
case. We hope to come out answering these questions in a near
future.

\appendix

\section{Quaternionic Kahler spaces in dimension higher than four}

   The generators $J^i$ of the Lie algebra
$sp(1)$ of $Sp(1)\simeq SU(2)$ have the multiplication rule
\be\lb{cape} J^{i}\cdot J^{j}=-\delta^{ij} I + \epsilon_{ijk}J^{k},
\ee which implies the $so(3)\simeq su(2)$ commutation rule
\be\lb{commutatus} [J^i, J^j]= \epsilon_{ijk}J^k. \ee We see that
$J^i J^i=-I$ and therefore $J^i$ will be called almost complex
structures. An useful $4n \times 4n$ representation is
$$
J^{1}=\left(\begin{array}{cccc}
  0 & -I_{n \times n} & 0 & 0 \\
  I_{n \times n} & 0 & 0 & 0 \\
  0 & 0 & 0 & -I_{n \times n} \\
  0 & 0 & I_{n \times n} & 0
\end{array}\right),\;\;\;\;
J^{2}=\left(\begin{array}{cccc}
  0 & 0 & -I_{n \times n} & 0 \\
  0 & 0 & 0 & I_{n \times n} \\
   I_{n \times n} & 0 & 0 & 0 \\
  0 & -I_{n \times n} & 0 & 0
\end{array}\right)
$$
\be\lb{reprodui} J^{3}=J^{1}J^{2}=\left(\begin{array}{cccc}
  0 & 0 & 0 & -I_{n \times n} \\
  0 & 0 & -I_{n \times n} & 0 \\
  0 & I_{n \times n} & 0 & 0 \\
  I_{n \times n} & 0 & 0 & 0
\end{array}\right).
\ee The group $SO(4n)$ is a Lie group and this means in particular
that for any $SO(4n)$ tensor $A^{a}_{b}$ the commutator $[A, J^{i}]$
will take also values in $SO(4n)$. We will say that $A$ belong to
the subgroup $Sp(n)$ of $SO(4n)$ if and only if \be\lb{commutatus2}
[A, J^{i}]=0. \ee Condition (\ref{commutatus2}) together with
(\ref{commutatus}) implies that a tensor $B^{a}_{b}$ belongs to the
subgroup $Sp(n)\times Sp(1)$ if and only if
$$
[B,J^{i}]=\epsilon_{ijk}J^{j}B_{-}^{k},
$$
being $B_{-}^{k}$ the components of $B$ in the basis $J^{k}$. Both
conditions are independent of the representation.

   We will write a metric over a $4n$ dimensional manifold $M$
as $g=\delta_{ab}e^{a}\otimes e^{b}$, being ${e^{a}}$ the $4n$-bein
basis for which $g$ is diagonal. Let us define the triplet of
$(1,1)$ tensors \be\lb{bobo} J^{i}=(J^{i})^a_b e_{a}\otimes e^{b},
\ee defined by the matrices (\ref{reprodui}). If the holonomy is in
$Sp(n)\times Sp(1)$, then from the beginning $\omega^a_b$ will take
values on its lie algebra $sp(n)\oplus sp(1)$. As we saw above, this
implies that \be\lb{alban} [\omega, J^{i}]=\epsilon_{ijk}J^{j}A^{k}.
\ee As usual, the connection $\omega^{a}_{b}$ is defined through
$$
\nabla_{X} e^{a}=-\omega^{a}_{b}(X)e^{b},
$$
together with the Levi-Civita conditions $\nabla g=0$ and
$T(X,Y)=0$. Using the chain rule $\nabla (A \otimes B)=(\nabla A)
\otimes B+ A \otimes (\nabla B)$ for tensorial products show us that
in the einbein basis \be\lb{alban2} [\omega, J^{i}]=\nabla_{X}J^{i}.
\ee Comparing (\ref{alban}) and (\ref{alban2}) we see that
quaternionic Kahler manifold are defined by the relation $$
\nabla_{X}J^{i}=\epsilon_{ijk}J^{j}A^{k},
$$
which is independent on the election of the frame $e^a$. This proves
that (\ref{rela2}) describe quaternion Kahler metrics
\cite{Ishihara} in dimension higher than four.

    The basis $e^a$ for a metric $g$ is defined up to an $SO(4n)$
rotation. Under this $SO(4n)$ transformation the tensors
(\ref{bobo}) are also transformed, but it can be shown that the
multiplication (\ref{cape}) is unaffected. In other words, given the
tensors $J^{i}$ one can construct a new set of complex structures
\be\lb{globos} J'^i= C^i_{j}J^{j},\qquad J'^{i}\cdot
J'^{j}=-\delta^{ij} I + \epsilon_{ijk}J'^{k}\qquad
\Longleftrightarrow\qquad C^i_{k} C^{k}_j=\delta^i_j \ee This can be
paraphrased by saying that a quaternionic Kahler manifold has a
bundle $V$ of complex structures parameterized by the sphere $S^2$.
Using the textbook properties of $\nabla$ it can be seen that
(\ref{rela2}) is unaltered under such rotations.

   Let us define three new tensors
$(\overline{J}^{i})_{ab}$ by
$(\overline{J}^{i})_{ab}=(J^{i})_{a}^{c}\delta_{cb}$. From
(\ref{reprodui}) it follows that
$$
(J^i)^a_b=-(J^i)^b_a \qquad\Longleftrightarrow\qquad
(\overline{J}^i)_{ab}=-(\overline{J}^i)_{ba}
$$
This show that $(\overline{J}^{i})_{ab}$ are the components of the
two-forms $\overline{J}^{i}$ defined by \be\lb{compo}
\overline{J}^{i}=(\overline{J}^{i})_{ab} e^{a}\wedge e^{b}. \ee The
forms (\ref{compo}) are known as the hyperKahler forms. From
(\ref{rela2}) it is obtained that
$$
\nabla_{X}J^{i}=\epsilon_{ijk}J^{j}A^{k}\qquad \Longrightarrow
\qquad d\overline{J}^{i}=\epsilon_{ijk}A^j \wedge \overline{J}^{k},
$$ being $d$ the usual exterior derivative.
The last implication proves relation (\ref{basta}).

  If we change the frame $e^a$ to a new one $x_{\mu}$ then the definition
$(\overline{J}^{i})_{ab}=(J^{i})_{a}^{c}\delta_{cb}$ should be
modified by the covariant one
$(\overline{J}^{i})_{\alpha\beta}=(J^{i})_{\alpha}^{\gamma}g_{\gamma\beta}$.
Here the greek index indicates the components in the new basis and
$g_{\gamma\beta}$ are the corresponding components of the metric.
Therefore
$$
(\overline{J}^i)_{ab}=-(\overline{J}^i)_{ba}
\qquad\Longleftrightarrow\qquad
(J^{i})_{\alpha}^{\gamma}g_{\gamma\beta}=(J^{i})_{\beta}^{\gamma}g_{\gamma\alpha}
$$
The last relation is equivalent to $$ g(J^{i} X,Y)=g(X,
J^{i}Y)\qquad\Longleftrightarrow \qquad g(X,Y)=g(J^{i} X, J^{i}Y)$$
for arbitrary vector fields $X$ and $Y$ in $TM$. Then the metric $g$
will be always quaternion hermitian with respect to the complex
structures. Relation (\ref{hermoso}) is also invariant under the
automorphism of the complex structures.

    In general, if in a given manifold there exist three complex structures
satisfying (\ref{cape}), and we take intersecting coordinate
neighborhoods $U$ and $U'$, then we have two associated basis $J^i$
and $J'^i$. Both basis should be related by an $SO(3)$
transformation in order to satisfy (\ref{cape}). This means that any
quaternion Kahler space is orientable \cite{Ishihara}. Consider now
the fundamental 4-form \be\lb{lafunda} \Theta=\overline{J}^1 \wedge
\overline{J}^1 + \overline{J}^2 \wedge \overline{J}^2 +
\overline{J}^3 \wedge \overline{J}^3, \ee and the globally defined
$(2,2)$ tensor \be\lb{22} \Xi= J^1 \otimes J^1 + J^2 \otimes J^2 +
J^3 \otimes J^3. \ee By means of the formula (\ref{globos}) it
follows that both tensors (\ref{lafunda}) and (\ref{22}) are
globally defined on the manifold M. For a quaternionic Kahler
manifold it is obtained directly from (\ref{rela2}) and
(\ref{basta}) that \cite{Ishihara}
$$
\nabla \Theta=0,\;\;\;\;\;\;\; \nabla \Xi=0.
$$
In $D=8$ for a quaternion Kahler manifold $d\Theta=0$ and if the
manifold is of dimension at least $12$ then $d\Theta$ determines
completely $\nabla \Theta$. In particular $d\Theta=0$ implies
$\nabla \Theta=0$ \cite{Swann}.

      One of the most important consequences of (\ref{rela2}) is
that quaternionic Kahler spaces are always Einstein with
cosmological constant  \cite{Wolf}. The proof is briefly as follows.
From the definition of the curvature tensor $R(X,Y)=[\nabla_{X},
\nabla_{Y}]-\nabla_{[X,Y]}$ together with (\ref{rela2}) it follows
in the einbein basis that \be\lb{una}
R_{ijm}^{l}(J^{a})^m_{k}-R_{ijk}^{m}(J^{a})^l_{m}=
\epsilon_{abc}(F^{b})_{ij}(J^{c})^{l}_{k}. \ee where $R_{ijm}^{l}$
are the components of the curvature tensor and the two form $F^{a}$
was defined as \footnote{In the physical literature sometimes the
three components $\omega_{-}$ are referred as an $SU(2)$ vector
potential and $F^a$ as the corresponding strength tensor}
$$
F^a=d\omega_{-}^{a}+\epsilon_{abc}\omega_{-}^{b}\wedge
\omega_{-}^{c}.
$$
We can rewrite (\ref{una}) as a commutator
$$
[R(X,Y), J^a]=\epsilon_{abc}F^{b}J^{c},
$$
being $X$ and $Y$ arbitrary vector fields. Multiplying (\ref{una})
by $(J^{a})_l^{s}$ and contracting indices, and then multiplying by
$(J^{b})^{k}_{l}$ and using the identity \footnote{Which is clearly
true in the representation (\ref{reprodui})}
$$
(J^{a})_l^{s}(J^{b})_s^{l}=4n \delta^{ab},
$$
gives the formula \be\lb{labuena}
F^{a}_{ij}=\frac{1}{2n}R_{ijk}^{l}(J^{a})_{l}^{k}. \ee Inserting
(\ref{labuena}) into (\ref{una}) yields
$$
R_{ijk}^{l}(J^{a})_{l}^{k}=\frac{2n}{2+n}R_{im}(J^{a})_{j}^{m},
$$
which can also be expressed as \be\lb{lame} R^i_{-}=\frac{2n}{2+n}R
\overline{J}^i, \ee being $R$ is the scalar curvature and $R^i_{-}$
are the $Sp(1)$ components of the curvature tensor. The second
Bianchi identities together with (\ref{lame}) shows that $R$ is
constant and thus $R_{ij}\sim g_{ij}$ \cite{Wolf}. Thus, in any
dimension, quaternionic Kahler spaces are always Einstein with non
zero cosmological constant $\lambda$.

        Because $R$ is a constant we see from (\ref{lame}) that
\be\lb{lamas} R^i_{-}=\Lambda \overline{J}^i, \ee being $\Lambda$
certain constant. We also have from (\ref{labuena}) that
\be\lb{rela} F^i=\Lambda' \overline{J}^i, \ee being $\Lambda'$
another constant.  In the limit $\lambda\rightarrow 0$ the constants
$\Lambda$ and $\Lambda'$ goes simultaneously to zero.

       If there exists a rotation of the local frame for which
$\omega_{-}=0$ then the complex structures are locally covariantly
constant, that is \be\lb{hypi} \nabla_{X}J^{i}=0. \ee In this case
$R^i_{-}=F^i=0$ thus the space has self-dual curvature, which
implies Ricci flatness. This spaces are called hyperKahler, and
(\ref{hypi}) shows that they are Kahler with respect of any of the
complex structures. Condition (\ref{hypi}) implies that the holonomy
is in $Sp(n)$ and that \be\lb{nocon} d\overline{J}^a= 0 \ee together
with the annulation of the Niejenhuis tensor given by \be\lb{Nie}
N(X,Y)=[X,Y]+J[X,JY]+J[JX,Y]-[JX,JY]. \ee  A complex structure for
which $N(X,Y)=0$ is called integrable.

\section{Quaternion Kahler manifolds in dimension four}

      As we saw starting the previous section, in four dimensions
the statement that the holonomy is $\Gamma \subseteq Sp(n)\times
Sp(1)$ is trivial due to the isomorphism $SO(4)\simeq SU(2)_L\times
SU(2)_R \simeq Sp(1)\times Sp(1)$. We will modify this definition
and we will say that a four dimensional manifold $M$ is quaternionic
Kahler if (\ref{lamas}) holds. This condition is not trivial, we
will show below that quaternion Kahler spaces in $d=4$ are Einstein
(as in the higher dimensional case) and with self-dual Weyl tensor.

     Let us consider a four dimensional metric
$g=\delta_{ab}e^a \otimes e^b$ and the connection $\omega^a_b$ given
by the first Cartan equation
$$
de^a + \omega^{a}_{b}\wedge e^b = 0,\;\;\;\; \omega_{ib}^a = -
\omega_{bi}^a.
$$
The notation $SU(2)_{\pm}$ denote the $SU(2)_L$ and $SU(2)_R$ groups
respectively. The $SU(2)_{\pm}$ components of the spin connection
are explicitly \be\lb{secon} \omega^{a}_{\pm}=\omega^a_{0}\pm
\epsilon_{abc}\omega^b_c. \ee The curvature tensor is given by the
second Cartan equation
$$
R^a_b=d\omega^a_b + \omega^a_s \wedge \omega^s_b = R^a_{b,st}e^s
\wedge e^t
$$
and the $SU(2)$ parts are \be\lb{secon2} R^{a}_{\pm}=R^a_{0}\pm
\epsilon_{abc}R^b_c. \ee The Ricci tensor is defined in the diagonal
basis by $R_{ij}=R^a_{i,aj}$ and the scalar curvature is $R_{ii}=R$.

    Instead of use the basis ${e^a \wedge e^b}$ we can use the basis
$\overline{J}^a_{\pm}=e^0 \wedge e^a \pm \epsilon _{abc} e^b \wedge
e^c$. Then it follows that $\overline{J}^a_{\pm}$ are separately
complex structure with definite self-duality properties, that is
$$
\ast \overline{J}^a_{\pm}=\pm \overline{J}^a_{\pm}.
$$
In this basis \be\lb{riemself} R^{a}_{+}=A_{ab}\overline{J}^{b}_{+}
+ B_{ab}\overline{J}^{b}_{-},\;\;\;
R^{a}_{-}=B_{ab}^{t}\overline{J}^{b}_{+} +
C_{ab}\overline{J}^{b}_{-} \ee where the matrices $A$ and $C$ are
symmetric. The components of the Ricci tensor are \be\lb{sericc}
R_{00}=Tr(A+B),\;\;\;
R_{0a}=\frac{\epsilon_{abc}}{2}(B_{bc}^t-B_{bc}),\;\;\;
R_{ab}=Tr(A-B)\delta_{ab} + B_{ab} + B_{ab}^t, \ee and the scalar
curvature is \be\lb{esca} R=4Tr(A)=4Tr(C). \ee It is clearly seen
from (\ref{sericc}) that the Einstein condition $R_{ij}=\Lambda
\delta_{ij}$ is equivalent to $B=0$ and $Tr(A)=Tr(C)=\Lambda$.

   The components of the Weyl tensor in the diagonal basis are given by
\be\lb{Weylo} W^{a}_{bcd}=
R^{a}_{bcd}-\frac{1}{2}(\delta_{ac}R_{bd}-\delta_{ad}R_{bc}
+\delta_{bd}R_{ac}-\delta_{bc}R_{ad})+\frac{R}{6}(\delta_{ac}\delta_{bd}
-\delta_{ad}\delta_{bc}). \ee The tensor $W$ is invariant under a
conformal transformation $g \rightarrow \Omega^2 g$ and the
associated two form is
$$
W^a_b=W^{a}_{bcd}e^c\wedge e^d.
$$
An explicit calculation shows that the $SU(2)_{\pm}$ of $W$ are
$$
W^{a}_{+}=W^a_{0}+ \epsilon_{abc}W^b_c=
(A_{ab}-\frac{1}{3}Tr(A)\delta_{ab})\overline{J}^b_{+},
$$
$$
W^{a}_{-}=W^a_{0}- \epsilon_{abc}W^b_c=
(C_{ab}-\frac{1}{3}Tr(C)\delta_{ab})\overline{J}^b_{-}.
$$
From this expressions we see that to say that an Einstein space is
self dual (i.e, $W^a_{-}=0$) is equivalent to \be\lb{lacond}
C_{ab}=\frac{\Lambda}{3}\delta_{ab}\;\;\;\Longleftrightarrow\;\;\;
R^{a}_{-}=\frac{\Lambda}{3}\overline{J}^a_{-}. \ee The second
(\ref{lacond}) is the same as (\ref{lamas}) in four dimensions. Thus
we conclude then that in $D=4$ quaternionic Kahler is the same as
self-dual Einstein.

\section{The solution generating technique for IIB backgrounds}

 The eleven supergravity backgrounds constructed in (\ref{as})
possess three commuting Killing vectors. We can obtain $T^2$ IIA
supergravity solutions by reduction along one isometry, say
$\phi_3$. Also, by making a T-duality along another isometry, say
$\phi_1$, we will obtain IIB supergravity backgrounds which are also
toric. Now if we make the $SL(2,R)$ deformation of the original
11-dimensional backgrounds and we make the IIB reduction we will
obtain a new background, the IIB deformed one. Comparison between
the resulting expression will give a technique in order to deform a
IIB background into another one. The result will be a one-parameter deformation.
This is a particular case of a two parameter deformation that
is known in the literature, which we will describe now.

    Recall that any IIB background can be casted in the form
$$
g_{IIB}=F\left[\frac{1}{\sqrt{\Delta}} (D\alpha_1-C(D\phi_2))^2
+\sqrt{\Delta}(D\alpha_2)^2\right] + \frac{e^{2\Phi/3} }{F^{1/3}}
\widetilde{g},
$$
$$
B=B_{12}(D\phi_1)\wedge (D\phi_2)+ D\phi_1\wedge B_1+D\phi_2\wedge
B_2 -\frac{1}{2}A_m \wedge B_{m}+ \frac{1}{2} {\tilde b}
$$
$$
C^{(2)}=C_{12}(D\phi_1)\wedge (D\phi_2) + D\phi_1\wedge
C_1+D\phi_2\wedge C_2 -\frac{1}{2}A_m\wedge C_{m}+
\frac{1}{2}{\tilde c},
$$
\be\lb{orlo} e^{2\Phi}=e^{2\phi},\qquad C^{(0)}=\chi, \ee
$$
C^{(4)}=-\frac{1}{2}( {\tilde d}+B_{12}{\tilde c}- \epsilon^{mn}B_{m
}\wedge C_{n}-B_{12}A_m\wedge C_{m}) \wedge D\phi_1\wedge D\phi_2
$$
$$
+\frac{1}{6}\left[C+3({\tilde b}+ A_1\wedge B_{1}-A_2\wedge
B_{2})\wedge C_{(1)}\right] \wedge D\phi_1 + d_{4}  +
 \hat d_{3} \wedge D\phi_2,
$$
where
$$
D\phi_2=d\phi_2+A_2,\qquad D\phi_1=d\phi_1+A_1.
$$
 The effect of the $SL(3,R)$ transformation over
these backgrounds is the following. We have three objects which
transform as vectors and tensors \be\lb{onefo}
V^{(1)}=\left(\begin{array}{c} -B_{2}\\A_1\\C_{2}
\end{array}\right),\qquad
V^{(2)}=\left(\begin{array}{c} B_{1}\\A_2\\-C_{1}
\end{array}\right):\qquad V^{(i)}\longrightarrow (\Lambda^T)^{-1}V^{(i)};
\ee
$$
W=\left(\begin{array}{c}
{\tilde c}\\{\tilde d}\\
{\tilde b}
\end{array}\right)\longrightarrow \Lambda W
$$
and one matrix
$$
M=g g^T ~,\qquad g^T= \left(\begin{array}{ccc}
e^{-\phi/3}F^{-1/3}&0&0\\
0&e^{-\phi/3}F^{2/3}&0\\
0&0&e^{2\phi/3}F^{-1/3}
\end{array}\right)
\left(\begin{array}{ccc}
1&B_{12}&0\\
0&1&0\\
\chi&-C_{12}+\chi B_{12}&1
\end{array}\right),
$$
with transformation law \be\lb{tralaw} M\longrightarrow \Lambda
M\Lambda^T. \ee The scalars $\Delta$, $C$ as well as the three form
$ C_{\mu\nu\lambda}$ stay invariant under these $SL(3,R)$
transformations. From these expression one can read the generic
transformation of any of the fields.

 We will restrict ourselves with a matrix of the form
$$
\Lambda=\left(\begin{array}{ccc}
          1 & \gamma& 0 \\
          0 & 1 & 0 \\
          0 & \sigma & 1
        \end{array}\right), \qquad \Lambda^T=\left(\begin{array}{ccc}
          1 & 0 & 0 \\
          \gamma & 1 & \sigma \\
          0 & 0 & 1
        \end{array}\right)
$$
$$
\Lambda^{-1}=\left(\begin{array}{ccc}
          1 & -\gamma& 0 \\
          0 & 1 & 0 \\
          0 & -\sigma & 1
        \end{array}\right), \qquad (\Lambda^T)^{-1}=\left(\begin{array}{ccc}
          1 & 0 & 0 \\
          -\gamma & 1 & -\sigma \\
          0 & 0 & 1
        \end{array}\right)
$$
Then the transformed fields are
$$
A_1'=A_1, \qquad A_2'=A_2-\sigma A_3, \qquad A_3'=A_3
$$
$$
\widetilde{c}'=\gamma \widetilde{d}, \qquad
\widetilde{d}'=\widetilde{d}, \qquad
\widetilde{b}'=\sigma\widetilde{d}
$$
In addition, the transformation law (\ref{tralaw}) implies that
$g^{T}$ should transform as $g^{T}\to R g^{T}\Lambda^{T}$ being $R$
an $SO(3)$ transformation. The Euler angles of this rotation should
be selected in order that the non diagonal matrix in the expression
for $g^T$ conserve its form, that is, the components $(2,1)$,
$(1,3)$ and $(2,3)$ should be zero. We have that
$$
g^{T}\Lambda^T=\left(\begin{array}{ccc}
e^{-\phi/3}F^{-1/3}&0&0\\
0&e^{-\phi/3}F^{2/3}&0\\
0&0&e^{2\phi/3}F^{-1/3}
\end{array}\right)
$$
$$
\times \left(\begin{array}{ccc}
1+\gamma B_{12}&B_{12}&\sigma B_{12}\\
\gamma&1&\sigma\\
\chi+\gamma (\chi B_{12}-C_{12})&\chi B_{12}-C_{12}& 1+\sigma (\chi
B_{12}-C_{12})
\end{array}\right)
$$
which is not of the desired form. We have to multiply this
expression for a rotation matrix $R(\alpha_1, \alpha_2, \alpha_3)$
and the condition that the components $(2,1)$, $(1,3)$ and $(2,3)$
vanish give the following system of equations
$$
\cot\alpha_2 B_{12}=-\cos\alpha_3
$$
$$
\sin\alpha_2 +\cos\alpha_3\cos\alpha_2 B_{12}=-(1+\sigma
G)\sin\alpha_3
$$
$$
(\sin\alpha_1\cos\alpha_2+\cos\alpha_3\sin\alpha_2\cos\alpha_1)(1+\gamma
B_{12})+(\sin\alpha_1\sin\alpha_2-\cos\alpha_3\cos\alpha_2\cos\alpha_1)\gamma
$$
$$
-(\chi+\gamma G)\cos\alpha_1\sin\alpha_3=0
$$
The first two equations involve only $\alpha_2$ and $\alpha_3$. The
angle $\alpha_1$ is then defined by the third equation, which turns
out to be
$$
\tan\alpha_1=-\frac{\cos\alpha_3\cos\alpha_2\gamma-\cos\alpha_3\sin\alpha_2(1+\gamma
B_{12})-(\chi+\gamma G)\sin\alpha_3}{(1+\gamma
B_{12})\cos\alpha_2+\gamma\sin\alpha_2}
$$
The transformation of $F$ and $\phi$ is then obtained by requiring
that for the non diagonal matrix in $g'^T$ the diagonal elements are
$g^T_{ii}=1$. The transformed components were worked out for
instance in \cite{Gursoy}, the result is
$$ g_{11}^T= \frac{e^{-\phi/3} \kappa}{\mu}, \;\; g_{12}^{T}=
\frac{e^{5/3\phi }}{\mu\kappa} (B_{12} +\gamma B_{12}^2 -
B_{12}C_{12}\sigma + F^2(\gamma -\chi \sigma)),\;\;g_{2,2}^T=
\frac{(e^{2\phi } F^2)^{1/3}}{\kappa},$$$$
g_{32}^T=\frac{e^{-\phi/3}}{\mu}(B_{12}\chi e^{2\phi} + C_{12}^2
\sigma e^{2\phi} + B_{12}^2\sigma (1+\chi^2 e^{2\phi})+ F^2\sigma -
C_{12}e^{2\phi} (1+ 2 B_{12}\chi\sigma) )$$$$ g_{31}^T=
\frac{e^{-\phi/3}}{\mu}(- C_{12}\gamma e^{2\phi} + C_{12}^2\gamma
\sigma e^{2\phi} + B_{12}\chi^2 e^{2\phi}\sigma(1+ B_{12}\gamma)
+\sigma(B_{12} + $$$$B_{12}^2\gamma + F^2\gamma) -\chi e^{2\phi} (-1
+ C_{12}\sigma + B_{12}\gamma (2 C_{12}\sigma
-1)))$$\be\lb{transformoso} g_{3,3}^T=
(\frac{e^{-\phi}}{F})^{1/3}\sqrt{(B_{12}^2 + F^2)\sigma^2 +
e^{2\phi}(1 - C_{12}\sigma + B_{12}\chi\sigma)^2}, \ee$$ \mu=
F^{1/3}\sqrt{(B_{12}^2 + F^2)\sigma^2 + e^{2\phi} (1 - C_{12}\sigma
+ B_{12}\sigma\chi)^2}$$$$ \kappa^2= F^2\sigma^2 + e^{2\phi}(\;
(B_{12}\gamma)^2 - 2 B_{12}\gamma(C_{12}\sigma -1) +
(C_{12}\sigma-1)^2 + F^2(\gamma -\sigma\chi)^2)
$$
The transformed fields are then $$
B_{12}'=\frac{g_{12}^T}{g_{11}^T}, \;\;
e^{\phi'}=\frac{g_{33}^T}{g_{11}^T}, \;\; \chi'=(\frac{g_{22}^T
g_{11}^T}{g_{33}^T})^{1/3}g^T_{31}
$$
\be\lb{transformoso2} C_{12}'= \chi' B_{12}' - g_{32}^T  g_{22}^T
 g_{11}^T,
\ee and are completely determined in terms of the fields of the
original IIB supergravity solution. The procedure is explicitly defined.

\section{IIB deformed superbackgrounds for the spherical case}

   Let us consider our now the IIA reduction of our $T^3$ supergravity
 backgrounds of section 5. In order to perform the reduction to
we need to make the decomposition \be\lb{decom} M_{ab}D\phi_a
D\phi_b=e^{-2\phi_{D}/3}\widetilde{h}_{mn}D\phi_m D\phi_n +
e^{4\phi_D/3}(D\phi_3 + N_m D\phi_m)^2 \ee with $m,n=1,2$. The
metric $\widetilde{h}$ should not be confused with the $h$ appearing
in (\ref{don})! We find that
$$
\phi_D=\frac{3}{4}\log (M_{33}),\qquad
N_m=\frac{M_{3m}}{M_{33}}\qquad
\widetilde{h}_{mn}=\frac{M_{mn}M_{33}-M_{3m}M_{3n}}{\sqrt{M_{33}}}
$$
and is straightforward to find the IIA reduced background. By making
a T-duality \cite{Buscher} to the resulting IIA background we obtain
the IIB solution
$$
g_{IIB}=\frac{1}{h_{11}} \left[ {1 \over \sqrt{\Delta} } (D\phi_1-C
D\phi_2)^2 +\sqrt{\Delta} (D\phi_2)^2 \right]
+e^{2\phi/3}\widetilde{g},
$$
$$
B=\frac{h_{12}}{h_{11}}D\phi_1 \wedge D\phi_2- D\phi_2\wedge
C_{(32)}+D\phi_1\wedge  A_1- \frac{1}{2}C_{(3)} +C_{(31)}\wedge A_1
$$
$$
C^{(2)}=-(N_2-\frac{h_{12}}{h_{11}}N_1)D\phi_1\wedge
D\phi_2-D\phi_2\wedge C_{(12)}- D\phi_1\wedge A_3 -\frac{1}{2} C_{1}
+ C_{(31)} \wedge A_3
$$
\be\lb{forom} e^{2\Phi}=\frac{e^{2\phi}}{h_{11}},\qquad C^{(0)}=N_1
\ee
$$
C^{(4)}=-\frac{1}{2}D\phi_2\wedge D\phi_1\wedge[  C_{(2)} + 2
C_{32}\wedge A_3 -\frac{h_{12}}{h_{11}}C_{(1)} +2 C_{(31)}\wedge
A_3)]
$$
$$
+ \frac{1}{6} (C + 3  C_{(3)} \wedge A_3 ) \wedge  D\phi_1  + d_4 +
\hat d_{3}\wedge D\phi_1
$$
$$
D\phi_1=d\phi_1-C_{(31)},\qquad D\alpha_2=d\phi_2+ A_2
$$
The forms $d_{4}$, $\hat d_{3} $ are determined by the self duality
conditions for the five form field strength.

  It will interesting how this procedure works for our example (\ref{chachal})
associated to $S^4$. The relevant quantities that we need are
$$
\phi_D=\frac{3}{4}\log M_{\tau\tau}=-\frac{1}{4}\log\Delta
$$
$$
N_1=-\sin\widetilde{\rho}\sin \theta \cos \varphi, \qquad N_2=-\sin
\theta(\sin\varphi\cos\widetilde{\rho}\sin\widetilde{\theta}
+\cos\varphi\cos\widetilde{\theta})
$$
\be\lb{rielo} \widetilde{h}_{11}=\frac{1-(\sin\widetilde{\rho}\sin
\theta \cos \varphi)^2}{\Delta^{1/2}}, \qquad
\widetilde{h}_{22}=\frac{1-\sin^2\theta(\cos\widetilde{\rho}\cos
\widetilde{\theta} \sin \varphi+\cos\widetilde{\theta}
\cos\varphi)^2}{\Delta^{1/2}} \ee
$$
\widetilde{h}_{12}=\frac{\sin\widetilde{\rho}\cos
\widetilde{\theta}+\sin\widetilde{\rho}\sin^2\theta \cos
\varphi\cos\widetilde{\rho}\cos \widetilde{\theta} \sin
\varphi+\cos\widetilde{\theta} \cos\varphi)}{\Delta^{1/2}}
$$
We have that $\widetilde{g}=\Delta^{1/6}(g_{AdS}+h)$ being $h$ given
in (\ref{cuntu}) and $\Delta$ given in (\ref{dotor}).
 The resulting IIB background is in this case
$$
g_{IIB}=\frac{1}{\widetilde{h}_{11}}[\frac{1}{\sqrt{\Delta}}(D\phi_1)^2+\sqrt{\Delta}(D\phi_2)^2]
+e^{2\phi_D/3}\widetilde{g}_{\mu\nu}dx^{\mu}dx^{\nu}
$$
\be\lb{IIB}
B=\frac{\widetilde{h}_{12}}{\widetilde{h}_{11}}D\phi_1\wedge
D\phi_2+D\phi_1\wedge A_1\qquad
e^{2\Phi}=\frac{e^{2\phi_D}}{\widetilde{h}_{11}}, \qquad C^{0}=N_1
\ee
$$
C^2=-(N_2-\frac{\widetilde{h}_{12}}{\widetilde{h}_{11}}N_1)D\phi_1\wedge
D\phi_2-D\phi_1\wedge A_3
$$
$$
C^4=\frac{1}{6}C^3\wedge D\phi_1 + d^4 +\widehat{d}^3\wedge D\phi_2
$$
$$
D\phi_1=d\phi_1
$$
The one forms $A_i$ are defined in (\ref{foron}). The 3-form
$\widehat{d}^3$ and the four form $d^4$ takes values in the
eight-dimensional metric and are determined by imposing that the
five form field strength $F_5=dC_4$ is self-dual, that is
$F_5=\ast_{10} F_5$. In our case we have that
$$
C^3= - \frac{k \sin u}{3} \sinh^3 \rho dt \wedge du\wedge dv ,
\qquad \frac{1}{6}d (C^3\wedge D\phi_1 )=-\frac{k}{6} \sin u \sinh^2
\rho d\rho\wedge dt \wedge du\wedge dv\wedge  d\phi_1
$$
and the self-duality condition is satisfied if $d^4$ is zero and
$$
C^4=\frac{1}{6}(C^3\wedge d\phi_1 +\sqrt{\Delta}\widehat{C}^3\wedge
D\phi_2)
$$
being $\widehat{C}^3$ such that $d\widehat{C}^3=\sqrt{\Delta}\ast_8
dC^3$. The hodge star operation $\ast_8$ concerns to the
8-dimensional metric $\widetilde{g}$. By comparing these expressions
with (\ref{orlo}) we obtain that
$$
F=\widetilde{h}_{11}, \qquad
B_{12}=\frac{\widetilde{h}_{12}}{\widetilde{h}_{11}}, \qquad
B_{1}=B_2=0,
$$
$$
\widetilde{b}=\widetilde{c}=0, \qquad
C_{12}=\frac{\widetilde{h}_{12}}{\widetilde{h}_{11}}N_1-N_2, \qquad
C_{1}=A_3, \qquad C=N_1
$$
and it follows that all these quantities are defined by formulas
(\ref{rielo}). From (\ref{transformoso}) and (\ref{transformoso2})
we obtain directly the explicit deformed fields. The procedure is
finished.

\end{document}